\documentclass[12pt]{article}
\usepackage{amssymb,amsmath}
\usepackage{fullpage}
\usepackage{graphicx}

\def\e{{\rm e}}

\def\d{\partial}

\newcommand{\be}{\begin{equation}}
\newcommand{\ee}{\end{equation}}
\newcommand{\bea}{\begin{eqnarray}}
\newcommand{\eea}{\end{eqnarray}}
\newcommand{\bg}{\begin{gather}}
\newcommand{\eg}{\end{gather}}
\newcommand{\bseq}{\begin{subequations}}
\newcommand{\eseq}{\end{subequations}}

\newcommand{\bpm}{\begin{pmatrix}}
\newcommand{\epm}{\end{pmatrix}}

\title{\Large \bf Signatures of unstable semiclassical trajectories
  in tunneling}
\date{}
\author{D.G.~Levkov$^{a}$\footnote{levkov@ms2.inr.ac.ru}, 
        A.G.~Panin$^{a}$\footnote{panin@ms2.inr.ac.ru},
        S.M.~Sibiryakov$^{b,a}$\footnote{sergey.sibiryakov@epfl.ch}\\
$^a${\small  Institute for Nuclear Research
  of the Russian Academy of Sciences,}\\
{\small 60th October Anniversary Prospect 7a, Moscow 117312, 
  Russia}\\
$^b${\small Institut de Th\'eorie des Ph\'enomenes Physiques,
EPFL, CH-1015, Lausanne, Switzerland}
}
\begin{document}
\maketitle
\begin{abstract}
It was found recently that processes of multidimensional 
tunneling are generally described at high energies by unstable
semiclassical trajectories. We study two observational signatures
related to the instability of trajectories. First, we find an
additional power-law dependence of the tunneling probability on the
semiclassical parameter as compared to the standard case of potential
tunneling. The second signature is substantial widening of the
probability distribution over final--state quantum numbers. These
effects are studied using modified semiclassical technique which
incorporates stabilization of the tunneling trajectories. The
technique is derived from first principles. We obtain expressions for
the inclusive and exclusive tunneling probabilities in the case of
unstable semiclassical trajectories.
We also investigate the ``phase transition'' between the cases of
stable and unstable trajectories across certain ``critical'' value
of energy.  
Finally, we derive the relation between the semiclassical
probabilities of tunneling from the low--lying and highly excited
initial states. This puts on firm ground a conjecture made
previously in the semiclassical description of collision--induced
tunneling in field theory.
\end{abstract}

\newpage
\section{Introduction}
\label{sec:introduction}
Tunneling in systems with several degrees of freedom is an
exceptionally rich subject of
investigation~\cite{Creagh:1998,Tomsovic:2001}.
The features and probability of multidimensional tunneling
depend crucially on the properties of underlying system,
or rather on the degree of complexity of its classical dynamics. In
particular, expressions for the tunnel splittings of energy levels
are
qualitatively different in the cases of
integrable~\cite{Miller:2001,Meyer:1991,Creagh:1994}
and near-integrable~\cite{Wilkinson:1986,Takada:1994,Creagh:2001,
Creagh:2006} dynamics. The other drastically different case, tunneling
in irregular (chaotic or mixed) systems, has been a subject of 
continuous theoretical~\cite{Bohigas:1993,Doron:1995,Shudo:1995,
Creagh:1999,Mouchet:2001,Ribeiro:2004,Levkov:2007e,
Backer:2008} and experimental~\cite{Dembowski:2000,Hensinger:2001,
Steck:2001,Backen:2008} research for the last few decades.

The basic concept in multidimensional tunneling is {\it dynamical
  tunneling}~\cite{Miller,Heller:1981}.
It is related to the classical dynamics and reflects the fact 
that transitions of a multidimensional
system between the in- and out- regions of phase space may be classically
forbidden even if there is {\it no} energy barrier separating the
regions. In this case the quantum probability ${\cal P}$ of transition
is on general grounds exponentially suppressed, 
\begin{equation}
  \label{eq:9}
  {\cal P} = A\, \mathrm{e}^{-F/\hbar}\;,
\end{equation}
where $F$ and $A$ are the suppression exponent and prefactor
respectively. The transition itself is called dynamical
tunneling~\cite{Heller:1981}, since the reasons for its exponential
suppression are hidden in the particularities of classical
dynamics.

A new mechanism of dynamical tunneling has been independently
discovered in Refs.~\cite{Onishi:2003,Takahashi:Ikeda}
and~\cite{Bezrukov:2003yf}. It governs tunneling in non--separable
systems with multiple degrees of freedom at energies exceeding certain
{\it critical energy} $E_c$. The value of the latter energy depends
on the details of the system dynamics but is always greater than the 
height of the potential barrier between the in- and out- states of the  
process. The new mechanism is {\it general}:
it is relevant for tunneling in
regular~\cite{Bezrukov:2003yf,Levkov:2007a,Levkov:2007prl} and
irregular~\cite{Onishi:2003,Levkov:2007e} scattering problems, for
transitions in time-dependent one--dimensional
potentials~\cite{Takahashi:Ikeda,Takahashi:2006,Takahashi:2008}, in
the case of chaotic tunneling\footnote{In chaotic case the new
  mechanism implies anomalously weak falloff of particle wave function
  in some parts of classically forbidden region (``plateau
  structure''~\cite{Shudo:1995,Onishi:2003}). This behavior leads to
  anomalously large tunneling probabilities, the effect
  known as {\it chaos--assisted
    tunneling}~\cite{Bohigas:1993}. }~\cite{Shudo:1995,Shudo:2008}. 
Another example  emerges in field theory where the new mechanism is
generically inherent in the processes of collision--induced tunneling
at high energies~\cite{Bezrukov:2003er,Levkov:2004tf}. 

The defining characteristics of the new mechanism have been given
within the semiclassical approach. It was noted that the semiclassical
trajectories describing tunneling transitions acquire qualitatively
new properties at $E>E_c$. Instead of connecting directly the in- 
and out- regions of the process, the trajectories end up performing
unstable motion on the boundary between the regions. In the simplest
case of two degrees of freedom this unstable motion proceeds along the
periodic orbit describing oscillations on top of the saddle point of
the potential. Following Ref.~\cite{Bezrukov:2003yf}, we call the
latter orbit {\it sphaleron}\footnote{This term is standard in
  field theory~\cite{Klinkhamer:1984di}; it is based on classic Greek
  adjective $\sigma\varphi\alpha\lambda\epsilon\rho o \varsigma$ ---
  ``ready to fall.''}  (or simply {\it unstable periodic
  orbit}).

In general case of systems with more than two degrees of freedom the
boundary between the in- and out- regions is normally hyperbolic
invariant manifold (NHIM)~\cite{Wiggins:2001}; the trajectories in the
new mechanism get attracted to this manifold. In this case we use the
term sphaleron in the sense equivalent to NHIM.

Due to the above property of the
trajectories, tunneling at $E>E_c$ proceeds in two stages. First, the
long-living sphaleron ``state'' gets created. Second, the sphaleron
decays into the final asymptotic region with the probability of order
one. The overall transition remains exponentially suppressed, since
creation of the sphaleron costs exponentially small
probability factor. We call the overall transition {\it
  sphaleron--driven tunneling}.

The aim of the present paper is twofold. First, we analyze the
possibility of direct experimental observation of the the mechanism of
sphaleron--driven tunneling. To the best of our knowledge, such
observation has not been performed so far. We 
study two signatures of the new mechanism which may be
helpful in future experiments. Second, we systematically develop
modified semiclassical method for the calculation of tunneling
probability in the sphaleron--driven case.

We discuss two  experimental signatures of the new tunneling mechanism. In 
Ref.~\cite{Levkov:2007prl} we have found that the probability of 
sphaleron--driven tunneling contains additional power-law dependence
on $\hbar$ as  compared to the ordinary case of potential tunneling.
The additional factor is $\hbar^{1/2}$ in the case of inclusive
tunneling processes, i.e. processes without specification of the
out-state. In this paper we review the result of
Ref.~\cite{Levkov:2007prl} and extend  the analysis to  the new case
of  exclusive processes, i.e. processes with fixed out-state quantum
numbers. We show that the additional factor  is  $\hbar$ in this
case. For example, consider two--dimensional inclusive
processes. Then, the prefactor $A$ in Eq.~(\ref{eq:9}) 
 is proportional to $\hbar^{1/2}$
and $\hbar$ in the cases of potential and sphaleron--driven tunneling
respectively. For exclusive processes this dependence
is $\hbar$ ($\hbar^2$) in the potential (sphaleron--driven) case.

It is important to stress that the dependence of the tunneling
probability on $\hbar$ can, in principle, be studied
experimentally. Indeed, the semiclassical parameter, which we denote
by $\hbar$ for convenience, is in fact a certain dimensionless
combination of the Planck constant and parameters characterizing
the system. Changing the latter parameters one varies the value of
effective $\hbar$. 

The second manifestation of the new mechanism is 
spreading of the out-state of the tunneling process over an 
anomalously wide range of quantum numbers. This effect was originally
observed in  Ref.~\cite{Takahashi:Ikeda} in the case of
a one-dimensional system with time--dependent  potential; here we
show that it is present in the multidimensional case,
cf. Ref.~\cite{Takahashi:2008}. Physically, the widening of the
out-state is related to the fact that the intermediate sphaleron 
orbit is classically unstable; thus, classical trajectories
describing sphaleron decay spread exponentially over phase space. In
quantum case this corresponds to the final state wave function which
is almost constant in some region of quantum numbers.

In the second part of this paper we develop the modified semiclassical
technique which is essential in the case of sphaleron--driven tunneling.
The motivation for the new technique becomes clear if we try to apply
the standard method of complex trajectories to the problem of
inclusive
sphaleron--driven tunneling in the scattering setup. Since 
the overall time interval of the scattering problem is infinite, 
one generically finds {\it two} different trajectories corresponding
to the two stages of the new tunneling mechanism: one trajectory
starts in the in-region and tends to the sphaleron orbit as
$t\to +\infty$, and the second trajectory starts at the sphaleron
at $t\to -\infty$ and arrives into the out-region. The first of these
trajectories is unstable: it can be destroyed by infinitesimally 
small changes in the initial Cauchy data.\footnote{Below we always
  refer to this kind of instabilities.} It is problematic to 
find unstable trajectories numerically. Besides, one wonders how to join
the two trajectories in order to describe the overall two--stage
process. Finally, it is not clear how to calculate the prefactor $A$ 
of the tunneling probability. Indeed, the standard formula for the
prefactor deals with the linear perturbations above the tunneling
trajectory. When the trajectory in question is unstable these
perturbations grow exponentially with time. Then the standard formula
gives $A = 0$, which is obviously incorrect.

Our modified semiclassical method overcomes the above difficulties. 
The main idea of the method was proposed in
Refs.~\cite{Bezrukov:2003yf,Levkov:2007prl}; here we present its
detailed derivation. The modified method is summarized as
follows. We evaluate the Feynman path integral for the tunneling
amplitude in two steps. First, we restrict the integral to paths
which arrive into the out-region in a {\it fixed} time interval
$\tau$. Second, we integrate 
over $\tau$. The integration at step $1$ can be  done by the standard
saddle-point method, since all trajectories at finite  $\tau$ are
stable and interpolate between the in- and out- regions. On the other
hand, the ordinary integral over $\tau$ at  step 2 should be evaluated
with care. In particular, we find that in the case of
sphaleron--driven tunneling this integral is  saturated in
the region $\tau \to +\infty$, rather than at the saddle point  at
finite $\tau$.

The above manipulations with the path integral lead to a
notably simple semiclassical description of sphaleron--driven 
tunneling. Namely, we show that the constraint in the path 
integral leads to the deformation of the semiclassical equations of motion
with the {\it imaginary} term proportional to the small parameter
$\epsilon = \epsilon(\tau)$. The evaluation of the integral over
$\tau$ corresponds to  taking the limit $\epsilon \to +0$ in both
cases of stable and unstable trajectories. However, the resulting
expressions for the tunneling probability are different in the two
cases, since the integral over $\tau$ is saturated in two different
regions. In particular, the probability formula in the case of
sphaleron--driven tunneling involves additional factor
$\hbar^{1/2}$ mentioned above. We call the modified semiclassical
technique by the {\it method of $\epsilon$--regularization}. 

The new mechanism of tunneling
is relevant only at sufficiently high energies, $E>E_c$. 
Below $E_c$ transitions proceed via the ordinary mechanism
of potential tunneling. In accordance with our results, the semiclassical
expression  for the prefactor $A$ changes discontinuously across the
critical energy. In particular, in the inclusive case in two
dimensions $A\propto 
\hbar^{1/2}$ and $\hbar$ at $E<E_c$ and  $E>E_c$ respectively. This
implies that both  expressions break down in a small vicinity of
$E_{c}$, where the correct {\it uniform} approximation should be
invoked. In the present paper we derive the required formula, which is
continuous and applicable in the entire energy range. At $|E-E_{c}|
\gg \hbar^{1/2}$ this formula coincides with the respective
``potential'' and ``sphaleron--driven'' semiclassical expressions.
In this regard it is similar to the uniform
approximation~\cite{Creagh:2006} for the tunnel level splitting at the
point of transition from integrable to near-integrable systems.

Next, we study semiclassically exclusive tunneling processes in the
sphaleron--driven case. We find that the new mechanism leads to 
proliferation 
 of complex
trajectories describing a given exclusive process. These trajectories
form an infinite sequence and have the following structure: they 
get attracted to the sphaleron orbit, follow it for an integer number
of periods and then slide away. The tunneling amplitude is the sum of
the contributions of all these trajectories. In analogy to the
case of inclusive probability
the sum is saturated by the trajectories which spend an
infinite time at the sphaleron. It is worth noting 
that, in contrast to the inclusive case, the individual trajectories
describing exclusive process are stable. Thus, a priori, there is
no need for the modified semiclassical technique in the
case of exclusive transitions. Still, in this paper we demonstrate
that our modified technique turns out to be useful in finding and
organizing the tunneling trajectories. It also provides the link
between the semiclassical descriptions of inclusive and exclusive
processes.

Finally, for the sake of completeness 
we study the processes of tunneling from low-lying
in-states. Naively, such states and hence the corresponding
tunneling processes cannot be described semiclassically; still, we show
that the probabilities of these processes are given by the
semiclassical formula~(\ref{eq:9}). In addition, we show that the
suppression exponent and 
prefactor of tunneling from the low--lying states can be obtained as
certain limits of the corresponding quantities in the case of highly
excited states. The limiting relation for the suppression exponent is
known in field theory as the Rubakov--Son--Tinyakov
conjecture~\cite{Rubakov:1992ec}; it plays an important role in the
semiclassical description of collision-induced
tunneling~\cite{induced}. We 
prove this conjecture in quantum mechanical setup. Our
limiting formula for the prefactor shows that the 
probability  of tunneling from the low--lying states contains
a factor $\hbar^{-1/2}$ as compared to the
case of highly excited in-states. 

We illustrate our findings by considering tunneling transitions in 
a simple model with two degrees of freedom. For this model we compare
predictions of the modified semiclassical technique with the exact
quantum mechanical results. The latter are extracted from the
numerical solution of the stationary Schr\"odinger equation. We find
perfect agreement between the two sets of results. 

The outline of the paper is as follows. After presenting the model in
Sec.~\ref{sec:model} we summarize the experimental signatures of 
sphaleron--driven tunneling in
Sec.~\ref{sec:summary-results}. In Sec.~\ref{sec:modif-semicl-techn}
we introduce the modified semiclassical technique: we
review the standard semiclassical method in Sec.~\ref{sec:semicl-form-tunn},
introduce $\epsilon$--regularization in
Sec.~\ref{sec:unst-traj} and derive the uniform formula in
Sec.~\ref{sec:unif-appr}. Application of the modified semiclassical
method  to the exclusive tunneling processes is discussed  
in Sec.~\ref{sec:nf}. Finally, we study tunneling from low--lying 
in-states in
Sec.~\ref{sec:limit-small-quantum}. Section~\ref{sec:discussion} 
contains discussion. Technical details are described in appendices.

\section{The model}
\label{sec:model}
\begin{figure}[t]
\centerline{\includegraphics[width=0.5\textwidth]{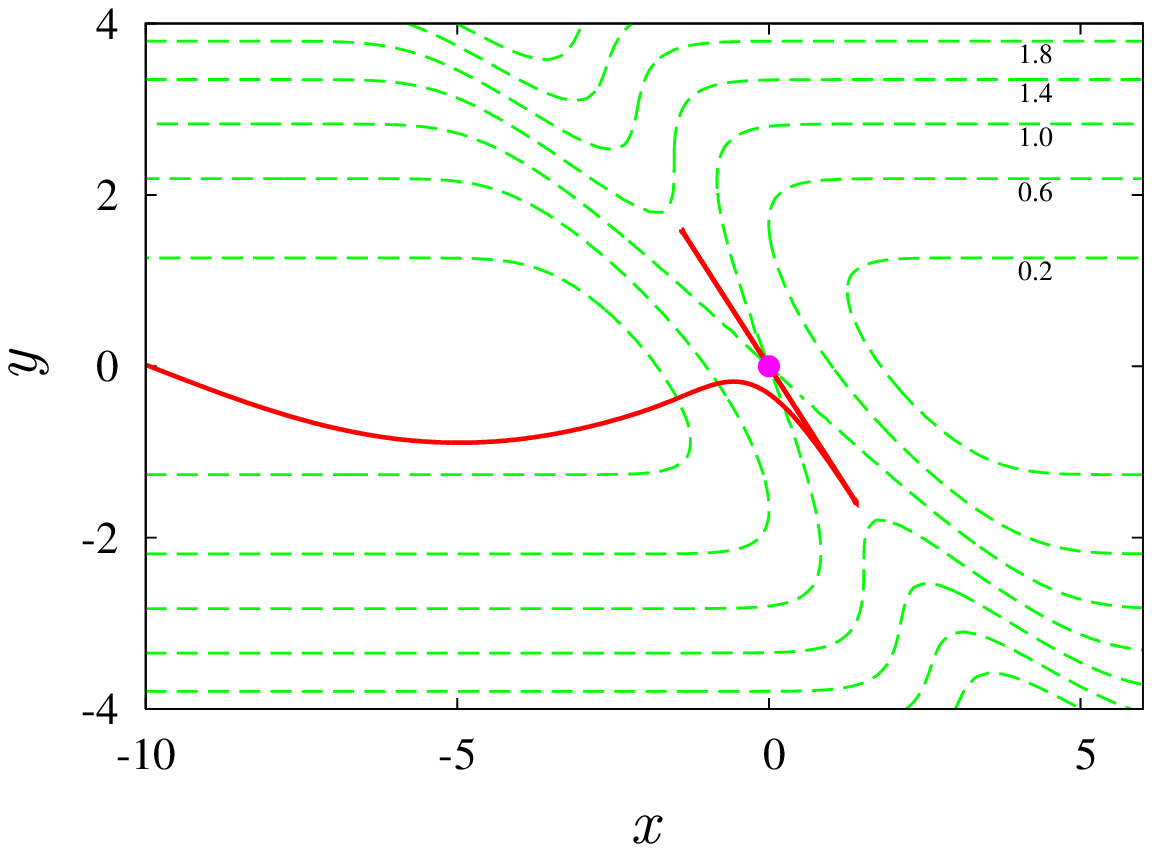}
  \includegraphics[width=0.5\textwidth]{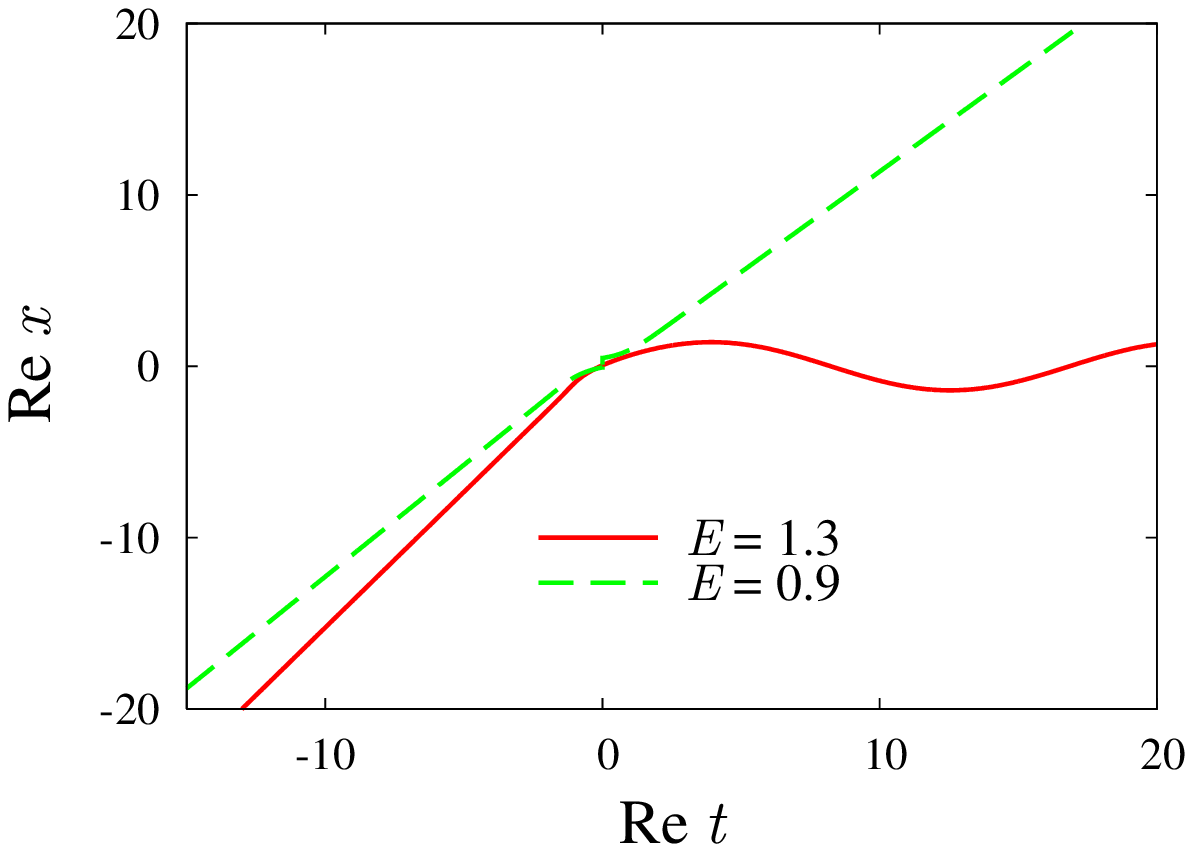}}
\hspace{4cm}(a)\hspace{8cm}(b)
\caption{\label{fig:0} (a) The contour plot of the potential
  (dashed lines) and the real part of the tunneling trajectory 
  at $E=1.3$, $E_y=0.05$ (solid line). The saddle point 
  is marked by the thick dot. (b) Time evolution of
  $\mathrm{Re}\, x$ for two complex trajectories with $E_y = 0.05$ and
  different values of total energy, $E = 0.9$ and $1.3$. 
Note that $E_c(E_y=0.05) \approx 1.1$.}
\end{figure}

We start by introducing the scattering model of
Refs.~\cite{Bonini:1999kj,Bezrukov:2003yf}.
It will be used throughout the paper for illustrative purposes.
The model describes motion of a particle with unit mass in the potential
\begin{equation}
\label{eq:24}
V(x,y) = \omega^2 y^2/2 + \mathrm{e}^{-(x+y)^2/2}. 
\end{equation}
The potential represents two--dimensional harmonic waveguide extended
along the $x$ direction and intersected at an angle by the
potential barrier. The contour plot of the potential is shown in 
Fig.~\ref{fig:0}a.  In this and other figures we use the value
$\omega=1/2$ for the waveguide frequency. Note that potentials similar
to (\ref{eq:24}) typically arise in the studies of collinear chemical 
reactions~\cite{Miller}.

We are interested in tunneling transitions of quantum particle
between the asymptotic regions $x\to -\infty$ and $x\to +\infty$ of
the potential (in- and out- regions respectively). In the in-region
the particle evolves with constant momentum in the $x$ direction 
oscillating along the $y$ axis. The corresponding in-state 
is fixed by the total energy $E$ and the energy of $y$ oscillations
$E_y$. Similarly, the out-state can be fully 
characterized by $E$ and $E_y^f$, where $E_y^f$ is the final oscillator
energy. In what follows we will often omit
the specification of the out-state and consider the total (inclusive)
probability of tunneling into the region $x\to +\infty$.

The height of the potential barrier separating  the in- and
out-  regions is $V_0 = 1$. It is given by the value of the potential
at the saddle point $(x,y) = (0,0)$. At $E<V_0$ the classical transitions
between the regions 
are forbidden energetically, and their underlying mechanism is
potential tunneling.  On the other hand, it is shown
in Ref.~\cite{Bonini:1999kj} that classical over--barrier
transitions between the asymptotic regions take place 
at $E>E_b(E_y)$, where $E_b(E_y)$ is larger than $V_0$. Hence, at
intermediate energies $V_0 < E < E_b(E_y)$ the transitions are in the
regime of dynamical tunneling, which we are interested in.

As we have already discussed in the Introduction, the multidimensional
processes of dynamical tunneling, such as ours, generically proceed via
sphaleron--driven mechanism at sufficiently high energies. Let us
illustrate the new mechanism in the model (\ref{eq:24}) comparing the
behavior  of semiclassical solutions at low and high energies
\cite{Bezrukov:2003yf}. Consider the inclusive tunneling transition
from the state $| E,E_y\rangle$ into the out-region $x\to +\infty$. 
We postpone the consistent formulation of the semiclassical method till
Sec.~\ref{sec:modif-semicl-techn}. The only fact we need here is that any
tunneling process is specified by a certain complex trajectory ---
solution to the (complexified) classical equations of motion. The
latter should interpolate between the in- and out- regions of the
process.

Fig.~\ref{fig:0}b shows the complex trajectories describing
tunneling transitions at $E_y = 0.05$ and two values of total energy,
$E=0.9$ and $1.3$. [The real part of the trajectory with $E=1.3$ is
also depicted in Fig.~\ref{fig:0}a.] The behavior of the two
trajectories is  drastically different.  While the  low--energy
trajectory interpolates between the asymptotic regions $x\to
\pm \infty$,  the solution with $E=1.3$ gets stuck at finite $x$
approaching the unstable periodic orbit as $t\to +\infty$.  The 
latter orbit is precisely the sphaleron discussed   
in the Introduction; it describes oscillations around the saddle point
of the potential, see Fig.~\ref{fig:0}a. Clearly, the high--energy
trajectory of 
Fig.~\ref{fig:0} describes only half of the transition
process, since it does not arrive into the out-region. Trajectory
corresponding to the other half can be 
obtained by adding to the unstable periodic orbit infinitesimally
small momentum in the direction of 
the out-region  and evolving the 
system classically. Thus constructed, the overall semiclassical
evolution involves {\it two} trajectories which describe
creation and subsequent decay of the sphaleron\footnote{There is
  another way to visualize the semiclassical
  evolution~\cite{Takahashi:Ikeda}. One 
  introduces stable and unstable manifolds of the sphaleron
  orbit. These are formed respectively by the trajectories arriving at
  the sphaleron at   $t\to +\infty$ and trajectories starting from it
  at $t\to   -\infty$. Then, the evolution describing  
  sphaleron--driven tunneling is guided in turn by trajectories
  belonging to the stable and unstable manifolds of the sphaleron.}. 
This evolution corresponds to the mechanism of sphaleron--driven
tunneling.

One finds \cite{Bezrukov:2003yf} 
that there exists the critical value $E=E_c(E_y)$ of
total energy which separates the regions of qualitatively different
behavior of tunneling trajectories. Namely, the trajectories
interpolate between the in- and out- regions at $E<E_c(E_y)$ and
approach the sphaleron orbit at $E_c(E_y)< E < E_b(E_y)$. This means
that the mechanism of transition changes from potential to
sphaleron--driven tunneling as the energy crosses the critical
value. From the physical viewpoint $E_c(E_y)$ can be understood as the
energy of ``phase transition'' between the two regimes of tunneling. 
We remark that the energies of sphaleron
orbits and thus the critical energy for the sphaleron--driven
tunneling exceed the height of the potential barrier. Therefore, the
new mechanism is relevant only in the case of dynamical tunneling. 

\begin{figure}
\centerline{\includegraphics[width=0.5\textwidth]{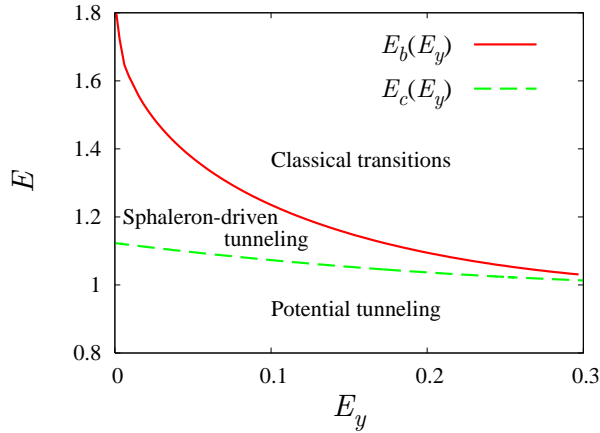}}
\caption{\label{fig:23} Regions in the plane of in--state quantum
  numbers corresponding to the potential and sphaleron--driven
  tunneling mechanisms.}
\end{figure}
The region $E_c(E_y) < E < E_b(E_y)$ corresponding to the 
sphaleron--driven tunneling in the model (\ref{eq:24}) is shown in
Fig.~\ref{fig:23}. The value of $E_c(E_y)$ is found numerically by
computing the complex trajectories at different energies and
investigating their stability.

\section{Experimental signatures}
\label{sec:summary-results}
In this Section we show that the mechanism of sphaleron--driven
tunneling leads to two observable effects which in principle can be
used for identification of the new mechanism in future
experiments. Both effects are related to the fact 
that the relevant semiclassical solutions are unstable. We  illustrate
our findings in the model (\ref{eq:24}) using the exact quantum
mechanical results. The exact calculations of this and the subsequent
sections are based  on the numerical solution of time--independent
Schr\"odinger equation, see
Refs.~\cite{Bonini:1999kj,Levkov:2007e,f90code} for the numerical 
method and Fortran 90 code.

The first signature of sphaleron--driven tunneling is the direct
consequence of the semiclassical analysis which will be presented in
Sec.~\ref{sec:modif-semicl-techn}. We find that the sphaleron--driven
mechanism changes the power--law dependence of the transmission
probability on $\hbar$ compared to the case of potential tunneling. 
To be concrete, let us discuss inclusive
tunneling transitions in the model (\ref{eq:24}). 
Then, the prefactor $A$ of  the probability is
proportional to $\hbar^{1/2}$ and $\hbar$ in the cases of potential
and sphaleron--driven tunneling respectively. 

The physics behind the additional power--law suppression becomes clear if
one uses the qualitative analogy with the classically allowed creation
of unstable state. The latter process considered at the classical
level requires fine tuning of the Cauchy data. As a
consequence, only a small part of the in-state wave function 
contributes into the amplitude of the process. This results in the 
additional suppression of the probability. On general
grounds one expects similar formal  suppression in the case of
sphaleron--driven tunneling. 

Experimentally, one can try to observe the unusual power--law
dependence on $\hbar$  by analyzing the probability graph ${\cal
  P}(\hbar)$. 
Note that the value of the semiclassical parameter which 
we denote by $\hbar$ is, in principle, adjustable in
experiments. Indeed, the magnitude of quantum fluctuations
is measured by the dimensionless ratio of the Planck constant to
a certain combination of parameters characterizing the
system. Changing the latter parameters in an appropriate way, one 
alters the value of the semiclassical parameter $\hbar$ without
affecting the classical dynamics of the system.

To illustrate this point consider the system (\ref{eq:24}). The
key quantity which enters into the semiclassical expansion is the
ratio of the action of the system to the Planck
constant. Restoring the dimensionful units we obtain
\[
\frac{S}{\hbar_0}=\frac{1}{\hbar_0}\int dt 
\left(\frac{M{\dot{\boldsymbol{x}}}^2}{2}-\frac{M\omega_0^2y^2}{2}
-V_0\e^{-(x+y)^2/2L^2}\right)\;,
\]
where $\hbar_0$ stands for the physical Planck constant. In terms of
dimensionless variables this expression reads
\[
\frac{S}{\hbar_0}=\frac{1}{\hbar}\int d\tilde t
\left(\frac{{\dot{\tilde{\boldsymbol{x}}}}^2}{2}
-\frac{\omega^2\tilde y^2}{2}-\e^{-(\tilde x+\tilde y)^2/2}
\right)\;,
\]
where $\hbar=\hbar_0/\sqrt{MV_0L^2}$,
$\omega^2=ML^2\omega_0^2/V_0$. The effective frequency $\omega$
completely determines the classical dynamics. On the other hand, the
effective Planck constant $\hbar$ is given by an independent
combination of parameters.

\begin{figure}[htb]
\centerline{\includegraphics[width=0.6\textwidth]{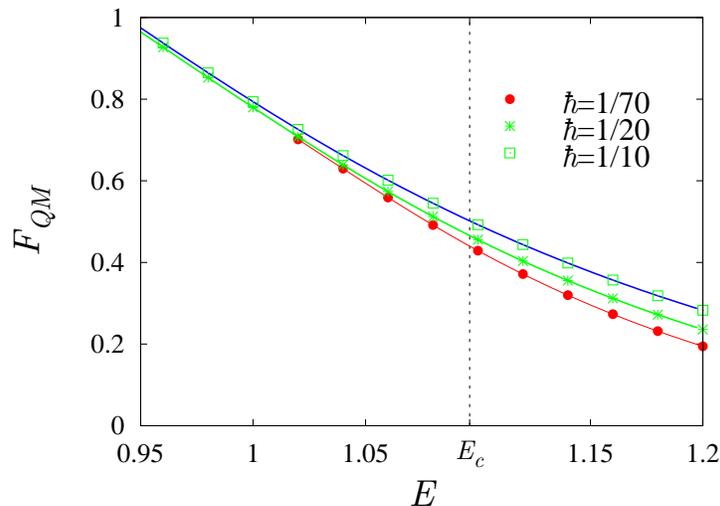}}
\caption{\label{fig:2} The probability logarithm $F_{QM} = - \hbar
  \log({\cal  P}/\hbar^{1/2})$ plotted as a function of total energy for 
  several values of $\hbar$ and $E_y = 0.05$.  Points represent
  the exact quantum mechanical results; the interpolating lines
  are drawn for convenience. The critical energy is shown by
  dashed vertical line.} 
\end{figure}

One can hardly hope to extract directly the additional factor $\hbar^{1/2}$
from the experimental data on transmission probability:  it is almost
impossible to identify the weak power--law dependence on top of the
leading semiclassical exponent.  We suggest an indirect method. Namely,
consider the quantity  
$$
  F_{QM} = - \hbar \log({\cal P}/\hbar^{1/2}).
$$
In the regime of potential tunneling ($A\propto \hbar^{1/2}$) 
$F_{QM}$ is almost independent of $\hbar$ at small values of the
latter. On the other hand, $F_{QM} \simeq -\frac12\hbar
\log \hbar + \mbox{const}$ whenever 
the new tunneling mechanism is involved. The difference between the
two cases is seen in Fig.~\ref{fig:2}, where the dependences of 
$F_{QM}$ on the total energy $E$ are shown for several values of
$\hbar$. The graphs in Fig.~\ref{fig:2} coincide at energies somewhat
smaller than $E_c$ (say, at $E\lesssim 1$), while at $E>E_c$ a clear
difference between the graphs appears. We remark that the change
in the behavior of the exact tunneling probability is gradual, in
spite of the fact that the complex trajectories have distinct
structure at $E<E_c$ and $E>E_c$. We discuss this point and
derive the appropriate uniform formula in Sec.~\ref{sec:unif-appr}. 

Another signature of the sphaleron--driven mechanism
was first pointed out in 
Refs.~\cite{Takahashi:Ikeda,Takahashi:2008}. 
One notes that the second stage of sphaleron--driven transition,
the decay of the sphaleron orbit, proceeds classically and does not 
affect
the leading suppression exponent $F$ of the probability.  In addition,
the sphaleron, being unstable, can evolve at the classical level into
the out-states with {\it different} values  of oscillator energy   
$E_y^{f}$. Classical trajectories corresponding to these
evolutions are obtained by adding small momentum in the direction
of the out-region at different points of the sphaleron orbit. One
concludes that in the case of sphaleron--driven tunneling the
distribution over final oscillator energies is almost constant in
some region $E_{y,1}^{f} < E_y^f < E_{y,2}^f$.  The latter region 
corresponds to the decays of the sphaleron along different
classical trajectories. 

Note that the above feature is in sharp contrast with the properties of
final states in the standard case of potential tunneling. Namely, in
a typical situation the complex trajectory describing transmission
through the barrier is unique, and the corresponding out-state wave
function forms sharply peaked Gaussian distribution around some
optimal value $E_y^f = \langle E_y^f \rangle$.

\begin{figure}[htb]
\centerline{\includegraphics[width=0.5\textwidth]{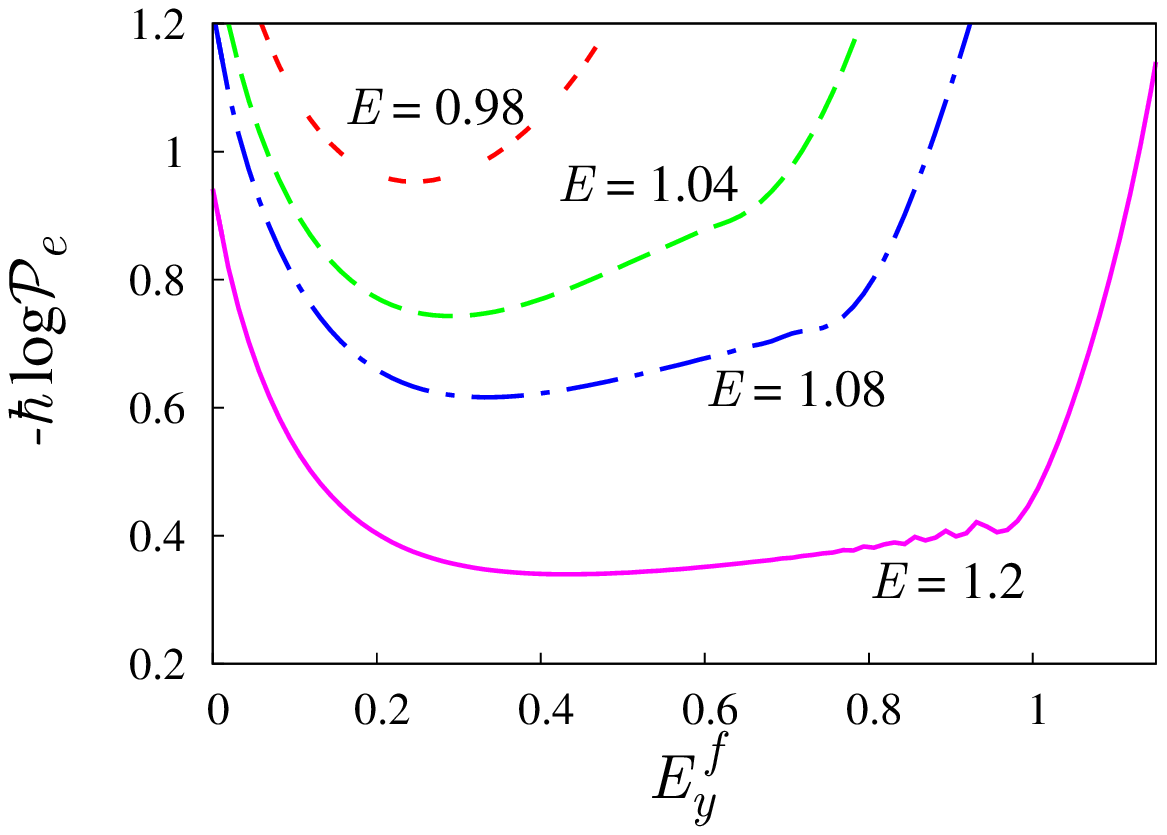}
\includegraphics[width=0.5\textwidth]{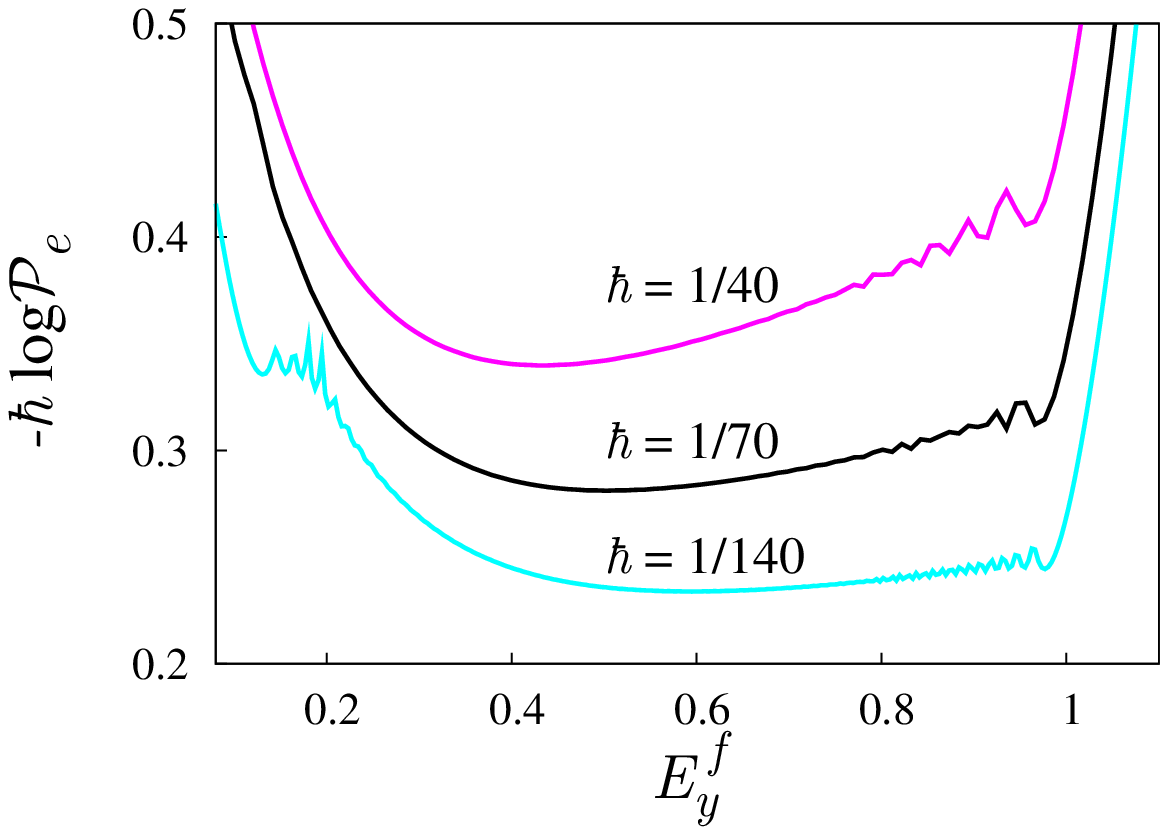}}
\hspace{4.2cm}(a)\hspace{8cm}(b)
\caption{\label{fig:3} Distributions of the logarithm of exclusive
  tunneling probability over the out-state quantum number $E_y^f$. The graphs
  are plotted at $E_y = 0.05$ and: (a) $\hbar = 1/40$ and 
  four values of total energy; (b) $E =
  1.2$ and different values of $\hbar$. Note that $E_c(E_y=0.05)
  \approx 1.1$.}
\end{figure}

To illustrate explicitly the effect of anomalously wide final
states in the case of sphaleron--driven tunneling,
we consider transitions between the {\it exclusive} in- and out-
states which have definite energies of $y$ oscillator, $E_y$ and
$E_y^{f}$ respectively, and the same total energy $E$. Then, we fix
the initial state ($E$ and  $E_y$) and analyze the dependence of the
exact exclusive probability ${\cal   P}_{e}$ on $E_y^f$. This
dependence is shown in Fig.~\ref{fig:3} in logarithmic scale for
several values of $E$. One
immediately sees in Fig.~\ref{fig:3}a that the width of the
out-state distribution grows as the value of total energy
approaches $E_c(E_y)$ from below. In particular, a flat plateau
gradually develops in the right side of the distribution. At
energies higher than critical the plateau is wide
and corresponds to the maximum probability of tunneling. Moreover,
the graphs become flatter as the value of $\hbar$
decreases, see Fig.~\ref{fig:3}b. 

One sees another feature of
the new tunneling mechanism: the short--scale fluctuations in the
right and left parts of the plateaux in Fig.~\ref{fig:3}b. This is the
hallmark of quantum interference phenomena, which seem to be important
for complete understanding of exclusive processes at $E>E_c(E_y)$.
We discuss this point in Sec.~\ref{sec:nf}. 

\section{Modified semiclassical technique}
\label{sec:modif-semicl-techn}
In this section we describe the semiclassical technique 
adapted to the analysis of sphaleron--driven tunneling. 
We start by reviewing the path integral derivation of 
the standard method of complex trajectories~\cite{Miller}.  
Then, we manipulate with the path integral and obtain
the modified semiclassical expressions in the case of
sphaleron--driven tunneling.

For simplicity we assume that the system undergoing tunneling
transition is similar to the model of Sec.~\ref{sec:model}. Throughout
this section we consider tunneling between the asymptotic in- and out-
regions of two--dimensional waveguide potential, where the in-state of
the process $|E,\, E_y\rangle$ is fixed and the final state is
inclusive. It is worth noting that both the standard and modified
semiclassical methods are completely general and the semiclassical
formulas of this section can be generalized to other systems. In
particular, the modified method was applied to the case of chaotic
tunneling in Ref.~\cite{Levkov:2007e} and to field theory in
Ref.~\cite{Bezrukov:2003er}.

\subsection{The standard method}
\label{sec:semicl-form-tunn}
Semiclassical calculations within the method of complex
trajectories proceed as follows. One reduces the problem of computing
the tunneling probability to a problem of finding the complex
trajectory $\boldsymbol{x}^{(s)}(t)$ --- complex solution to the
classical equations of motion with certain boundary conditions. In
practice this solution is obtained numerically. Then, tunneling
probability is given by Eq.~(\ref{eq:9}) where $F$ and $A$ are certain
functionals of $\boldsymbol{x}^{(s)}(t)$. In this section we derive
the boundary conditions for $\boldsymbol{x}^{(s)}(t)$ and expressions
for the functionals $F$, $A$ in the standard case of potential
tunneling.

In order to compute the inclusive tunneling probability we first
obtain the semiclassical expression for the final state of the
tunneling process. The wave function $\Psi_f$ of the final state has
the form,
\begin{equation}
\label{eq:1}
\Psi_f(\boldsymbol{x}_f) = \langle \boldsymbol{x}_f |\,
\mathrm{e}^{-iH (t_f - t_i)/\hbar}\, | E,\,E_y \rangle = 
\int d\boldsymbol{x}_i \, \langle \boldsymbol{x}_f |\,
\mathrm{e}^{-iH (t_f - t_i)/\hbar} \,| \boldsymbol{x}_i \rangle 
\,\Psi_i(\boldsymbol{x}_i) \;,
\end{equation}
where $\boldsymbol{x} = (x,\, y)$, while $\Psi_i(\boldsymbol{x}_i) =
\langle \boldsymbol{x}_i | E,\, E_y\rangle$ is the in-state
wave function. Below we assume implicitly 
that $\Psi_i$ and $\Psi_f$ have support in the in- and out-
asymptotic regions respectively. One uses the path integral
representation  for the quantum propagator in Eq.~(\ref{eq:1}) and
writes,
\begin{equation}
  \label{eq:7}
  \Psi_f(\boldsymbol{x}_f) = \left. \int
  d\boldsymbol{x}_i \, \Psi_i(\boldsymbol{x}_i)\,
  \int [d\boldsymbol{x}]\right|_{\boldsymbol{x}_i}^{\boldsymbol{x}_f} \,   
  \mathrm{e}^{iS[\boldsymbol{x}]/\hbar}\;,
\end{equation}
where $S$ stands for 
the classical action of the system. One observes that at small $\hbar$
the integrand in Eq.~(\ref{eq:7}) contains fast--oscillating exponent;
thus, the respective integral can be evaluated by the saddle--point
method. To keep the discussion short, we defer the details of the
saddle--point integration to appendix~\ref{sec:eval-pre-expon}; here
we quote the result. One finds the extremum of the leading exponent in
Eq.~(\ref{eq:7}), which is 
represented by the trajectory $\boldsymbol{x}^{(s)}(t)$ going between  
the in- and out- regions. This trajectory is generically complex. It
satisfies the
classical equations of motion $\delta S/\delta \boldsymbol{x}(t) = 0$
and arrives at a given point $\boldsymbol{x}=\boldsymbol{x}_f$ at $t =
t_f$. The boundary conditions at $t = t_i$ for
$\boldsymbol{x}^{(s)}(t)$ are  obtained from the saddle--point
integration over $\boldsymbol{x}_i$; they fix the values of 
in-state quantum numbers, 
\begin{equation}
  \label{eq:6}
  E_y = (\dot{y}_i^2  + \omega^2 y_i^2)/2\;, \qquad \qquad
  E = \dot{x}_i^2/2 + E_y\;,
\end{equation}
where the subscript $i$ marks the quantities evaluated at $t = t_i$.
For brevity we omit the superscript $(s)$
of the semiclassical trajectory in Eq.~(\ref{eq:6}) and in what follows.

As the result of integration in Eq.~(\ref{eq:7}), one finds the
semiclassical wave function of the final state,
\begin{equation}
  \label{eq:8}
   \Psi_f(\boldsymbol{x}_f) =
   D^{-1/2}\cdot\exp\left\{{\frac{i}{\hbar}(S[\boldsymbol{x}] 
    + B_i[\boldsymbol{x}]) +  \frac{i\pi}{4}}\right\}\;,
\end{equation}
where 
$B_i$ is the in-state contribution to the exponent and $D$
represents the prefactor determinant, see Eqs.~(\ref{eq:3}),
(\ref{eq:B5}) in appendix~\ref{sec:eval-pre-expon} for explicit
expressions. 
Note that the leading
exponent $S+B_i$ in Eq.~(\ref{eq:8}) is evaluated on the saddle--point
trajectory $\boldsymbol{x}(t)$.

The inclusive probability ${\cal P}$ of transmission is equal to the
total flux\footnote{We use the in-state with the unit flux
  normalization.} of the out-wave~(\ref{eq:8}) through the distant line  
$x_f =  x_f^{(0)}$, where $x_f^{(0)}$ is large
and positive. Semiclassically, one writes,
\begin{equation}
  \label{eq:19}
   {\cal P} = \int dy_f \, |\Psi_f(\boldsymbol{x}_f)|^2 \,\mathrm{Re}\,
   \dot{x}_f\;, 
\end{equation} 
where we used the fact that $\partial S/\partial x_f =
\dot{x}_f$. The integral in the above expression is again computed
by the saddle--point technique. In appendix~\ref{sec:eval-pre-expon} we
show that the extremum of the leading exponent in Eq.~(\ref{eq:19}) is
achieved when 
\begin{equation}
  \label{eq:20}
  x_f = x_f^{(0)}\;, \qquad\qquad \mathrm{Im}\, \dot{y}_f =
  \mathrm{Im}\, y_f = 0\;. 
\end{equation}
Equations~(\ref{eq:20}) fix the boundary conditions at $t = t_f$
for the semiclassical trajectory. 

After the saddle--point integration in Eq.~(\ref{eq:19}) one finally
arrives at the familiar semiclassical expression~(\ref{eq:9}) for the
tunneling probability, where the leading exponent is 
\begin{equation}
  \label{eq:26}
F_{pot} = 2\, \mathrm{Im}(S + B_i)\;.
\end{equation}
Note that we mark all the standard
semiclassical expressions with the subscript $pot$ which stands for ``potential
tunneling''.

The prefactor $A_{pot}$ is computed
as follows (see appendix~\ref{sec:eval-pre-expon} for the
derivation). 
One finds two independent perturbations $\delta
\boldsymbol{x}^{(1)}(t)$ and $\delta \boldsymbol{x}^{(2)}(t)$ in the
background of the complex trajectory $\boldsymbol{x}(t)$. These
perturbations satisfy the linearized classical equations of motion,
\begin{equation}
  \label{eq:4}
  \delta\ddot{\boldsymbol{x}}^{(n)} + \hat{V}'' (\boldsymbol{x}(t)) \delta
  \boldsymbol{x}^{(n)} = 0\;, \qquad \qquad n=1,2
\end{equation}
with certain Cauchy data\footnote{First, the perturbations are real at
  $t = t_f$. Second, they do not change the value of total energy,
  $\delta  E[\delta\boldsymbol{x}^{(n)}] = 0$. Third, $\Omega(\delta  
  \boldsymbol{x}^{(1)}, \delta \boldsymbol{x}^{(2)}) = 1$, where
  $\Omega$ is the canonical symplectic form.} at $t = t_f$. After  
evolving $\delta \boldsymbol{x}^{(n)}(t)$ back in time from $t = t_f$ 
to $t = t_i$, one computes the prefactor by the
formula\footnote{As discussed in appendix~\ref{sec:eval-pre-expon},
  this formula is canonically covariant.}
\begin{equation}
  \label{eq:21}
  A_{pot} =  \frac{\hbar^{1/2}\omega }{\sqrt{4 \pi\,
      \mathrm{Im} (\delta E_y[\delta\boldsymbol{x}^{(1)}]\cdot \delta E_y^*
             [\delta\boldsymbol{x}^{(2)}] ) }}\;,
\end{equation} 
where the linear functional 
\begin{equation}
\label{eq:67}
  \delta E_y[\delta \boldsymbol{x}] = \dot{y}_i \delta \dot{y}_i +
  \omega^2 y_i \delta y_i
\end{equation}
measures the change in the initial oscillator energy $E_y$ due to
the perturbation $\delta \boldsymbol{x}(t)$. We stress that $\delta
E_y[\delta \boldsymbol{x}^{(n)}]$ involves perturbations in the
in-region, while the  
Cauchy data for $\delta \boldsymbol{x}^{(n)}(t)$ are set at $t = t_f$. 
We also note that the prefactor~(\ref{eq:21}) is explicitly
proportional to $\hbar^{1/2}$; this fact was used in the previous
section.

The standard semiclassical calculation is summarized as
follows. One finds the complex trajectory $\boldsymbol{x}(t)$
satisfying the classical equations of motion with  the boundary
conditions~(\ref{eq:6}),~(\ref{eq:20}). Our numerical method for
finding the trajectory is presented in
appendix~\ref{sec:numerical-method}. The suppression  exponent
$F_{pot}$  of the probability is given by the value of the functional 
(\ref{eq:26}) on the trajectory $\boldsymbol{x}(t)$. Then, one
considers the linear perturbations around the semiclassical trajectory
and finds the prefactor $A_{pot}$ using the expression (\ref{eq:21}).

\begin{figure}[htb]
  \centerline{\includegraphics[width=0.5\textwidth]{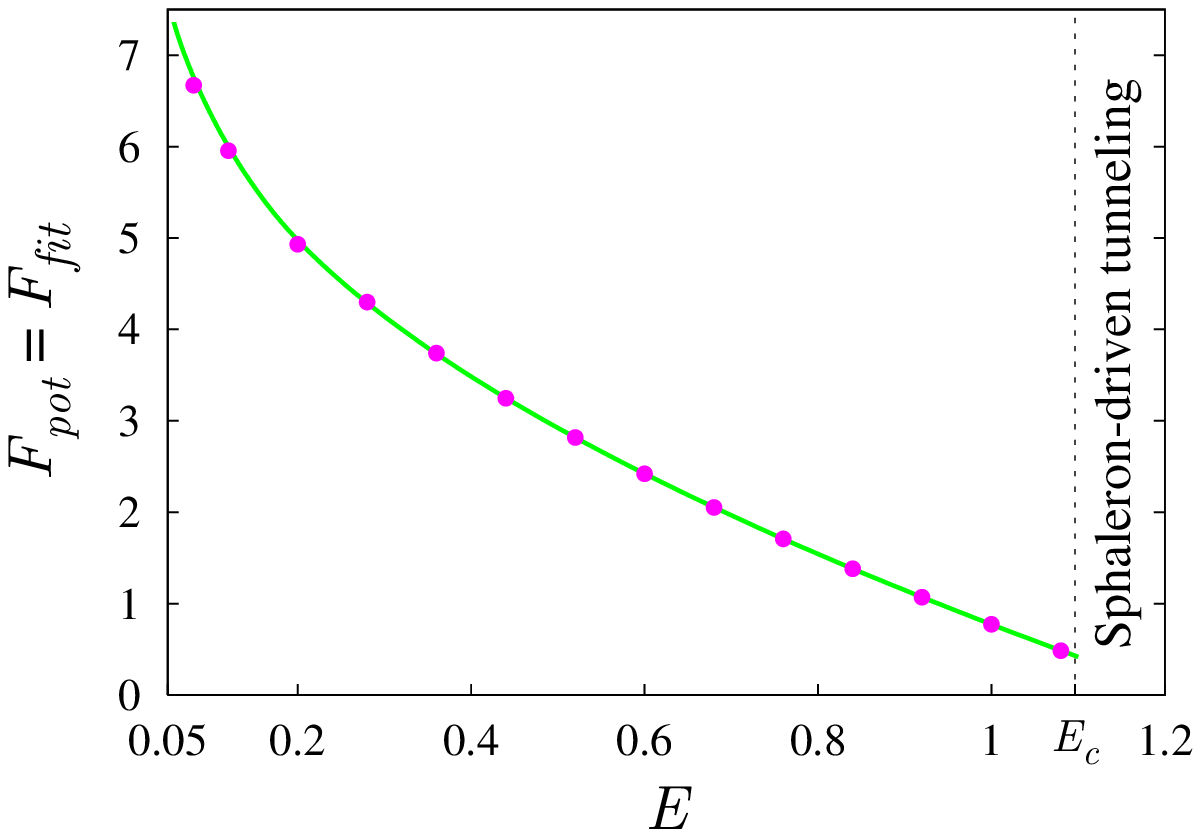}
    \includegraphics[width=0.5\textwidth]{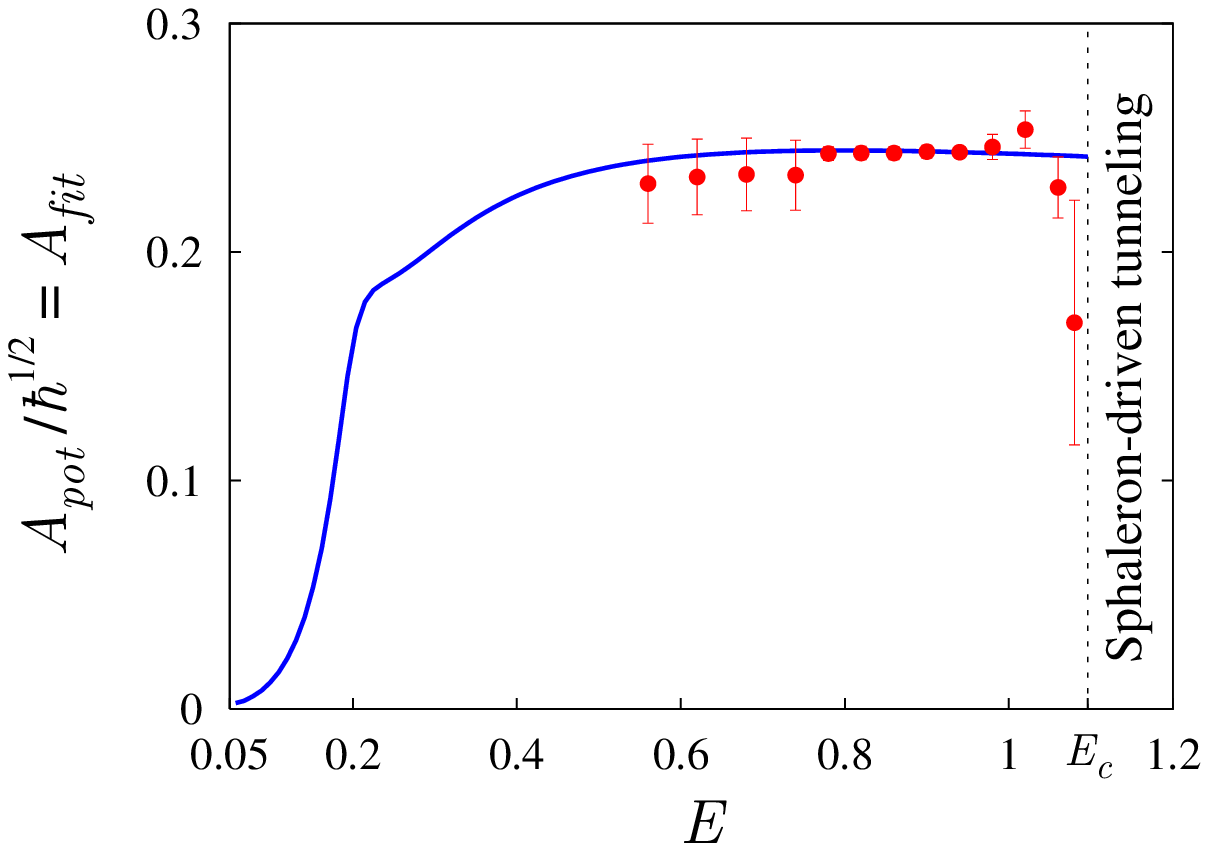}}
  \hspace{4.2cm}(a)\hspace{8cm}(b)
  \caption{\label{fig:6} The semiclassical (lines) and exact quantum
    mechanical (points) results for (a) leading suppression exponent
    and (b) prefactor; $E_y =   0.05$. Errorbars represent uncertainty
    of the fit (\ref{eq:32}).} 
\end{figure}

Before proceeding to the case of sphaleron--driven tunneling, we
demonstrate explicitly that the semiclassical expressions
(\ref{eq:26}), (\ref{eq:21}) produce correct values of 
suppression exponent and prefactor. To this end, we calculate the
exact probability of transition by solving numerically the
stationary Schr\"odinger equation (see
Refs.~\cite{Bonini:1999kj,Levkov:2007e} for the numerical method). 
The exact
values of $\cal P$ are computed at several\footnote{To be precise, we
  use three values $\hbar = 1/70,\; 1/40,\; 1/30$ at $E > 0.75$, two
  values $\hbar = 1/20,\; 1/10$ at $0.55<E<0.75$ and only one value 
$\hbar=1/10$ at $E<0.55$. This choice is dictated by the limitations
of the numerical method which does not allow to perform computations
when the value of the tunneling probability is too small.}
$\hbar$. Then, the dependence ${\cal 
  P}(\hbar)$ is fitted\footnote{At $E < 0.75$ only two values of
  $\hbar$ were considered, and we set $C_{fit} = 0$. At $E<0.55$ (one value
  of $\hbar$) we were unable to extract the prefactor from the quantum
  mechanical simulation.}
with the formula 
\begin{equation}
  \label{eq:32}
  -\hbar \log({\cal P}/\hbar^\gamma) = F_{fit} - \hbar \log A_{fit}  +
  \hbar^2 C_{fit} \;, 
\end{equation}
where $\gamma = 1/2$ and the last term accounts for the higher-order
semiclassical corrections. The fit produces the ``exact'' values
$F_{fit}$, $A_{fit}$ of the suppression exponent and prefactor; they should
coincide with the corresponding semiclassical quantities. 
In Fig.~\ref{fig:6} we compare the semiclassical results computed
by Eqs.~(\ref{eq:26}),~(\ref{eq:21}) with those extracted from the
fit~(\ref{eq:32}). One observes remarkable
agreement. It is worth noting that the fit (\ref{eq:32})
is extremely sensitive to the assumed $\hbar$-dependence of the
prefactor. In particular, if 
one erroneously uses $\gamma = 0$ or
$\gamma = 1$ in Eq.~(\ref{eq:32}), the value of prefactor extracted
from the fit becomes close to zero or extremely large. Hence, the
graph in Fig.~\ref{fig:6}b confirms, in particular, the qualitative
formula $A_{pot} \propto \hbar^{1/2}$. 

\subsection{Modification}
\label{sec:unst-traj}
At high energies tunneling proceeds by the new mechanism based on
qualitatively new properties of semiclassical trajectories. Namely, at
$E>E_c(E_y)$ the trajectories get attracted to the unstable 
sphaleron orbit and thus become unstable themselves. 

The instability of complex trajectories sets
obstacles for the semiclassical description.
The most important 
difficulty is related to the calculation of the prefactor
$A_{pot}$. Equation~(\ref{eq:21})
implies that  $A_{pot}$ is inversely proportional  to the values
of linear perturbations $\delta \boldsymbol{x}^{(n)}$ at $t = t_i$,
while the Cauchy data for $\delta \boldsymbol{x}^{(n)}$ are set at $t = 
t_f$. On the other hand, linear perturbations in the background of
unstable trajectory contain exponentially growing part. 
Thus, at $E>E_c(E_y)$ when the complex  
trajectory spends infinite time interval in the vicinity of the
sphaleron, the formula (\ref{eq:21}) gives 
$A_{pot} = 0$. This means that Eq.~(\ref{eq:21}) is incorrect 
in the case of sphaleron--driven tunneling and suggests that the
prefactor is suppressed by an additional power of $\hbar$.

The main idea of the modified semiclassical method was proposed in
Ref.~\cite{Levkov:2007prl}; it is close in spirit to the constrained
instanton technique of Ref.~\cite{Affleck:1980mp}. Namely, we evaluate
the path integral (\ref{eq:7}) for the tunneling amplitude
in two steps. First, we integrate over paths spending a {\it given
  time} $\tau$ in the vicinity of the sphaleron. Second, we integrate
over $\tau$.

The above manipulations with the path integral lead to the following
method. At the first step we obtain certain modified boundary value
problem for a family of complex trajectories labeled by the
parameter $\tau$. These trajectories are stable and interpolate
between the asymptotic regions $x\to \pm \infty$. The second step
produces expressions for the suppression exponent $F_{sph}$ and
prefactor $A_{sph}$ in the sphaleron--driven case. These expressions
relate the values of $F_{sph}$ and $A_{sph}$ to limits $\tau \to
+\infty$ of certain functionals evaluated on the modified
trajectories.

One introduces the functional $\tau = T_{int}[\boldsymbol{x}]$ which,
roughly speaking, 
measures the time spent by the path $\boldsymbol{x}(t)$ in the region
of non--trivial dynamics.
We call $T_{int}$ {\it interaction time}. It has the
following properties. First, $T_{int}$ is positive--definite for real
paths. 
Second,
it is finite for any real path satisfying $x\to \pm \infty$ as  $t
\to \pm\infty$ and infinite otherwise. The simplest choice is 
\begin{equation}
  \label{eq:15}
  T_{\mathrm{int}}[\boldsymbol{x}] = \int dt\, f(\boldsymbol{x}(t))\;,
\end{equation}
where the function $f(\boldsymbol{x}) > 0$ vanishes at $x \to \pm
\infty$. We use
$$
f(\boldsymbol{x}) = \mathrm{exp} \,\left\{-(x+y)^2/2\right\}
$$
in the model~(\ref{eq:24}).

Consider the path integral (\ref{eq:7}) for the final state. One
inserts into the integrand of Eq.~(\ref{eq:7}) the unity factor
\begin{equation}
\label{eq:unity}
1 = \int_0^{+\infty} d\tau \, \delta (T_{\mathrm{int}}[\boldsymbol{x}]
- \tau) = \int_0^{+\infty} d\tau \int_{i\infty}^{-i\infty}
\frac{id\epsilon}{2\pi \hbar} \, \mathrm{e}^{-\epsilon 
T_{\mathrm{int}}[\boldsymbol{x}]/\hbar + \epsilon \tau  /\hbar}\;,
\end{equation}
where the Fourier representation of the $\delta$--function was used in the
second equality. Expression (\ref{eq:7})
takes the form,
\begin{equation}
  \label{eq:28}
  \Psi_f(\boldsymbol{x}_f) =
  \int_0^{+\infty} d\tau \int_{i\infty}^{-i\infty}
  \frac{id\epsilon}{2\pi \hbar} \,\mathrm{e}^{\epsilon \tau/\hbar}
  \left\{ \left. \int d\boldsymbol{x}_i \,
      \Psi_i(\boldsymbol{x}_i)\int
      [d\boldsymbol{x}]\right|_{\boldsymbol{x}_i}^{\boldsymbol{x}_f}
       \mathrm{e}^{i(S[\boldsymbol{x}]+i\epsilon
         T_{int}[\boldsymbol{x}])/\hbar} \right\}\;,
\end{equation}
where we changed the order of integrations. One
notes that the integral in brackets is exactly the same as
in Eq.~(\ref{eq:7}) up to the substitution 
\begin{equation}
  \label{eq:5}
  S[\boldsymbol{x}] \to S_\epsilon[\boldsymbol{x}] \equiv S[\boldsymbol{x}]
  +   i\epsilon T_{\mathrm{int}}[\boldsymbol{x}]\;.
\end{equation}
This integral is evaluated by the saddle--point method in
the same way as the integral in Eq.~(\ref{eq:7}). Namely, one
finds the {\it regularized} trajectory $\boldsymbol{x}_{\epsilon}(t)$
which extremizes the modified action $S_{\epsilon}$ and arrives at
the point $\boldsymbol{x}_f$ at $t = t_f$. The initial 
conditions for the trajectory are still given by Eqs.~(\ref{eq:6}),
since the evolution in the in-region is not affected by the
functional $T_{int}$. The result of integration in Eq.~(\ref{eq:28})
is 
\begin{equation}
  \label{eq:2}
  \Psi_f(\boldsymbol{x}_f) =
  \int_0^{+\infty} d\tau \int_{i\infty}^{-i\infty}
  \frac{id\epsilon}{2\pi \hbar} \,\mathrm{e}^{\epsilon \tau/\hbar} \cdot
     D_{\epsilon}^{-1/2} \cdot \exp \left\{{\frac{i}{\hbar}
         (S_{\epsilon}[\boldsymbol{x}_\epsilon]
    + B_i[\boldsymbol{x}_\epsilon]) +  \frac{i\pi}{4}}\right\}\;,
\end{equation}
cf. Eq.~(\ref{eq:8}). The prefactor $D_{\epsilon}$ in this equation
is given by the same determinant formula as $D$, but with the
substitution $S\to S_\epsilon$, $\boldsymbol{x}(t) \to
\boldsymbol{x}_\epsilon(t)$. 

Let us remark on the representation (\ref{eq:2}). One keeps in
mind that the integrand in Eq.~(\ref{eq:2}) accounts for the
contribution of paths which spend a given time $\tau$ in the region of
finite $x$ (interaction region). 
In particular, this is true for the saddle--point trajectory
$\boldsymbol{x}_{\epsilon}(t)$. The latter interpolates 
directly between the in- and out- regions $x\to \pm \infty$ and thus
is stable. Note that
the stabilization of complex trajectory is achieved by the modification of
the classical equations of motion. Namely, the substitution~(\ref{eq:5})
modifies the potential of the system 
$$V(\boldsymbol{x}) \to
V(\boldsymbol{x}) - i\epsilon f(\boldsymbol{x})\;.$$ 
We will see below that the relevant values of 
$\epsilon$ are real; thus, $\boldsymbol{x}_{\epsilon}(t)$ describes
evolution in {\it complex} potential. 

The rest of the calculation proceeds as follows. 
One evaluates the saddle--point
integral with respect to $\epsilon$. The integral over interaction
time is kept in front of the formula. This ensures stability of
complex trajectories. The resulting expression for $\Psi_f$ is
substituted into the tunneling probability
(\ref{eq:19}). A subtle point is that, since ${\cal  P}$
involves the square of the out-state, one obtains at this stage {\it
  two} integrals over interaction times $\tau$,  $\tau'$, where the latter
comes from $\Psi_f^*$. One of these integrals can be computed by the
saddle--point technique. Indeed, returning to the original expression
for the tunneling probability in terms of the integral over real
paths, one sees that fixing the sum $\tau_+ = (\tau +
\tau')/2$ is sufficient to make both interaction times $\tau$ and
$\tau'$  finite. Thus, we change the
integration variables to $\tau_+$ and $\tau_- = \tau-\tau'$ and
evaluate the saddle--point integrals over $\tau_-$ and over the final
state. In this way we are left with the single integral over $\tau_+$.

We leave the details of the above computation to
appendix~\ref{sec:saddle-integr-modif} and discuss the result. First,
one arrives at the saddle--point conditions
\begin{equation}
  \label{eq:10}
  \mathrm{Re}\, T_{int}[\boldsymbol{x}_{\epsilon}] = \tau_+ \;, \qquad
  \epsilon = \epsilon^{*}\;,
\end{equation}
which come from the integrals over $\epsilon$ and $\tau_-$
respectively. The integral over out-states
produces, as before, the boundary conditions~(\ref{eq:20}) at $t = t_f$ for 
$\boldsymbol{x}_{\epsilon}(t)$. Note that the first of
Eqs.~(\ref{eq:10}) implies, in particular, 
that $\boldsymbol{x}_{\epsilon}(t)$ is stable. The result for the 
probability is 
\begin{equation}
  \label{eq:44}
  {\cal P} = \int_{0}^{+\infty} \frac{d\tau_+}{\sqrt{\pi\hbar}}
  \left[ -\frac{d\epsilon}{d\tau_+}\right]^{1/2} \cdot
  A_{pot,\, \epsilon} \;\mathrm{e}^{-(F_{pot,\epsilon} -
    2\epsilon\tau_+)/\hbar}\;,
\end{equation}
where the suppression exponent $F_{pot,\epsilon}$ and prefactor
$A_{pot,\epsilon}$ are computed by the same formulas~(\ref{eq:26}) and
(\ref{eq:21}) as before, but with the substitution $S\to S_\epsilon$. 
Note that the latter substitution implies that both the 
classical equations of motion and linearized
equations (\ref{eq:4}) are modified. 

We now proceed to the second step of the calculation and consider the integral
over the interaction time $\tau_+$. One makes an important
observation: the values of $\tau_+$ and $\epsilon(\tau_+)$ are related
by the Legendre transformation. Indeed, by construction the configuration
$\{\boldsymbol{x}_{\epsilon}(t),\; \epsilon(\tau_+)\}$ corresponds to
the extremum of the leading 
exponent $F_\epsilon = F_{pot,\epsilon} - 2\epsilon\tau_+$ in
Eq.~(\ref{eq:44}), and the respective derivatives of $F_{\epsilon}$
are equal to zero. Thus,
\begin{equation}
  \label{eq:11}
  \frac{dF_\epsilon}{d\tau_+} = 
  \frac{\partial} {\partial\tau_+} (F_{pot,\, \epsilon} - 2\epsilon
  \tau_+) = -2\epsilon\;,
\end{equation}
where only the explicit dependence of $F_{\epsilon}$ on $\tau_+$ was
taken into account in 
the last equality. Due to the property~(\ref{eq:11}), the integral in
Eq.~(\ref{eq:44}) is saturated at $\epsilon=0$. This point corresponds
to the original 
semiclassical equations: recall that the modification term in the
classical action, Eq.~(\ref{eq:5}), is proportional to $\epsilon$. 
One concludes that the integral for the tunneling probability is 
saturated in the vicinity of the original complex trajectory at
$\epsilon=0$.  

So far in our calculation we did not make any reference 
to the particular tunneling mechanism.
Thus, Eq.~(\ref{eq:44}) can be used in cases of both potential and
sphaleron--driven tunneling. The difference between the 
two mechanisms becomes crucial in the evaluation of 
the integral over $\tau_+$. In the standard  case of  
potential tunneling the trajectory at $\epsilon=0$ is stable and
corresponds to the finite value of interaction time $\tau_+$; one
takes the integral in Eq.~(\ref{eq:44}) by the saddle--point method
and arrives at the expressions~(\ref{eq:26}),~(\ref{eq:21}) from the
previous subsection. The case of sphaleron--driven tunneling is
considerably 
different, because the time interval spent by the trajectory in
the vicinity of the sphaleron tends to infinity as
$\epsilon\to +0$. Thus, the integral in  Eq.~(\ref{eq:44}) is
saturated by the end--point of the integration  interval $\tau_+\to
+\infty$. Using the appropriate asymptotic expression\footnote{This
  expression is derived as follows. One moves
  the leading exponent in Eq.~(\ref{eq:44}) under the differential
  using the relation $2\epsilon \cdot\exp\{-
    F_\epsilon/\hbar\}\, d\tau_+ = \hbar \,d\exp\{-
    F_\epsilon/\hbar\}$ and  integrates by parts. After
    integration the leading semiclassical approximation is given by
    the boundary 
  term at $\tau_+ \to   +\infty$; the boundary  term at
  $\tau_+=0$ and the remaining integral over $\tau_+$ are 
  exponentially and power--law suppressed respectively.} for the
integral, 
one  obtains
Eq.~(\ref{eq:9}) with
\begin{subequations}
  \label{eq:14}
\begin{align}
\label{eq:14a}
  &F_{sph} = \lim_{\epsilon\to +0} F_\epsilon \;,\\
\label{eq:14b}
  &A_{sph} = \hbar^{1/2} \lim_{\epsilon\to +0}
  \frac{A_{pot,\,\epsilon}}{\epsilon\sqrt{-4\pi
      \frac{d\mathrm{Re}\,T_{int}[\boldsymbol{x}_\epsilon]} {d\epsilon}}}\;,
\end{align}
\end{subequations}
where we mark the quantities corresponding to the new mechanism 
with the subscript $sph$. Note that the prefactor $A_{pot,\,
  \epsilon}$ is computed by the formula~(\ref{eq:21}) with
modification~(\ref{eq:5}), while the exponent  
$$
F_{\epsilon} =
F_{pot,\,\epsilon} - 2\epsilon\tau_+ =
2\mathrm{Im}(S[\boldsymbol{x}_{\epsilon}] +
B_i[\boldsymbol{x}_\epsilon])
$$
 is given by the value of the {\it 
  original} action on the {\it modified} trajectory. 
Let us remark that the limit $\epsilon\to +0$ in Eqs.~(\ref{eq:14})
does actually exist; this is shown analytically in
appendix~\ref{sec:linear}.
One observes that the expression~(\ref{eq:14b}) for the prefactor is
very different from that in the case of potential tunneling. In
particular, $A_{sph} \propto \hbar^{1/2} A_{pot}$.

To summarize, we derived the following method of calculating the
probability of sphaleron--driven tunneling. One modifies the classical
action of the system by adding purely imaginary term proportional to
the small regularization parameter $\epsilon>0$, see Eq.~(\ref{eq:5}). Then 
one solves the modified equations of motion with the original boundary
conditions~(\ref{eq:6}), (\ref{eq:20}) and finds the modified
complex trajectory $\boldsymbol{x}_{\epsilon}(t)$. This trajectory  
interpolates between the asymptotic regions $x\to \pm\infty$ and is
stable. The modified values of the suppression exponent
$F_{pot,\,\epsilon}$ and prefactor $A_{pot,\,\epsilon}$
are computed by the same formulas as before, Eqs.~(\ref{eq:26})
and~(\ref{eq:21}), but with the substitution $S\to S_\epsilon$.  The
final result for 
the tunneling probability is obtained\footnote{In practice the limit
  in Eqs.~(\ref{eq:14}) is taken by considering small values
  of the regularization parameter, $\epsilon\sim 10^{-6}$. At these
  $\epsilon$ the values of the suppression exponent $F_{sph}$ and
  prefactor $A_{sph}$ stabilize at the level of accuracy $10^{-5}$.} in the  
limit $\epsilon\to +0$, see Eqs.~(\ref{eq:14}).  We call the above
modified semiclassical method by {\it $\epsilon$--regularization
technique}~\cite{Bezrukov:2003yf}.

\begin{figure}[htb]
  \centerline{\includegraphics[width=0.5\textwidth]{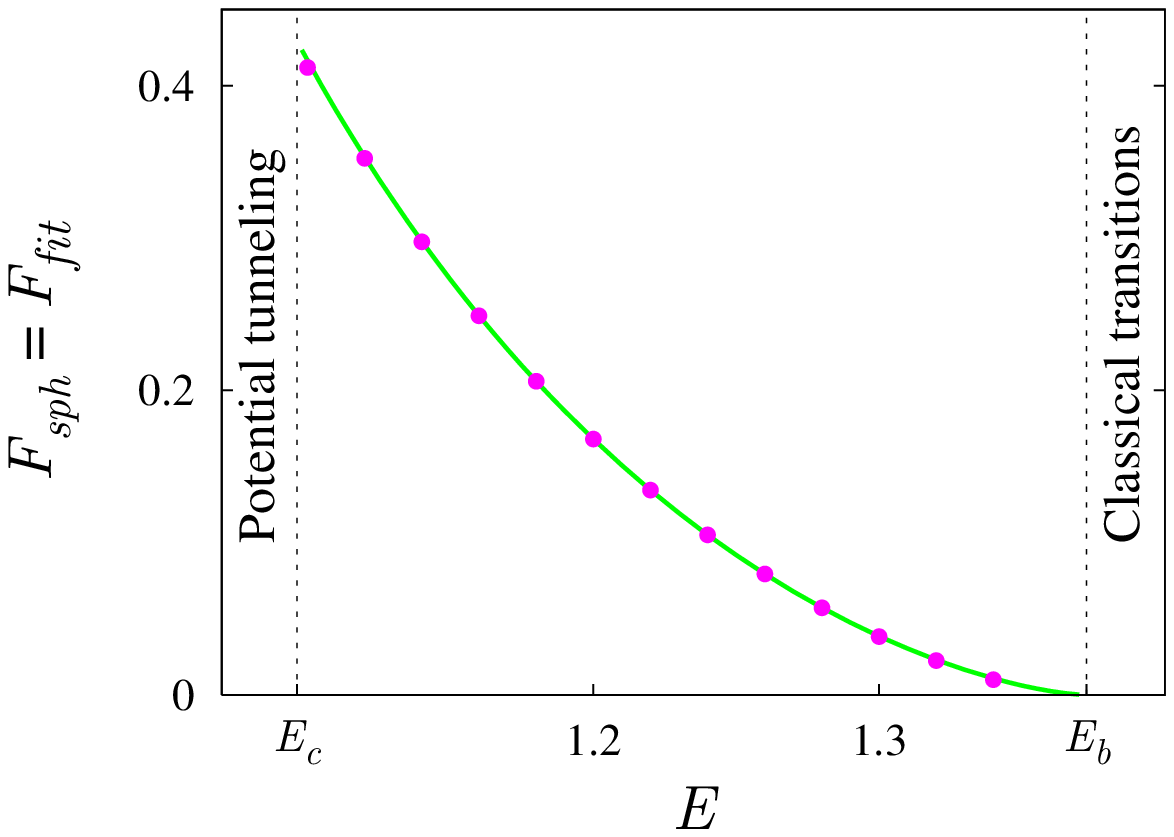}
    \includegraphics[width=0.5\textwidth]{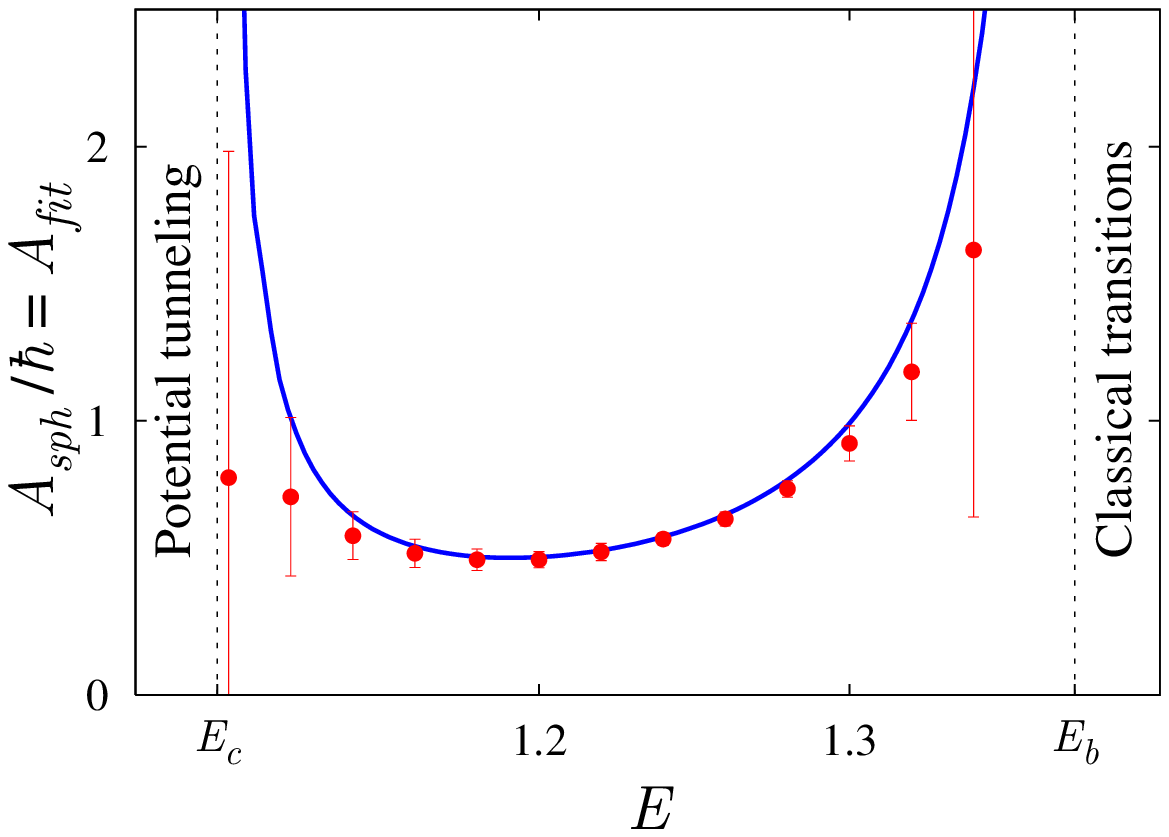}}
  \hspace{4.2cm}(a)\hspace{8cm}(b)
  \caption{\label{fig:8} (a) Suppression exponent and (b) prefactor
    in the case of sphaleron--driven tunneling, $E > E_c(E_y)$; $E_y =
    0.05$. The points are extracted from the fit (\ref{eq:32}),
    while the lines stand for the semiclassical results,
    Eqs.~(\ref{eq:14}). The vertical dotted lines bound the range of
    energies  for sphaleron--driven tunneling.} 
\end{figure}

We perform straightforward check of the modified semiclassical
method by comparing the semiclassical predictions (\ref{eq:14}) with
the results of the exact quantum mechanical computations. The latter 
are used to extract the values of the suppression exponent and
prefactor 
by the fitting procedure described in the previous
subsection, where  $\gamma = 1$ in
Eq.~(\ref{eq:32}). The comparison is shown in Fig.~\ref{fig:8}. The 
observed agreement between the semiclassical and quantum mechanical
results supports the 
modified semiclassical technique. In particular, we checked that the
fit (\ref{eq:32}) produces unacceptably large values of the prefactor if
one erroneously assumes the same
$\hbar$--dependence $\gamma=1/2$
as in the case of potential tunneling.
Thus, the scaling $A_{sph} \propto \hbar$ is confirmed.

\subsection{Uniform approximation}
\label{sec:unif-appr}
Our expressions for $A_{pot}$ and $A_{sph}$ imply
apparent discontinuity of the semiclassical probability across the
critical energy; on the other hand, the exact quantum probability is a smooth
function of energy. As a consequence,  the
$\hbar$--dependences $A_{pot} \propto \hbar^{1/2}$ and $A_{sph}
\propto \hbar$ fail to describe the quantum mechanical data in the
immediate vicinity of $E_c(E_y)$. [This is seen in
Figs.~\ref{fig:6}b,~\ref{fig:8}b, where the quality of the fit
(\ref{eq:32}) becomes worse as $E\to E_c(E_y)$.] One observes that
both the standard and modified formulas are invalid at $E\approx
E_c(E_y)$.

In this section we derive the uniform asymptotic formula for the
tunneling probability which is applicable in the
vicinity of $E_c(E_y)$. Our formula has the form
(cf. Ref.~\cite{Creagh:2006}),
\begin{equation}
  \label{eq:29}
  {\cal P}_{uni} = \left\{ \begin{array}{ll}
      {\cal M}_{pot} \cdot {\cal P}_{pot} 
       & \;\; \mbox{at} \;\; E<E_c(E_y)\;, \\
      {\cal M}_{sph}\cdot  {\cal P}_{sph} 
       & \;\;\mbox{at} \;\; E>E_c(E_y)\;,
    \end{array} \right.
\end{equation}
where ${\cal M}_{pot}$ and ${\cal M}_{sph}$ are the correction factors
in the cases of potential and sphaleron--driven
tunneling respectively. We will find that ${\cal M}_{pot,\, sph}
\approx 1$ at $|E-E_c| 
\gg \hbar^{1/2}$; thus, the formula (\ref{eq:29}) is relevant in the
small region of width $\Delta E \sim \hbar^{1/2}$ around the
critical point.  We stress that the uniform probability ${\cal
  P}_{uni}$ is {\it continuous} at $E
= E_c(E_y)$.  

We obtain the desired approximation by examining the integral
over $\tau_+$ for the tunneling probability, Eq.~(\ref{eq:44}). Recall
that Eq.~(\ref{eq:44}) is applicable in both cases of potential and
sphaleron--driven tunneling. To make the discussion transparent, we change
the integration variable to
\begin{equation}
\label{eq:16}
w(\tau_+) = \frac1{\sqrt{\pi\hbar}}  \int_{\tau_+}^{+\infty}  d\tau_+'
\,\sqrt{-d\epsilon'/d{\tau'_+}} \cdot A_{pot,\, \epsilon'}\;.
\end{equation}
Note that the limiting values $\tau_+ = 0$ and  $\tau_+ \to +\infty$
correspond respectively to $w=w_0 > 0$ and $w\to +0$. 
In new terms the integral (\ref{eq:44}) takes a particularly
simple form, 
\begin{equation}
\label{eq:54}
{\cal P} = \int_0^{w_0} dw \, \mathrm{e}^{-F_{\epsilon}(w)/\hbar}\;,
\end{equation}
where the leading semiclassical exponent is now considered as 
function of $w$. 

The difference between the two mechanisms of tunneling is now
understood as follows. At small energies the integral (\ref{eq:54}) is
saturated by the saddle point $w=w_s>0$. The value of $w_s$
decreases with energy, so that at $E = E_c(E_y)$ the saddle point
crosses  the boundary $w=0$ and leaves the integration interval. 
At $E>E_c(E_y)$ the saddle point $w_s$ is situated
outside the region of integration.

Consider the Taylor series expansions of the semiclassical exponent
$F_\epsilon$ around the points $w=w_s$ and $w=0$,
\begin{align}
\label{eq:30}
& F_\epsilon(w) = F_{pot} + F''(w_s) \cdot (w-w_s)^2/2 +
O((w-w_s)^3)\;,\\
\label{eq:31}
& F_\epsilon(w) = F_{sph} + F'(0) \cdot w  + F''(0) \cdot w^2/2  +
O(w^3)\;,
\end{align}
where the primes denote derivatives with respect to $w$ and 
we marked by $F_{pot}$, $F_{sph}$ the values of the exponent at
$w=w_s$ and $w=0$. The semiclassical expressions of
Secs.~\ref{sec:semicl-form-tunn} and~\ref{sec:unst-traj} are obtained
from the expansions (\ref{eq:30}) and (\ref{eq:31})
respectively. Namely, at $E<E_c(E_y)$ one implements the saddle--point 
method, i.e. substitutes Eq.~(\ref{eq:30}) into
Eq.~(\ref{eq:54}) and extends the interval of integration to the
entire $w$ axis. At energies higher than critical
the minimum value of the exponent $F_{\epsilon}$ is achieved at $w=0$,
and one uses the expansion (\ref{eq:31}), where only the zeroth- and
first-order terms are kept. It is straightforward to check 
that in this way one obtains the expressions (\ref{eq:26}),
(\ref{eq:21}) of Sec.~\ref{sec:semicl-form-tunn} and (\ref{eq:14}) of
Sec.~\ref{sec:unst-traj}.

One observes that the above 
two integration methods are not applicable
if the saddle point $w_s$ is close to the end--point
$w=0$. Indeed, in the saddle--point technique at $E<E_c(E_y)$ the
interval of integration cannot be extended to the entire $w$ axis,
since the contribution from the additional interval $w<0$ is not
negligible.  In the end--point integration at $E>E_c(E_y)$ the third
term in Eq.~(\ref{eq:31}) is not small in  comparison with the second
term, because the first derivative $F'(0)$ vanishes in the limit
$w_s\to 0$. Given these reasons, one easily remedies the formulas
keeping the finite integration interval at $E<E_c(E_y)$ and three
terms in the expansion (\ref{eq:31}) at energies higher than
critical. The resulting expressions for the correction factors are
\begin{subequations}
\label{eq:37}
\begin{align}
\label{eq:37a}
&{\cal M}_{pot}  =  \frac12  \left\{ 1 + \mathbf{\Phi}
  (\varkappa_{pot}) \right\} \;, & \mbox{where}~~ &
\varkappa_{pot} = w_s \sqrt{F''(w_s)/2\hbar}\;,\\ 
\label{eq:37b}
&{\cal M}_{sph}  = \sqrt{\pi} \varkappa_{sph} \left\{ 1 -
  \mathbf{\Phi}(\varkappa_{sph})\right\} \cdot
\mathrm{e}^{\varkappa_{sph}^2}\;, &\mbox{where}~~&
\varkappa_{sph} = F'(0)/\sqrt{2\hbar F''(0)}\;.
\end{align}
\end{subequations}
Here $\mathbf{\Phi}$ 
is the Fresnel integral, $\mathbf{\Phi}(z) = 2/\sqrt{\pi}
\int_0^z dt \, \mathrm{e}^{-t^2}$.

It is straightforward to check that
the factors (\ref{eq:37}) have the required
properties. First, the uniform formula (\ref{eq:29}) is continuous 
at $E = E_c(E_y)$ by construction.
Indeed, in this case $w_s$ and $F'(0)$ are equal to
zero and the expressions (\ref{eq:30}), (\ref{eq:31}) used in the
integration coincide. Second, at $|E - E_c(E_y)| \gg
\hbar^{1/2}$ the arguments of the Fresnel integrals are 
large. Using the asymptotics of
$\mathbf{\Phi}(z)$, one finds that ${\cal M}_{pot,\, sph} \approx 1$
outside the immediate vicinity of the critical energy. Third, one notes
that in the region $|E - E_c(E_y)| \sim \hbar^{1/2}$ the ``potential'' and
``sphaleron--driven'' parts of the uniform formula coincide up to
higher--order semiclassical corrections. Indeed, in this region the
central points $w = w_s$ and $w = 0$ of respective Taylor expansions
are parametrically close to each other, $w_s \sim \hbar^{1/2}$; thus,
the results obtained from Eqs.~(\ref{eq:30}) and (\ref{eq:31}) are
close as well.

Numerically, one extracts the quantities entering the correction
factors (\ref{eq:37}) by studying the dependence of $F_\epsilon$ 
and $A_{pot,\, \epsilon}$ on $\epsilon$. We discuss this calculation in
appendix~\ref{sec:calc-unifrom}.  

\begin{figure}[htb]
\centerline{\includegraphics[width=0.7\textwidth]{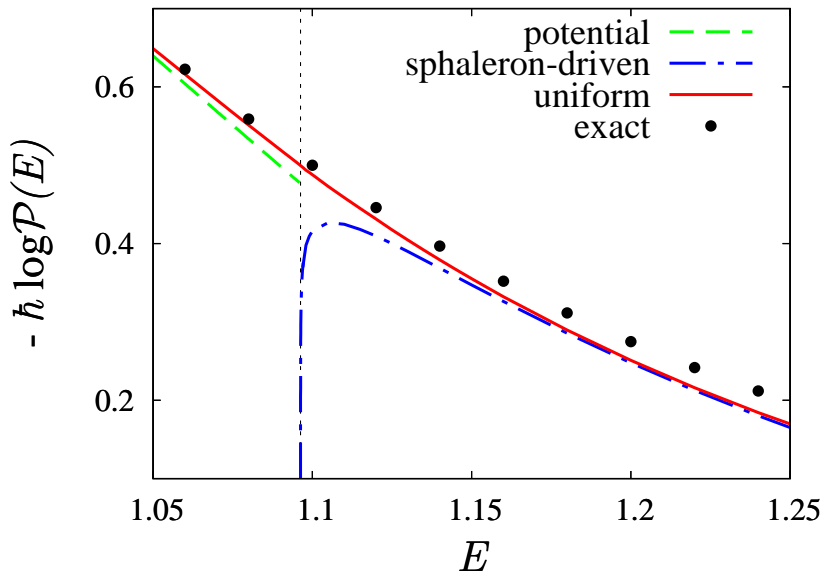}}
\caption{\label{fig:10} Uniform approximation for the tunneling 
probability (solid line) shown
  at $\hbar=1/30$, $E_y = 0.05$ in the vicinity of the critical energy
  $E_c(E_y) \approx 1.1$. We also plot the semiclassical results of
  Secs.~\ref{sec:semicl-form-tunn} and~\ref{sec:unst-traj} (dashed
  lines) and exact quantum probability (points). The critical energy
  is marked with the vertical dotted line.}
\end{figure}

To summarize, we derived the continuous asymptotic formula for the
tunneling probability,
Eq.~(\ref{eq:29}), which works at energies close to critical and
interpolates between the two semiclassical expressions corresponding
to the cases of potential and sphaleron--driven tunneling.  
In Fig. \ref{fig:10} we compare the uniform approximation 
(\ref{eq:29}) (solid line) with the semiclassical
probabilities ${\cal P}_{pot}$ and ${\cal P}_{sph}$ (dashed lines), as
well as with the exact quantum probability (points).

\section{Exclusive processes}
\label{sec:nf}
Here we study semiclassically the effect of the new
tunneling mechanism on exclusive processes, i.e. processes with
completely fixed out-states. We discuss the application of the
modified semiclassical technique to the exclusive case 
and obtain expressions, analogous to Eqs.~(\ref{eq:14}), for the
suppression exponent and prefactor of exclusive probability. 
We show that in the semiclassical limit of vanishingly small $\hbar$
the exclusive prefactor $A_{e,sph}$ is proportional to $\hbar^2$ in
the sphaleron--driven case. This  should be compared with the
dependence $A_{e,pot}\propto \hbar$ in the case of potential
tunneling.

\subsection{Exclusive trajectories}

We consider tunneling transitions between the exclusive states $|E,\,
E_y\rangle$ and $|E,\, E_{y}^f \rangle$ specified by the same value of
total energy $E$ and definite energies $E_y$, $E_{y}^f$ of 
$y$-oscillator. The standard semiclassical method in the case of exclusive
transitions is formulated in Ref.~\cite{Miller}. Its derivation is
completely analogous to that carried out in
Sec.~\ref{sec:semicl-form-tunn} for inclusive processes. Fixation of
the out-state changes the final boundary conditions for 
the complex trajectory: instead of Eqs.~(\ref{eq:20}) one has,
\begin{equation}
  \label{eq:34}
  x_f = x_f^{(0)}\;, \qquad \qquad E_{y}^f = (\dot{y}_f^2 +\omega^2
  y_f^2)/2 \;.
\end{equation}
The initial conditions remain the same,
Eqs.~(\ref{eq:6}). The exclusive suppression exponent 
is given by the action functional 
\begin{equation}
\label{eq:38}
F_{e,pot} = 2\mathrm{Im}(S[\boldsymbol{x}] + B_i[\boldsymbol{x}] -
B_f[\boldsymbol{x}])\;,
\end{equation}
computed on the trajectory, cf. Eq.~(\ref{eq:26}). Note that the new
term $B_f$ in Eq.~(\ref{eq:38}) is related to the out--state of the 
process; it is given by the same expression as $B_i$, but at $t = t_f$
and  with the out-state quantum numbers $E$, $E_{y}^f$. 
We do not write  here the formula for the prefactor $A_{e,pot}$; it can be
found in Ref.~\cite{Miller}. Importantly, this formula implies that
$A_{e,pot} \propto \hbar$.

\begin{figure}
\centerline{\includegraphics[width=0.5\textwidth]{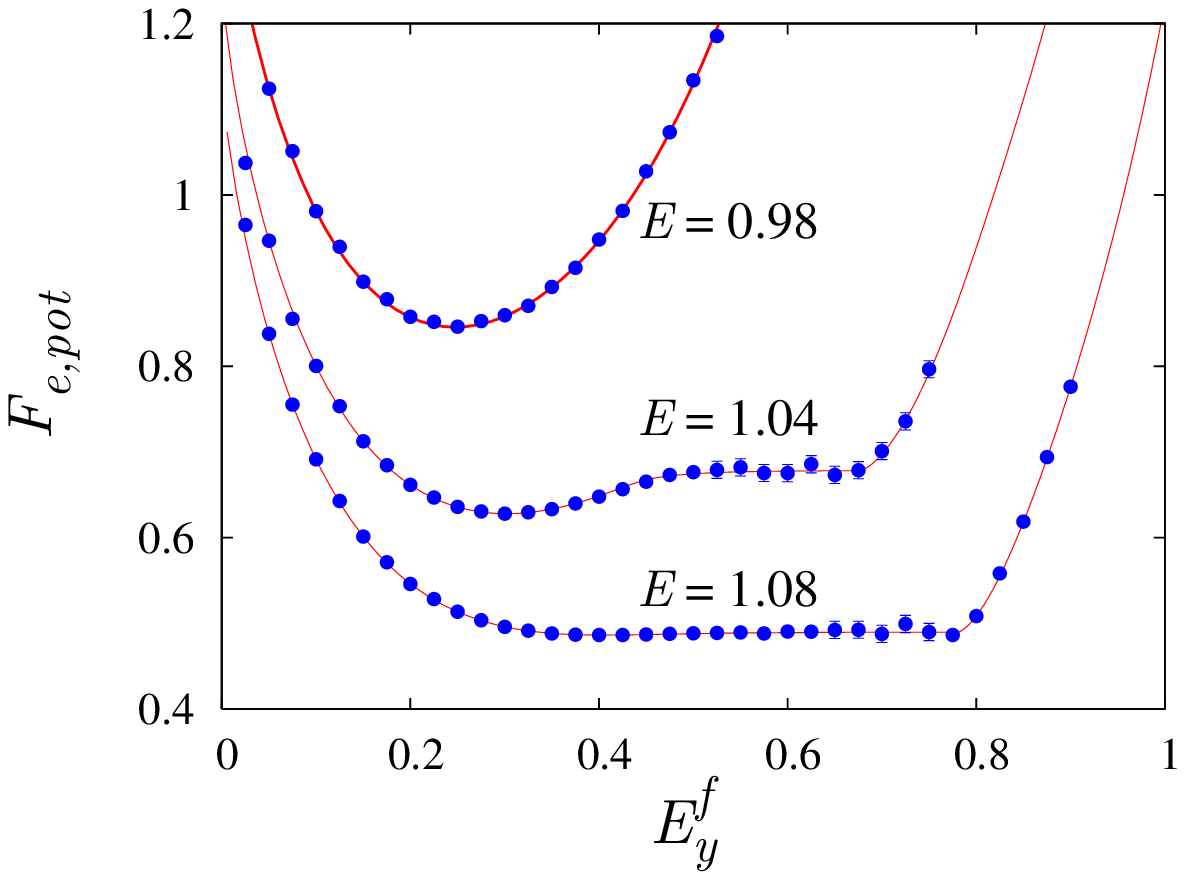}
\includegraphics[width=0.5\textwidth]{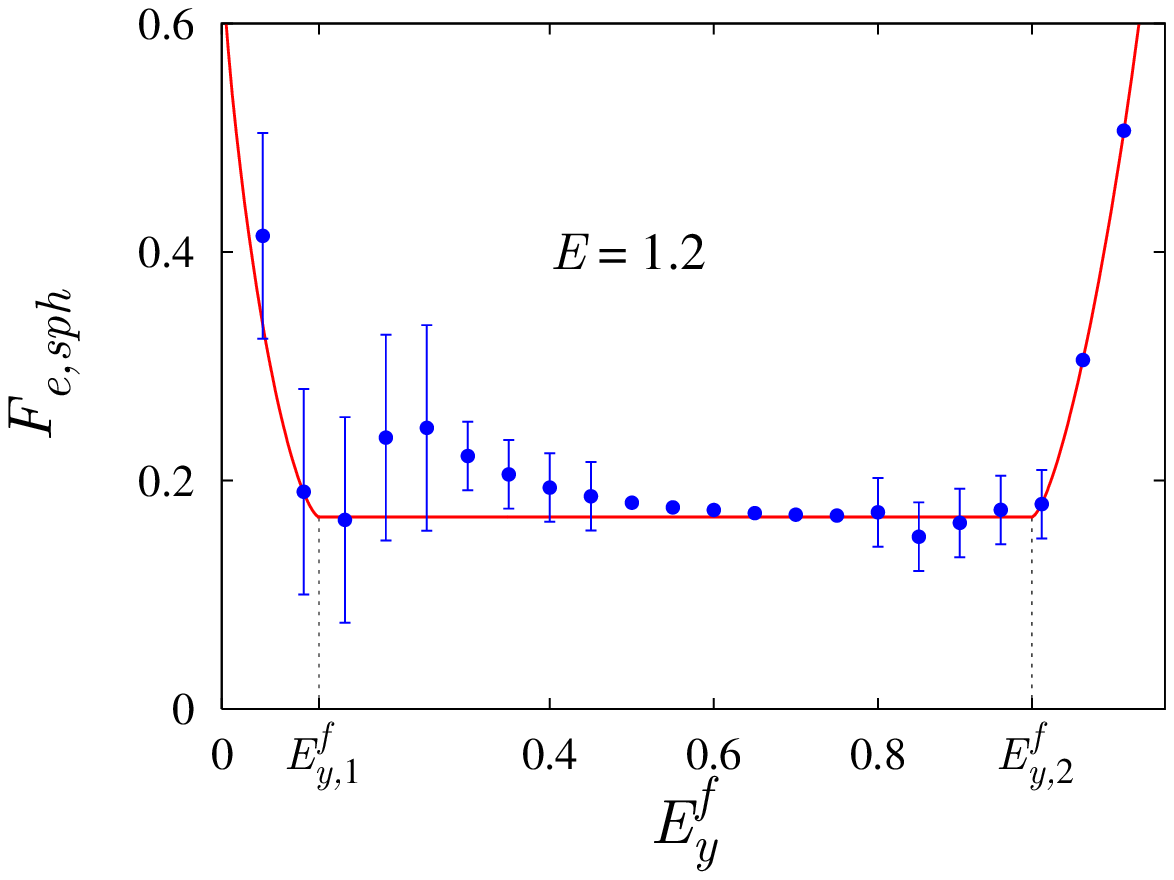}}
\hspace{4.3cm}(a)\hspace{8cm}(b)
\caption{\label{fig:14} Exclusive suppression exponent $F_e$ in the
 cases of (a) potential and (b) sphaleron--driven tunneling; $E_y =
 0.05$. The semiclassical results for $F_{e}$ (lines) are compared
 with the exact data (points). Errorbars represent
 the uncertainty of the fit (\ref{eq:32}).} 
\end{figure}

We apply the above method in the case of potential tunneling,
$E<E_c(E_y)$.
In Fig.~\ref{fig:14}a (lines) we plot the out--state distributions of
the exclusive exponent (\ref{eq:38}) for several values of
energy $E$. The exact results (points) are
extracted from the  fit~(\ref{eq:32}) with $\gamma = 1$.  The
semiclassical and exact data coincide. 

One observes that well below the critical 
energy $E_c(E_y)\approx 1.1$ the function $F_{e,pot}(E_{y}^f)$ has a clear
minimum corresponding to a sharp maximum of the quantum
probability. As the energy tends to $E_c(E_y)$, a flat plateau
develops in the right side of the graph.
As discussed in Sec.~\ref{sec:summary-results}, this behavior is
copied by the exact quantum probability, cf. Fig.~\ref{fig:3}a.

Before introducing the semiclassical method for exclusive
tunneling in the sphaleron--driven case,  we preview the result for
the suppression 
exponent $F_{e,sph}$ in Fig.~\ref{fig:14}b (solid line). At
$E>E_c(E_y)$ the exclusive exponent is exactly constant in the 
region $E_{y,1}^{f} < E_{y}^f < E_{y,2}^{f}$;
clearly, this feature corresponds to a wide and flat maximum of
quantum probability, cf. the exact graphs in Fig.~\ref{fig:3}b.  
Thus, the distribution of the exclusive probability 
over the out-state quantum numbers becomes   anomalously wide when the
sphaleron--driven mechanism is involved. So far the
semiclassical study of this property was restricted
to one--dimensional systems with non--autonomous
potentials~\cite{Takahashi:Ikeda,Takahashi:2008}. Here we find the
same effect
in the
two--dimensional setup of Sec.~\ref{sec:model}.

\begin{figure}[h]
\centerline{\includegraphics[width=0.5\textwidth]{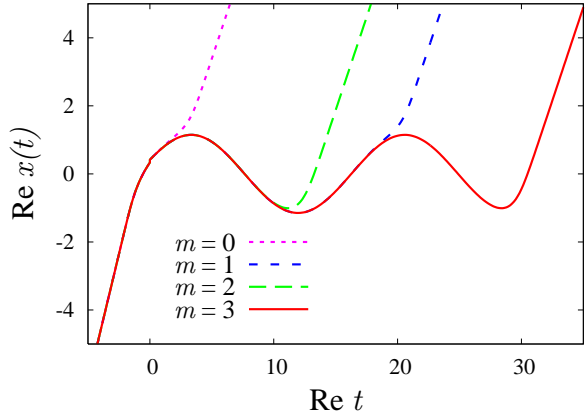}}
\caption{\label{fig:17}  Four exclusive trajectories
  at $E_{y}^f=0.6$. The initial parts of the trajectories
  are indistinguishable on the graph. The in-state quantum
  numbers are $E=1.2$, $E_y = 0.05$.}
\end{figure}

Returning to the semiclassical description of exclusive tunneling
processes, we find the following manifestation of the
sphaleron--driven mechanism.  In contrast to the case of potential
tunneling where the exclusive trajectory is unique, at $E>E_c(E_y)$
there is an infinite sequence of complex trajectories corresponding to
the same final oscillator energy $E_y^f$. In Fig.~\ref{fig:17} we
plot the first four trajectories for $E_y^f=0.6$.
\begin{figure}[h]
\centerline{\includegraphics[width=0.6\textwidth]{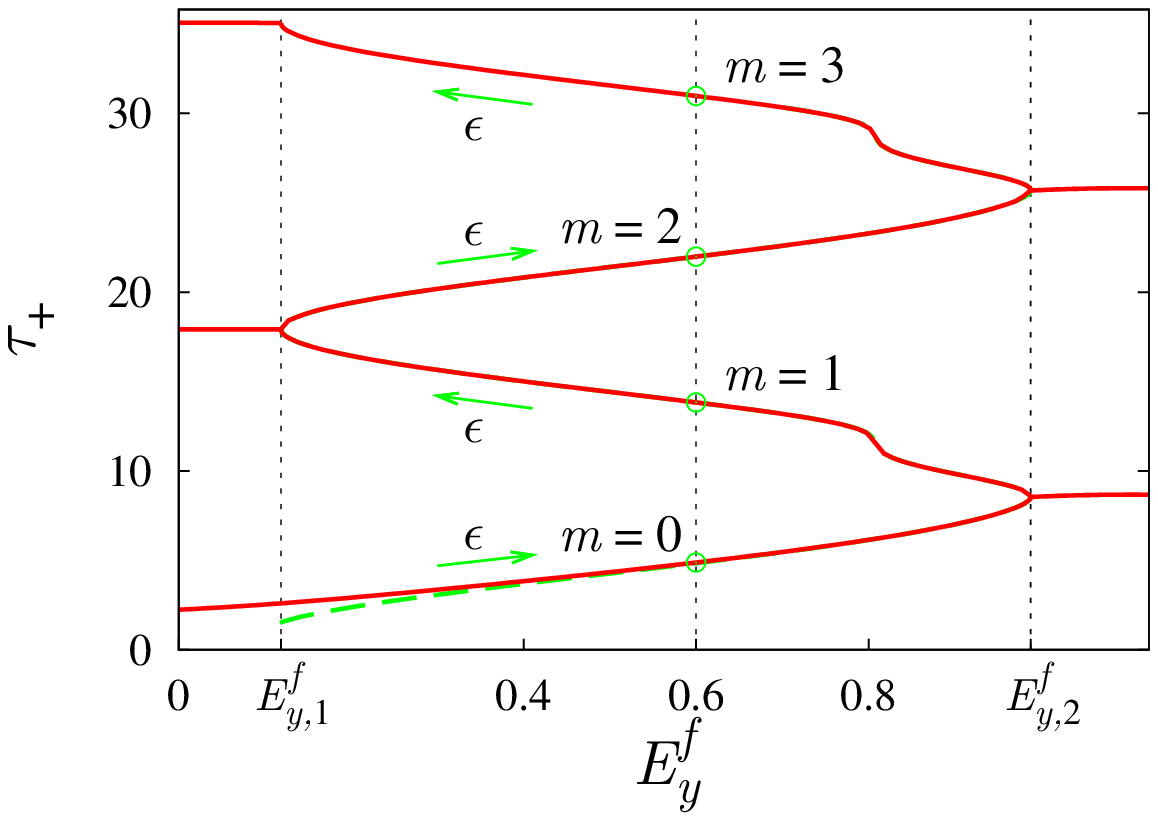}}
\caption{\label{fig:16} The curve in $(E_y^f,\tau_+)$--plane  (solid
  line) representing exclusive tunneling trajectories. Circles
  correspond to the trajectories from Fig.~\ref{fig:17}. The
  dashed line represents the modified inclusive
  trajectories. It is almost coincident with the sin--like part of the
  solid line. Arrows indicate the direction of decreasing 
  $\epsilon$. The in-state quantum  numbers are the same as in
  Fig.~\ref{fig:17}. } 
\end{figure}
One observes the following behavior: the trajectories 
reach the unstable periodic orbit (sphaleron), perform several
oscillations there (i.e. around 
the point $x=0$) and then slide off describing the sphaleron decay
into the final state. Importantly, in order to arrive into the
out-state with given $E_y^f$, the trajectory must leave\footnote{This
  notion can be given precise meaning by saying that the trajectory
  leaves the sphaleron once the distance $|\boldsymbol{x}(t) -
  \boldsymbol{x}_{sph}(t)|$ between the trajectory and the sphaleron
  orbit reaches a certain value $\delta\ll 1$.} the sphaleron at a
particular oscillation phase $\varphi$. More precisely, there are two
choices\footnote{This follows from the fact that the final oscillator 
  energy $E_y^f$ is a periodic function of $\varphi$; thus, equation
  $E_y^f(\varphi) = \mbox{const}$ ~has (at least) two solutions.} of 
phase per sphaleron period. We conclude that the interaction time
$\tau_+$ spent by the exclusive tunneling trajectories at given
$E_y^f$ is restricted to two values plus an integer number of
sphaleron periods. This gives rise to an infinite family of tunneling 
trajectories which describe the same  exclusive process but differ by
the number of ``half--period'' oscillations on top of the unstable
periodic orbit.

To investigate the properties of exclusive trajectories in
the sphaleron--driven case, we proceed as follows. For each trajectory we
compute the value of the interaction time functional $\tau_+ =
\mathrm{Re}\,T_{int}[\boldsymbol{x}]$,  Eq.~(\ref{eq:15}). In this
way we characterize the trajectories by points in the plane
$(E_y^f,\tau_+)$, see Fig.~\ref{fig:16}.

A comment is in order. The exclusive tunneling trajectories are stable
even in the sphaleron--driven case. Thus, they can be found without
$\epsilon$--regularization. Still, as we will discuss shortly,
it is convenient to use the modified semiclassical technique at the
intermediate steps of the computation and remove the
regularization afterwards. 
To avoid confusion, let us stress that the solid line
in Fig.~\ref{fig:16} corresponds to trajectories which are obtained
{\em after} removal of the regularization. Consequently, 
the functional $T_{int}[\boldsymbol{x}]$ does not enter the
equations of motion for these trajectories and is used only to
characterize their temporal behavior.

From Fig.~\ref{fig:16} one sees that the exclusive trajectories are
naturally divided into two classes. The
trajectories from the first class lie in the interval $E_{y,1}^{f} <
E_{y}^f < E_{y,2}^{f}$ corresponding to the plateau in 
Fig.~\ref{fig:14}b. In Fig.~\ref{fig:16} they form a
sin--like curve extended to the infinite values of $\tau_+$.
All these trajectories
describe creation and subsequent decay of the sphaleron. Moreover, we
find that the latter decay proceeds {\em classically}  
since the imaginary part of the trajectories becomes small after one
sphaleron oscillation. As a consequence, the value of
the functional (\ref{eq:38}) is almost independent of the 
individual trajectory from the first class. Besides, it is clear from
the figure that these trajectories form an infinite sequence of 
branches marked with the integer number $m$ of ``half--period''
oscillations in the vicinity of the sphaleron. 
 
The trajectories from the second class represent the ``wings'' $E_{y}^f <
E_{y,1}^{f}$, $E_{y}^f>E_{y,2}^{f}$ of the out-state distribution in
Fig.~\ref{fig:14}b. They correspond to the
case when the decay of the sphaleron orbit into the out-state with given
$E_{y}^f$ cannot proceed classically. Consequently, 
the probability of this decay
is exponentially suppressed. Due 
to the additional suppression, the exponent (\ref{eq:38}) strongly
depends on  the out-state at $E_{y}^f < E_{y,1}^{f}$,
$E_{y}^f>E_{y,2}^{f}$. Note, however, that the sphaleron still serves as
the mediator of the two-stage tunneling process; hence, the
second--class trajectories with fixed $E_{y}^f$ form an infinite sequence
marked with the topological number $m$, see Fig.~\ref{fig:16}. The
values of the suppression exponents $F_{e,pot}^{(m)}$ calculated on
trajectories with different topology and given $E_y^f$ are almost
degenerate.

In practice the exclusive trajectories are conveniently found using
the regularization method of Sec.~\ref{sec:unst-traj}. 
The procedure is based on the following observation. Consider 
$\epsilon$--regularized trajectories corresponding to the
{\em inclusive}
tunneling process. They describe creation and subsequent classical
decay of the sphaleron. The final state of the decay depends on the
value of $\epsilon$. Changing $\epsilon$ one covers the whole range of
final oscillator energies $E_{y,1}^f\leq E_y^f\leq E_{y,2}^f$
accessible in the classical sphaleron decay. 
This consideration is
illustrated by the dashed curve in Fig.~\ref{fig:16} which represents
the modified inclusive trajectories at different values of $\epsilon$
in the $(E_y^f,\tau_+)$ -- plane. We see
that the graph 
closely follows the sin--like curve  of exclusive trajectories from
the first class, and the value of $\epsilon$ decreases towards large
$\tau_+$ (along arrows). Thus, the  modified solutions with different
$\epsilon$ form a {\it single} branch which smoothly interpolates
between the branches of exclusive trajectories.  

Numerically, we exploit the above property by applying the deformation
procedure of appendix~\ref{sec:numerical-method}. Namely, we start with
the modified trajectory at a given $E_{y}^f = E_{y,0}^{f}$. Suppose 
it has topology $m$. Then, the 
trajectory with topology $m+1$ ($m-1$) is  obtained
by decreasing (increasing) the value of $\epsilon$ until
the final oscillator energy  arrives to $E_{y,0}^{f}$ again
(see Fig.~\ref{fig:16}). Repeating this procedure, we find the
sequence of modified trajectories at $E_{y}^f =
E_{y,0}^{f}$. Finally, we impose the boundary conditions~(\ref{eq:34})
and set $\epsilon=0$. In this way we find all exclusive trajectories
from the first class sorted by the topological number $m$. The solutions
at the ``wings'' are obtained by taking the trajectories
corresponding to $E_y^f = E_{y,1}^f$ ($E_{y,2}^f$) and deforming them by
decreasing (increasing) $E_y^f$.

\subsection{Exclusive probability}

Let us derive the expression of the form (\ref{eq:9}) for the
exclusive tunneling probability in the sphaleron--driven case. We
start with the semiclassical formula
\begin{equation}
\label{eq:51.3}
{\cal P}_e = \sum_{m=0}^{\infty} A_{e,pot}^{(m)} \cdot
\mathrm{e}^{-F_{e,pot}^{(m)}/\hbar} \;, 
\end{equation}
where the sum runs over all complex trajectories describing the same
process. Note that the terms due to interference between different
trajectories are neglected in Eq.~(\ref{eq:51.3}); we will discuss
them later. One recalls that the number $m$ of
exclusive trajectory increases with the time interval
$\tau_+$ spent by the trajectory in the vicinity of the sphaleron
orbit. In accordance  with the new tunneling mechanism this implies
that the sum 
in Eq.~(\ref{eq:51.3}) is saturated at $m\to+\infty$: the individual
suppressions $F_{e,pot}^{(m)}$ decrease  with $m$ and reach the minimum 
at $m\to +\infty$. This minimum is the overall suppression exponent of
the process,
\begin{equation}
\label{eq:51.5}
F_{e,sph} = \lim_{m\to +\infty} F_{e,pot}^{(m)}\;.
\end{equation}
Note that the value of $F_{e,sph}$ is the same for all trajectories from
the first class\footnote{One proves this by noting that 
  exclusive trajectories at large $m$ are close to the respective
  modified trajectories, and the limit $m \to +\infty$ in
  Eq.~(\ref{eq:51.5}) can be substituted with $\tau_+\to
  +\infty$. Since the modified trajectories sweep the interval 
  $E_{y,1}^f\leq E_y^f\leq E_{y,2}^f$ as $\tau_+$ grows, the limiting
  value $F_{e,sph}$ does not depend on $E_y^f$ within this 
  interval.} and equal to the suppression $F_{sph}$ of inclusive
tunneling probability. This property gives rise to the plateau in the
dependence $F_{e,sph}(E_y^f)$ in Fig.~\ref{fig:14}b. 

Let us now turn to the prefactor. Since the suppressions
$F_{e,pot}^{(m)}$ change at large $m$ in small steps, one may be
tempted to replace the sum in Eq.~(\ref{eq:51.3}) by the
integral and evaluate it in a straightforward way. However, this
replacement is in general incorrect: even  for small change of
$F_{e,pot}^{(m)}$ the change in the exponent
$\e^{-F_{e,pot}^{(m)}/\hbar}$ can be large. 

We proceed carefully. In what follows we restrict our attention to
the plateau case $E_{y,1}^f < E_y^f < E_{y,2}^f$. One starts by
relating the limit $m\to +\infty$ of exclusive quantities to
$A_{sph}$,
\begin{equation}
\label{eq:limem}
\lim_{m\to  +\infty} A_{e,pot}^{(m)} \left| \frac{dF_{e,pot}^{(m)}}
    {dE_y^f}\right|^{-1} =\omega A_{sph}\;.
\end{equation}
This formula is obtained as follows. One changes the
integration variables from $\tau_+$ to $E_y^f$ in the expression
(\ref{eq:44}) for inclusive probability, 
$$
\int d\tau_+ = \int dE_y^f  \sum_m \left|\frac{d\tau_+}{dE_y^f}\right|\;,
$$
where the derivative is taken along the dashed line in 
Fig.~\ref{fig:16}. Comparing the resulting integral with the
relation
$$
{\cal P} = \int \frac{dE_y^f}{\hbar\omega} \,{\cal P}_e
$$
between the inclusive and exclusive probabilities, one expresses the
modified suppression exponent and prefactor  in terms of
$F_{e,pot}^{(m)}$, $A_{e,pot}^{(m)}$,
\begin{equation}
\label{eq:43}
F_\epsilon = F_{e,pot}^{(m)}\;, \qquad\qquad 
A_{pot,\epsilon} = A_{e,pot}^{(m)}\cdot \frac{\sqrt{\pi}}{\omega
  \sqrt{\hbar}} \left[-\frac{d\tau_+}{d\epsilon}\right]^{1/2}
\left|\frac{dE_y^f}{d\tau_+}\right| \;.
\end{equation}
Note that the value of $\epsilon$ in these formulas is fixed by 
the specification of the final oscillator energy $E_y^f$ and topological
number $m$ of the respective trajectory. Now, one notes that the
limit in Eq. (\ref{eq:14b}) can be computed by considering the
subclass of modified trajectories with fixed $E_y^f$. These are close
to the respective exclusive solutions; one uses the latter in  
the r.h.s. of Eq.~(\ref{eq:14b}) and substitutes the limit $\epsilon
\to +0$ with $m\to +\infty$. Then,  Eqs.~(\ref{eq:43}) and the Legendre
transformation (\ref{eq:11}) imply Eq.~(\ref{eq:limem}). 

Now, we exploit the dependence of the individual suppressions
$F_{e,pot}^{(m)}$ on $m$ at large $m$. It is shown in appendix
\ref{sec:linear} that the suppressions approach the limiting value
$F_{sph}$ exponentially,
\begin{equation}
\label{eq:appr}
F_{e,pot}^{(m)}-F_{sph}=
\begin{cases}
\alpha_{even}(E_y^f)\,\e^{-\beta n}~, & m=2n\\
\alpha_{odd}(E_y^f)\,\e^{-\beta n}~~, & m=2n+1
\end{cases}
\end{equation}
where the coefficient $\beta = \tilde{\beta} T_{sph}$ is related to the
positive Lyapunov exponent $\tilde{\beta}$ and period $T_{sph}$ of the
sphaleron orbit. Clearly, $\beta$ does not depend on the final oscillator
energy. Substituting Eqs.~(\ref{eq:appr}), (\ref{eq:limem}) into the
formula (\ref{eq:51.3}), we find,
\begin{equation}
\label{eq:sums}
\begin{split}
{\cal P}_e=\e^{-F_{sph}/\hbar}\cdot\omega A_{sph}
\bigg\{&\left|\frac{d\alpha_{even}}{d E_y^f}\right|
\sum_{n=0}^\infty\exp\left(-\beta n-\frac{\alpha_{even}}{\hbar}
\e^{-\beta n}\right)\\
&+
\left|\frac{d\alpha_{odd}}{d E_y^f}\right|
\sum_{n=0}^\infty\exp\left(-\beta n-\frac{\alpha_{odd}}{\hbar}
\e^{-\beta n}\right)
\bigg\}\;.
\end{split}
\end{equation}
Let us concentrate on the first term in braces, the second term is treated in
the same way. The sum is saturated near the point $n_0$ corresponding
to the maximum of the exponent,
\begin{equation}
\label{eq:51}
n_0=-\frac{1}{\beta}\log\frac{\hbar}{\alpha_{even}}\;.
\end{equation}
Generically, $n_0$ is not integer. Factoring out the value of the
summand at $n=n_0$, one writes, 
\begin{equation}
\label{eq:41}
\sum_{n=0}^\infty\exp\left(-\beta n-\frac{\alpha_{even}}{\hbar}
\e^{-\beta n}\right)=\frac{\hbar}{\alpha_{even}}
\sum_{n=-\infty}^\infty\exp\left(-\beta (n-n_0)-
\e^{-\beta (n-n_0)}\right)\;,
\end{equation}
where in the r.h.s. we extended the sum to all integer $n$ by noting
that the terms at $n<0$ are negligibly small. Substituting this
relation into Eq.~(\ref{eq:sums}), one finally obtains expression
for the exclusive prefactor, 
\begin{equation}
\label{eq:expref}
A_{e,sph}=\hbar\omega A_{sph}\bigg\{\frac{1}{\alpha_{even}}
\left|\frac{d\alpha_{even}}{dE_y^f}\right|
~s\Big(\beta,\log{\frac{\hbar}{\alpha_{even}}}\Big)
+
\frac{1}{\alpha_{odd}}
\left|\frac{d\alpha_{odd}}{dE_y^f}\right|
~s\Big(\beta,\log{\frac{\hbar}{\alpha_{odd}}}\Big)
\bigg\}\;, 
\end{equation}
where 
\begin{equation}
\label{eq:s}
s(\beta,\xi)\equiv\sum_{n=-\infty}^{\infty}
\exp\left(-\beta n-\xi-\e^{-\beta n-\xi}\right)
\end{equation}
is a periodic function of $\xi$ with period $\beta$.

Let us discuss our result. The dependence of the exclusive prefactor
(\ref{eq:expref}) on $\hbar$ is different in the cases $\beta\ll 1$,
$\beta \sim 1$, $\beta \gg 1$. At $\beta \ll 1$ the sum in
Eq.~(\ref{eq:s}) can be replaced by the integral and one obtains
$s(\beta,\xi) \approx 1/\beta$. Then the $\hbar$--dependence of
$A_{e,sph}$ reduces to the simple proportionality law\footnote{Recall
  that $A_{sph}\propto \hbar$.} $A_{e,sph}\propto \hbar^2$. In the
generic case $\beta \sim 1$ one 
observes, besides 
the overall scaling $A_{e,sph}\propto \hbar^2$, the modulation of the
prefactor by the periodic function of $\log\hbar$. The latter
modulation is elusive, however, in models with $\beta \gg
1$. Namely, the periodic nature of $s(\beta,\xi)$ becomes apparent
only at $|\xi| = |\log(\hbar/\alpha_{even})| \sim \beta$ which
corresponds to {\it exponentially} small values of $\hbar\sim
\mathrm{e}^{-\beta}$. 

Realistically, at large $\beta$ one works in the regime $\beta \gg
|\xi|$. In this case the formulas
(\ref{eq:51.5}), (\ref{eq:expref}) are not applicable, since they are
derived under  the assumption $n_0 \gtrsim 1$, see Eq.~(\ref{eq:51}). At
$n_0\ll 1$ one uses the original expression (\ref{eq:51.3}), where the
sums over even/odd $m$ are saturated by the first terms. Then
$A_{e,sph} \propto \hbar$. Let us roughly estimate the relative size
of the terms with $m=0$ and $m=1$.  One takes
$$
F_{e,pot}^{(1)} - F_{sph} =
\mathrm{e^{-c\beta}}(F_{e,pot}^{(0)}-F_{sph})
\;,$$ 
where $c$ is a coefficient of order  $1$, and uses
Eq.~(\ref{eq:limem}) to estimate the prefactors. This yields that the
trajectory with $m=1$ is relevant when
\begin{equation}
\label{eq:52}
\hbar \lesssim \frac{F_{e,pot}^{(0)}-F_{sph}}{c\beta}
\end{equation}
and is negligible at larger $\hbar$. 
One concludes that, depending on the value of $\hbar$, 
the term with $m=0$ or $m=1$ dominates.

The characteristic values of the parameter $\beta$ are related
to the  properties of the unstable periodic orbits, which are fixed in
the model under consideration. As estimated in appendix
\ref{sec:linear}, in the setup (\ref{eq:24}) $\beta  \sim 24$. On the
other hand, the 
numerical quantum mechanical computations are feasible only down to
$\hbar \gtrsim 10^{-2}$. Thus, we are in the regime $\hbar \gg
\mathrm{e}^{-\beta}$, where the exclusive probability is saturated
by the trajectories with $m=0,1$.

\begin{figure}
\centerline{\includegraphics[width=0.5\textwidth]{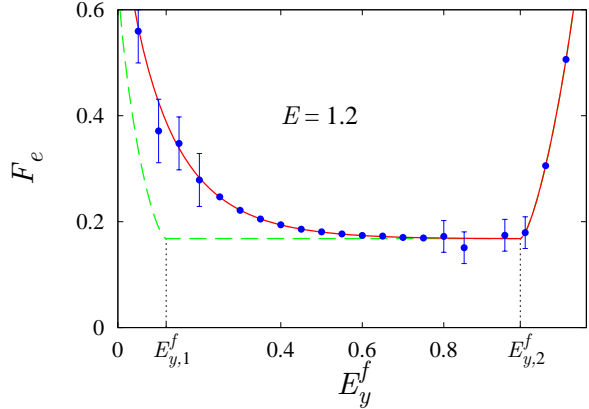}}
\caption{\label{fig:22} Exclusive suppression exponents in the
  sphaleron--driven case: the first exponent $F_{e,pot}^{(0)}$
  (line), the limiting value $F_{e,sph}$ (dashed line) and the fit of
  the exact data with $\hbar \ge 1/120$ (errorbars). The   in--state
  quantum  numbers are the same as in Figs.~\ref{fig:14}b,~\ref{fig:3}b.}
\end{figure}

To see this explicitly, we compare the first suppression exponent
$F_{e,pot}^{(0)}$ and the limiting exponent\footnote{In our case
  $F_{e,pot}^{(1)} = F_{e,sph}$ with good accuracy.} $F_{e,sph}$ with
the exact suppression. Since in our case $A_{e,sph}\propto
\hbar$, one extracts the exact suppression from the fit
(\ref{eq:32}) with $\gamma=1$. We consider separately the exact data
with $\hbar \geq 1/120$ and $\hbar < 1/120$. Using the first set of data,
we obtain the points in Fig.~\ref{fig:22}. The results of the fit
closely follow $F_{e,pot}^{(0)}$ and differ substantially from the
limiting exponent $F_{e,sph}$ in the left part of the graph. One
concludes that at $\hbar\geq 1/120$ the trajectory with $m=0$ saturates
the tunneling probability. 

Second, we analyze the exact quantum data with $\hbar<1/120$. Consider
Fig.~\ref{fig:3}b where the logarithm of the exact tunneling
probability is plotted at several values of $\hbar$. One sees that the
graph at $\hbar = 1/140$ is notably different from the graphs at
larger $\hbar$. We attribute this difference to the contribution of
the trajectory with $m=1$. Indeed, at $E_y^f \approx E_{y,1}^{f}$ the
difference in the suppressions is large, $F_{e,pot}^{(0)} - F_{e,sph}
\sim 0.2$. Then, the estimate (\ref{eq:52}) implies that at $\hbar
\lesssim 1/120$ the first odd trajectory enters into the game. 
The results of the fit of exact data with small $\hbar$ are shown in
Fig.~\ref{fig:14}b. As expected, they coincide
with $F_{e,sph}$ in the leftmost part of the graph. 

Finally, let us briefly discuss interference between
exclusive trajectories. Consider first the case $\beta \sim
1$. The sum (\ref{eq:s}) for the prefactor is then saturated by the fixed
number of terms, $\Delta n=O(\hbar^0)$. Since each term corresponds to
the complex trajectory, the number of trajectories giving substantial 
contribution into the probability is finite, and the interference
between the trajectories is important. This gives rise to oscillations
in the dependence of the probability on the in-- and out--state
quantum numbers $E$, $E_y$,
$E_y^f$. The period of these oscillations tends to
zero as $\hbar\to 0$; thus, they become indiscernible in the
semiclassical limit. However, at finite $\hbar$ the interference is
important. 

In our case of large $\beta$ the exclusive probability is dominated by
two complex trajectories, and the interference picture is seen whenever
the contributions of these trajectories are comparable. In accordance
with the above discussion, this happens at $\hbar\sim 1/140$ and
$E_y^f \approx E_{y,1}^f$, see Fig.~\ref{fig:3}b. The small--scale
oscillations in the right part of the plateau in Fig.~\ref{fig:3}b
are explained as follows. Let us take a look at 
Fig.~\ref{fig:16}. One observes that at $E_{y}^f \approx E_{y,2}^f$ the
trajectory with $m=1$ is almost coincident with the dominant one.
Thus, this trajectory gives substantial contribution into the
probability in the vicinity of $E_{y,2}^f$, and the interference
between the two trajectories is seen in this region.

\section{The limit of small quantum numbers}
\label{sec:limit-small-quantum}
According to the common lore low--lying quantum states are ``not
semiclassical.'' Indeed, one cannot use the semiclassical
expressions for the wave functions of these states  in the majority of
applications: at $E\sim\hbar$ the momentum is parametrically small
and the semiclassical approximation is  not justified. One finds,
however, that tunneling processes are very special in regard of
low--lying states. Namely, the semiclassical tunneling probability
depends only on exponentially small tails of in- and out-state
wave functions; these tails can be computed semiclassically even at
small values of respective quantum numbers.

In this section we generalize the semiclassical method to the 
case of tunneling from the
low--lying in-states of $y$-oscillator, $E_y \sim \hbar $. At the
same time the total energy is assumed to be semiclassically large,
$E\sim 1$.  To be concrete, we take $E_y = \hbar \omega/2$, which
corresponds to 
the oscillator ground state. Note, however, that 
the method of this section can be used for other low--lying oscillator
states as well. 

Let us address the following questions: \\
(i) Is it legitimate to use the
semiclassical approximation for the 
wave function of the oscillator deep inside the classically
forbidden region, $|y| \gg \hbar^{1/2}$ ?\\
(ii) Is the integral over initial states in Eq.~(\ref{eq:7}) saturated
deep inside the classically forbidden region at small $E_y$?\\
If the answers to the above questions are positive, one can
use the semiclassical expressions  for the
probability of tunneling from the ground state of
$y$-oscillator (e.g. Eqs.~(\ref{eq:26}), 
(\ref{eq:21}) in the case of potential tunneling).

To answer the first question, we compare the semiclassical and exact
oscillator wave functions in the case of ground state, 
$E_y = \hbar\omega/2$: 
\begin{align}
\label{eq:39}
\psi_{y,s}(y) &= \left(\frac{\omega}{2\pi p_{y,i}(y)}\right)^{1/2}
\cdot \mathrm{exp}\left(\frac{i}{\hbar}\int_{\sqrt{2E_y}/\omega}^{y} 
p_{y,i}(y') \,dy' + \frac{i\pi}{4}\right)\;,\\
\label{eq:40}
\psi_{y,0}(y) &= {\left( \frac{\omega}{\pi \hbar} 
\right)}^{1/4}\cdot \mathrm{exp}\left(-\omega y^2/2\hbar\right)\;.
\end{align}
In the
above expressions $|y|$ is large and $p_{y,i}(y) = \sqrt{2E_y - \omega^2
  y^2}$. Equations~(\ref{eq:39}) and
(\ref{eq:40}) look  quite  
different: the exact wave function involves the factor
$\hbar^{-1/4}$ which is not present in the semiclassical expression.
However, substituting $E_y = \hbar \omega/2$ into the leading exponent
of Eq.~(\ref{eq:39}), one finds,
$$
\int_{\sqrt{\hbar/\omega}}^{y} dy' p_i(y') = i\omega y^2/2 -
i \hbar/4 - i\hbar/2
\,\log\left(2 y\sqrt{{\omega}/{\hbar}}\right) + O(\hbar^2)\;. 
$$
Thus, up to high--order semiclassical corrections
\begin{equation}
\label{eq:6.1}
\psi_{y,0}(y)  =
\left(\frac{\pi}{\mathrm{e}}\right)^{1/4}\cdot \psi_{y,s}(y)\Big|_{E_y
  = \hbar \omega/2}\;.
\end{equation}
One concludes that the two wave functions are related  by the simple
renormalization factor $(\pi/\mathrm{e})^{1/4}$. 

The relation (\ref{eq:6.1}) is not surprising. Indeed, the standard
derivation of the semiclassical wave function (\ref{eq:39}) proceeds in two
steps. First, one solves the Schr\"odinger equation with 
$\psi_{y,s}(y) = C\cdot\mathrm{e}^{i\sigma(y)/\hbar}$ considering $\sigma/\hbar
\gg 1$. This is certainly valid deep inside the classically forbidden
region, even for $E_y = \hbar\omega/2$. The second step is the evaluation
of the normalization constant $C$ by taking the integral $\int
|\psi_{y,s}(y')|^2 dy' = 1$. At small $E_y$ the
latter integral is saturated at $y\approx 0$, i.e. right in the
vicinity of the turning points, where the semiclassical expression
(\ref{eq:39}) is not applicable. Consequently, the semiclassical
calculation produces an incorrect value for the constant $C$ at 
$E_y = \hbar\omega/2$. Equation (\ref{eq:6.1}) shows that the correct
value is $(\pi/e)^{1/4}$ times larger than the one obtained
semiclassically. 

We have the following answer to the question (i): the semiclassical
expression (\ref{eq:4}) can be used at $E_y = \hbar\omega/2$ deep
inside the classically forbidden region; however, the final result for the
probability should be multiplied by the correction factor 
$(\pi/\mathrm{e})^{1/2}$. 

Let us address the question (ii). Consider the complex trajectory
$\boldsymbol{x}(t)$ in the in-region. One finds,
\begin{equation}
\label{eq:6.2}
x(t) \to p_{x,i}(t - t_i) + x_i\;, \qquad y(t)
\to a \mathrm{e}^{-i\omega t} + \bar{a} \mathrm{e}^{i\omega t} \qquad \mbox{as}
\qquad t\to -\infty\;.
\end{equation}
The initial boundary conditions (\ref{eq:6}) guarantee that the
quantities $p_{x,i} = \sqrt{2(E - E_y)}$ and $a\bar{a} = E_y/2\omega^2$ are
real. Therefore, one can define two {\it real} parameters $T$, $\theta$
by the relations
\begin{equation}
\label{eq:6.3}
\mathrm{Im}\, x_i = -p_{x,i} T\;, \qquad a^* = \bar{a}
\mathrm{e}^{-2\omega T-\theta}\;. 
\end{equation}
As discussed in Refs.~\cite{Bonini:1999kj,Bezrukov:2003yf}, 
these parameters are in
one-to-one correspondence with the in-state quantum numbers $E$,
$E_y$. In other words, $T$ and $\theta$
provide an alternative parameterization of tunneling trajectories. 
Note that $T = \theta = 0$ represent 
classically allowed transitions, $\boldsymbol{x}(t) \in
\mathbb{R}$. On the other hand, the limit $\theta \to +\infty$ 
corresponds to $E_y\to 0$. Indeed, in this limit one obtains $a\to 0$
and $\bar a$ finite \cite{Bonini:1999kj,Bezrukov:2003yf}, which are
the Feynman boundary
conditions for tunneling from the ground state. 
From Eq. (\ref{eq:6.2}) one finds that
$|y_i| \to |\bar{a}| \gg \hbar^{1/2}$.
Thus, the integral over initial states in Eq. (\ref{eq:7})
is saturated deep inside the classically forbidden region, where the
semiclassical expression for the in-state wave function is
trustworthy. 

One concludes that, apart
from the additional multiplier $(\pi/\mathrm{e})^{1/2}$, the
semiclassical expressions 
for the tunneling probability (\ref{eq:9}),
such as Eqs.~(\ref{eq:26}), (\ref{eq:21}), are still
applicable  at $E_y =  \hbar \omega /2$. 

Note that in the considered case of tunneling from the ground state
the expressions 
(\ref{eq:26}), (\ref{eq:21}) depend on $\hbar$ in non-trivial
way through 
$E_y = \hbar \omega/2$. It is convenient to extract this dependence
explicitly and bring the expression for the tunneling probability into
the form (\ref{eq:9}) with $F$ independent of $\hbar$ and $A$ having
only the power-law dependence. This is done in 
appendix~\ref{sec:state-parameters}, the result is
\begin{equation}
  \label{eq:6.5}
  F_{pot,0} = \lim_{E_y\to +0} F_{pot}\;, \qquad A_{pot,0} =
  \left(\pi/\hbar\right)^{1/2}\,\mathrm{e}^{\theta_0/2}
  \lim_{E_y\to +0} A_{pot}\;.
\end{equation}
where $F_{pot}$ and $A_{pot}$ are the standard semiclassical
expressions for the suppression exponent and prefactor. The quantity 
$\theta_0$ entering Eq.~(\ref{eq:6.5}) is extracted from the
small--$E_y$ asymptotic of the leading exponent 
\begin{equation}
\label{eq:50}
F_{pot} = F_{pot,0}+ \frac{E_y}{\omega} \log(2E_y/\omega)-
\frac{\theta_0+1}{\omega}E_y + O(E_y^2)\;.
\end{equation}
Let us remark on Eqs.~(\ref{eq:6.5}). First, note that $A_{pot,0}$
contains the additional factor $\hbar^{-1/2}$ as compared to
the case of highly excited in-states. Second, we did
not use the dynamical properties of complex trajectories in the
derivation of  Eqs.~(\ref{eq:6.5}). Thus, the above expressions are
valid both for inclusive and exclusive processes. They also hold 
in the case of sphaleron--driven tunneling, where one substitutes
$F_{pot} \to F_{sph}$, $A_{pot}\to A_{sph}$ in Eqs.~(\ref{eq:6.5}). In
particular, for the prefactors of 
inclusive processes one has $A_{pot,0} \propto
\hbar^0$ and $A_{sph,0} \propto \hbar^{1/2}$. 

\begin{figure}
\centerline{\includegraphics[width=.45\textwidth]{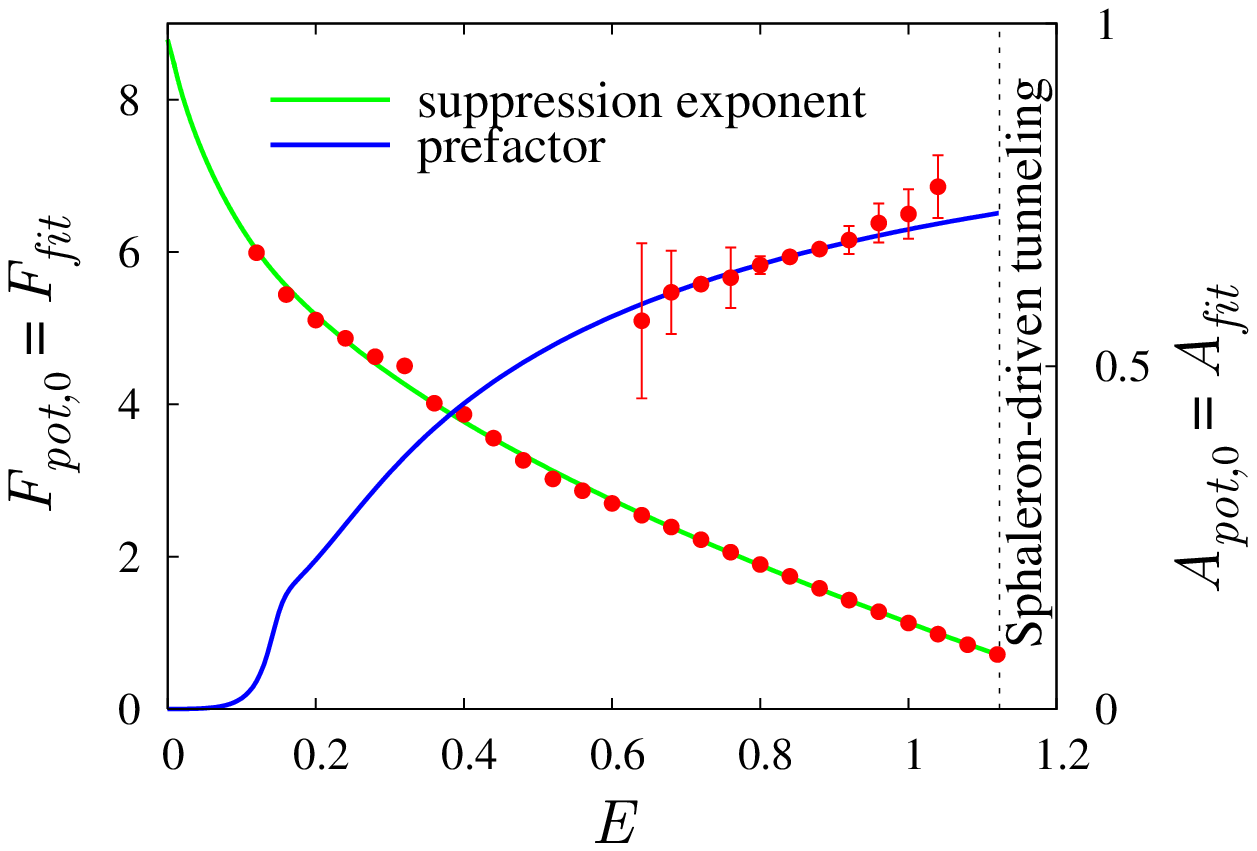}~~~~
\includegraphics[width=.45\textwidth]{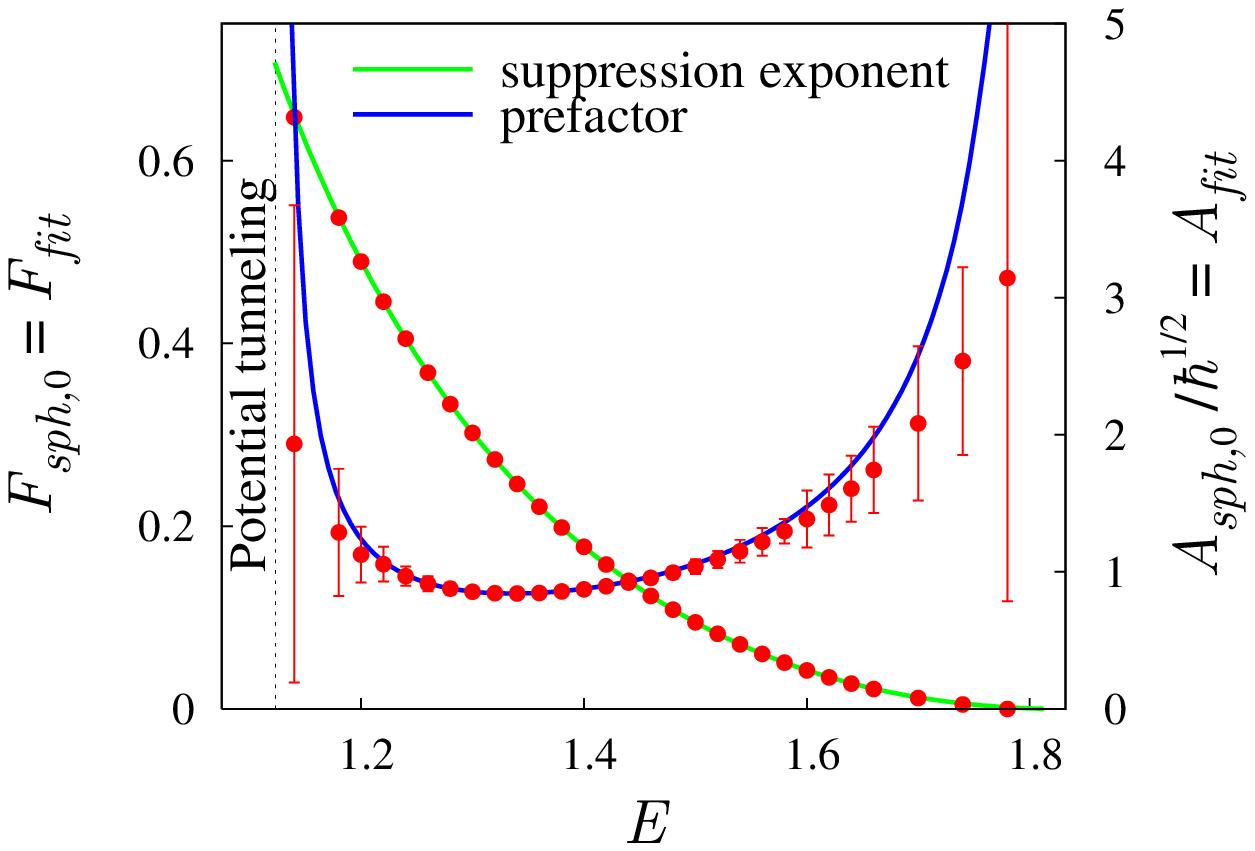}}
\hspace{3.9cm}(a)\hspace{7.8cm}(b)
\caption{\label{fig:18} Comparison between the exact (points) and
  semiclassical (lines) results for the suppression exponent and
  prefactor at $E_y = \hbar \omega/2$ in the cases of (a) potential
  and (b) sphaleron--driven tunneling. Note that $E_c (0) \approx
  1.1$. Errorbars   represent uncertainty of the fit (\ref{eq:32}).}  
\end{figure}

Finally, it is worth mentioning that the first of Eqs.~(\ref{eq:6.5}),
namely, the limiting relation between the suppression exponents of
tunneling from the low--lying and highly excited in-states is known in
field theory as the Rubakov--Son--Tinyakov
conjecture~\cite{Rubakov:1992ec}. We proved this conjecture in 
quantum mechanical setup.

\begin{figure}
\centerline{\includegraphics[width=.5\textwidth]{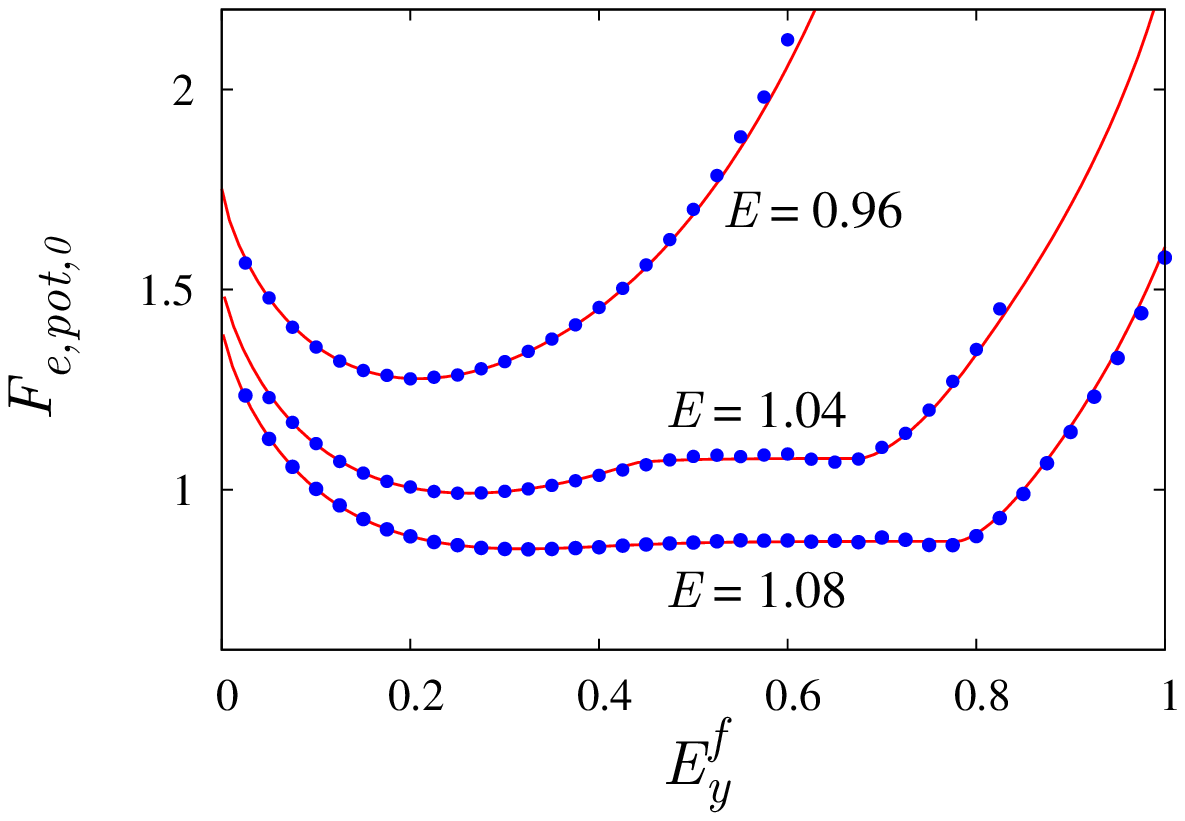}~~~~
\includegraphics[width=.5\textwidth]{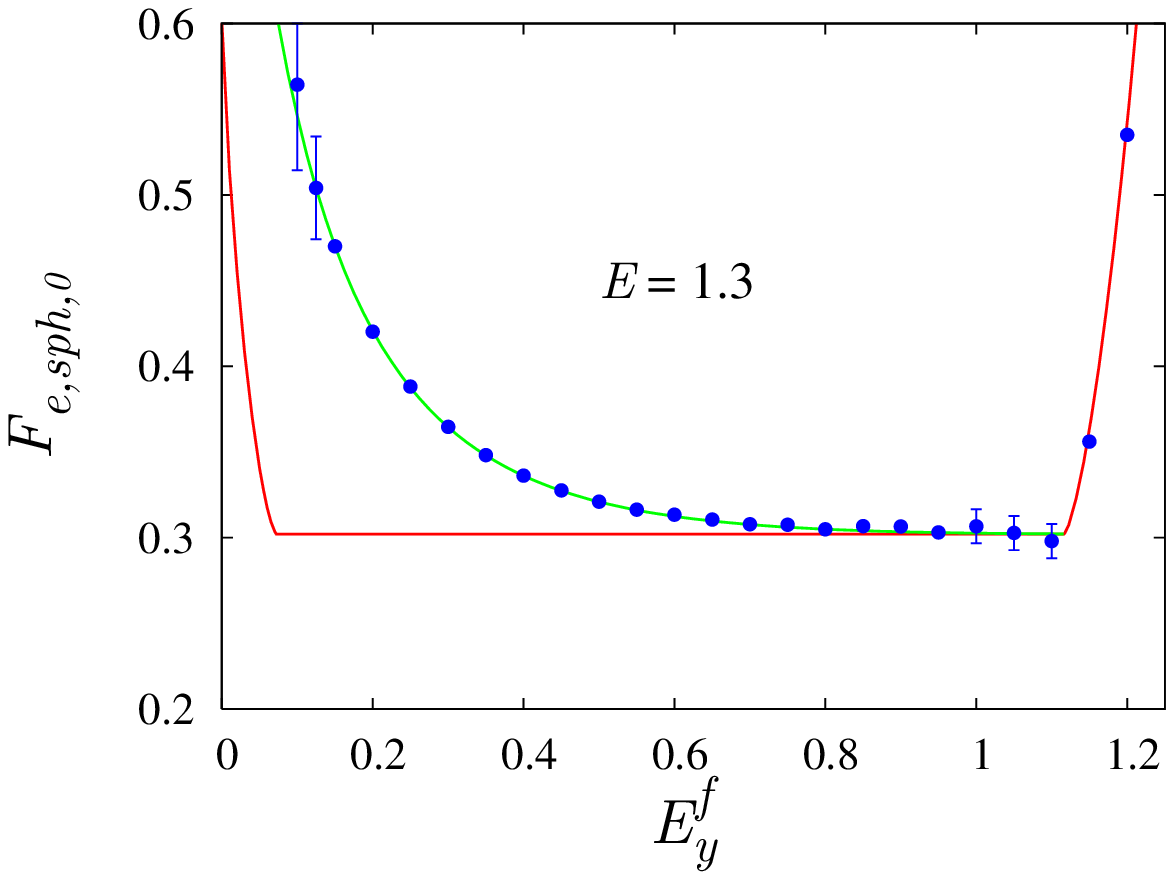}}
\hspace{4.0cm}(a)\hspace{8.3cm}(b)
\caption{\label{fig:20} Comparison between the exact (points) and
  semiclassical (lines) results for the suppression exponent of
  exclusive processes at $E_y = \hbar \omega/2$ in the cases of (a) potential
  and (b) sphaleron--driven tunneling. 
In the case (b) the exact data coincide with the first exponent
  $F_{e,pot,0}^{(0)}$. Errorbars represent uncertainty of the fit
  (\ref{eq:32}).}  
\end{figure}

We close this section by comparing the semiclassical
results for the suppression exponent and prefactor,
Eqs. (\ref{eq:6.5}), with the results extracted by the fit
Eq.~(\ref{eq:32}) from the solution of
the Schr\"odinger equation. The comparison in the
cases of potential and sphaleron--driven tunneling is presented in
Figs.~\ref{fig:18}a, \ref{fig:18}b for inclusive and in
Figs.~\ref{fig:20}a, \ref{fig:20}b for exclusive processes. In the
latter case we compare the exact suppression exponent with the
suppression of the first exclusive trajectory.
One observes nice agreement. 

\section{Summary and Discussion}
\label{sec:discussion}
In this paper we investigated the mechanism of tunneling via unstable
semiclassical solutions (sphaleron--driven tunneling) which  
governs the processes of multidimensional tunneling at energies higher
than some critical value $E_c$. 
There were two aspects in our study. First, we analyzed the
experimental signatures of sphaleron--driven tunneling. These are
suppression of the tunneling probability by the additional power of the
semiclassical parameter $\hbar$ and substantial
widening of the final--state distributions 
as compared to the case of ordinary barrier
tunneling. 

The second aspect of this paper was related to the development of the
modified semiclassical technique (the method of
$\epsilon$--regularization), which is applicable in the case of
sphaleron--driven tunneling. This method is completely general;
it was derived from first principles using the formal operations with
the path integral. Similar modified technique has been  
implemented in several quantum
mechanical~\cite{Bezrukov:2003yf,Levkov:2007e} and field
theoretical~\cite{Bezrukov:2003er} tunneling problems. Using the
modified method, we obtained expressions for the inclusive and
exclusive tunneling probabilities in the
case of sphaleron--driven mechanism, investigated the ``phase
transition'' between the cases of potential and sphaleron--driven
tunneling. We also derived relation between the
probabilities of tunneling from the low--lying and highly excited
in-states. 

Our results for the power--law dependences of the semiclassical
prefactor are summarized in Table~\ref{tab:1}. 

\begin{table}[htb]
\centerline{\begin{tabular}{|ll||l|l|}
    \hline
    & &Potential & Sphaleron--driven \\
    \hline\hline
    Inclusive, &$E_y\sim 1$           & $\hbar^{1/2}$ & $\hbar$ \\ 
    Exclusive, &$E_y\sim 1$            & $\hbar$ & $\hbar^2$ \\
    Inclusive, &$E_y\sim \hbar $ & $\hbar^0$ & $\hbar^{1/2}$\\
    Exclusive, &$E_y \sim \hbar $ & $\hbar^{1/2}$ & $\hbar^{3/2}$\\
    \hline
  \end{tabular}}
\caption{\label{tab:1} Summary: the power--law dependences of
  the tunneling probability in two dimensions. } 
\end{table}

Let us comment on the relation between the mechanism of
sphaleron--driven tunneling and Wilkinson formula for the tunnel
energy splitting~\cite{Wilkinson:1986,Takada:1994,Creagh:2001,Creagh:2006}.  
The latter formula is applicable in the cases of near--integrable (as
opposed to completely integrable) systems with double--well potentials.
It is based on the following property of near--integrable dynamics:
tunneling trajectories stemming from the wells of near--integrable
system do not end up in opposite wells  (as in the integrable
case), but rather get attracted to a certain  unstable periodic
orbit\footnote{This orbit is the intersection of the Lagrange
  manifolds associated with the two wells~\cite{Wilkinson:1986}.}.
Due to this feature the 
splitting in Wilkinson formula is suppressed by the additional
factor $\hbar^{1/2}$ as compared to the case of completely integrable
system. One observes that, technically, the reason for this factor
is similar to that in the  mechanism of
sphaleron--driven tunneling considered in this paper. However,
the two cases are physically different: transition to the
sphaleron--driven regime is unrelated to the transition from
integrable to  non--integrable 
dynamics. In addition, the relevant periodic
orbit in the Wilkinson formula is complex while the sphaleron 
orbit is real. 

We finish this paper with remarks on the recent
observation~\cite{Takahashi:2006} that the new tunneling mechanism
generically leads to anomalously large times of tunneling. Indeed,
the semiclassical trajectories describing sphaleron--driven
transitions spend infinite time interval in the
vicinity of the sphaleron orbit. Clearly, the time scale $\Delta t$ of
such transitions should be large, in particular, one expects $\Delta
t\to +\infty$ as $\hbar\to 0$. A rough estimate of $\Delta t$ can be
obtained as follows.
Due to quantum fluctuations the system cannot approach the
sphaleron orbit in the phase space closer than at the distance
determined by the uncertainty principle,
 $\Delta \boldsymbol{p} \Delta
\boldsymbol{x} \sim \hbar$. The semiclassical trajectories starting in the
vicinity of unstable sphaleron go away from it {\it exponentially}
with time; thus it takes them the time 
$\Delta t\sim \log |\Delta \boldsymbol{p}|,
\log|\Delta\boldsymbol{x}|$ to leave the sphaleron neighborhood. 
This translates into the characteristic lifetime of the sphaleron
$\Delta t \propto |\log \hbar|$, which sets the characteristic time
scale for sphaleron--driven tunneling. 
The dependence of tunneling time on $\hbar$ provides another
possible experimental signature of the new tunneling mechanism. Yet more
signatures can be found by analyzing the probability
distribution over tunneling time. The modified semiclassical
method proposed in this paper allows comprehensive study of these
issues which will be published elsewhere~\cite{Levkov:2008}.

\paragraph*{Acknowledgments.} We are indebted to 
F.L.~Bezrukov, S.V. Demidov, D.S. Gorbunov, M.V.~Libanov, N.S.~Manton,
V.V.~Nesvizhevsky and V.A.~Rubakov  for useful and stimulating
discussions. This work was  supported in part by the RFBR grant
08-02-00768-a, Grants of the President of Russian Federation
NS-1616.2008.2 and MK-1712.2008.2 (D.L.), Grant of the Russian Science
Support Foundation (A.P.), the Fellowships of the ``Dynasty''
Foundation (awarded by the Scientific board of ICPFM) (D.L. and A.P.)
and the Tomalla Foundation (S.S.). The numerical calculations 
were performed on the Computational cluster of the Theoretical
division of INR RAS.

\appendix
\section{Semiclassical tunneling probability}
\label{sec:eval-pre-expon}
In this appendix we give details of the standard method of
complex trajectories. The main idea of the method is presented in
Sec.~\ref{sec:semicl-form-tunn}. 

Our starting point is the path integral representation (\ref{eq:7})
for the out-state wave function $\Psi_f$. This representation contains
two main ingredients, the in-state 
$\Psi_i$ and the quantum propagator written as a path integral.
In accordance with the discussion in the main body of the paper, the
in-state has definite values of the total
energy $E$ and $y$-oscillator energy $E_y$. One writes
$\Psi_i(x,\,y)$ as a product $\psi_x(x)  \cdot
\psi_{y}(y)$, where $\psi_x$ is a
plane wave  with momentum $p_{x,i} = \sqrt{2(E-E_y)}$ and unit flux
normalization, while $\psi_{y}$ represents the semiclassical
wave function of the oscillator with energy $E_y$. Combining $\psi_x$ and
$\psi_y$, one obtains,
\begin{equation}
  \label{eq:47}
  \Psi_i(x,y)  = \left(\frac{\omega}{2\pi p_{y,i}(y)p_{x,i}} \right)^{1/2}
  \cdot \exp\left(\frac{i}{\hbar} B_i(x,y) + \frac{i\pi}{4}\right)\;.
\end{equation}
In this formula $p_{y,i}(y) = \sqrt{2E_y - \omega^2y^2}$ is
the $y$ component of the momentum in the in-region $x\to -\infty$, 
while
\begin{equation}
  \label{eq:3}
  B_i(x,y) = p_{x,i} x + \int_{\sqrt{2E_y}/\omega}^{y} p_{y,i}(y') \,
  dy'
\end{equation}
stands for the classical action in this region. 
Note that in Eq.~(\ref{eq:47}) we keep 
only one of the two exponents
entering the standard expression for the oscillator wave function. The
reason is that 
$\Psi_i(x,y)$  will be used deep inside the classically
forbidden region, where the omitted exponent is
negligible\footnote{We assume appropriate choice of the branch of
  $p_{y,i}(y)$, see e.g. Ref.~\cite{Berry}.}. 

At small $\hbar$ the path integral for the quantum propagator is
evaluated by the saddle--point technique. The result is given by the
Van Vleck formula~\cite{VanVleck,Miller}, 
\begin{equation}
  \label{eq:66}
  \left.\int [d\boldsymbol{x}]\right|_{\boldsymbol{x}_i}^{\boldsymbol{x}_f} \,   
  \mathrm{e}^{iS[\boldsymbol{x}]/\hbar} =
  \frac{\mathrm{e}^{iS[\boldsymbol{x}^{(s)}]/\hbar}}{2\pi i \hbar} 
  \cdot  \left[ \det \frac{\partial^2 S}{\partial
        \boldsymbol{x}_i \partial\boldsymbol{x}_f}\right]^{1/2}\;.
\end{equation}
We refer the interested reader to Ref.~\cite{Kleinert} for
derivation. The formula (\ref{eq:66}) is written in terms of
the semiclassical trajectory $\boldsymbol{x}^{(s)}(t)$, which has the meaning
of a saddle--point path saturating the path integral. This trajectory satisfies
the classical equations of motion; it starts from  $\boldsymbol{x} =
\boldsymbol{x}_i$ at $t = t_i$ and arrives to $\boldsymbol{x} =
\boldsymbol{x}_f$ at $t = t_f$. Below we omit the superscript $(s)$ of
the semiclassical trajectory. 

We substitute the semiclassical expressions 
(\ref{eq:47}) and (\ref{eq:66}) into Eq.~(\ref{eq:7}) and take 
the saddle--point integral over $\boldsymbol{x}_i$. The result for the
out-state wave function has the exponential form~(\ref{eq:8}),
where $D^{-1/2}$ collects all prefactors including the determinant
due to the saddle--point integration; we will evaluate $D$ below.
Note that integration over $\boldsymbol{x}_i$ changes initial
  conditions for the trajectory $\boldsymbol{x}(t)$. Namely, the
extremum of the leading exponent $S+B_i$ with respect to
$\boldsymbol{x}_i$ is achieved when 
\begin{equation}
  \label{eq:13}
  \dot{x}_i = p_{x,i}\;, \qquad \qquad
  \dot{y}_i = p_{y,i}(y_i)\;.
\end{equation}
One finds that these conditions are equivalent to the fixation of the
in-state quantum numbers, Eqs.~(\ref{eq:6}).

A remark is in order. We consider the case of classically
forbidden transitions which implies that there is {\it no} real
solutions starting in the in-region with fixed $E$, $E_y$ and arriving
into the out-region at $t = t_f$. Accordingly, the saddle--point trajectory
$\boldsymbol{x}(t)$ is {\it complex}. 

Given the final state wave function, one evaluates the inclusive
probability of tunneling performing the saddle--point integration over
$y_f$ in  Eq.~(\ref{eq:19}). One obtains the familiar
semiclassical formula~(\ref{eq:9}) for the inclusive tunneling
probability, where the leading exponent $F_{pot}$ is given by the
value of the action functional (\ref{eq:26}) evaluated on the complex
trajectory $\boldsymbol{x}(t)$. The prefactor will be discussed
shortly.

Let us comment on the final boundary
conditions~(\ref{eq:20}) obtained after integration over $y_f$.  One
finds that all of them have different origin. Namely,
the final value of $x_f$ is already fixed in the probability
formula~(\ref{eq:19}); the  condition $\dot{y}_f = \dot{y}_f^*$
corresponds to the extremum  of the leading
semiclassical exponent with respect to $y_f$. The third condition,
namely, reality of $y_f$, follows from uniqueness of complex
trajectory, which is assumed\footnote{The condition $y_f = y^*_f$ should
  be relaxed if several complex trajectories contribute into
  the out-state of the process. In this case the
  trajectories with complex $y_f$ give rise to interference 
  terms in the tunneling probability.}. One also notes
that the trajectory $\boldsymbol{x}(t)$ is {\it real} in the 
out-region. Indeed, $x_f$, $y_f$, $\dot{y}_f$ are real due to the boundary
conditions at $t = t_f$, while $\dot{x}_f\in \mathbb{R}$ 
due to conservation of total real energy $E$.

\begin{figure}[htb]
\centerline{\includegraphics[width=0.6\textwidth]{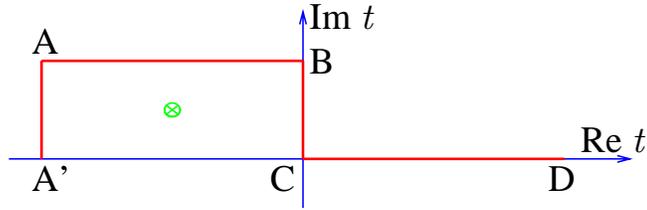}}
\caption{\label{fig:5} The  tunneling solution is obtained along the
  contour $A'ABCD$ in the complex time plane. The branch point 
  of the solution is shown by the cross.}
\end{figure}

We finish the discussion of the
complex trajectory $\boldsymbol{x}(t)$ by remarking shortly on the
important issue of non--trivial contour in complex
time~\cite{Miller,Bonini:1999kj}. We noted already that
$\boldsymbol{x}(t)$ is real in the out-region. Solving the classical
equations of motion  
backwards in time, one concludes that at $t\in \mathbb{R}$ the
trajectory is real as well. Thus, $\boldsymbol{x}(t)$ in real time
corresponds to classically allowed reflection from the barrier;
clearly this solution is not relevant for the description of
tunneling. One observes, however, that the semiclassical trajectory
has a branch point in the complex time plane, see
Fig.~\ref{fig:5}. The solution describing tunneling is obtained along
the contour winding around this branch point (the contour A'ABCD in
the figure).

Let us evaluate the prefactor $A_{pot}$. There are two non--trivial
contributions into $A_{pot}$ coming from the saddle--point
integrations in Eqs.~(\ref{eq:7}) and (\ref{eq:19}). 
We start with the prefactor $D^{-1/2}$ of the out-state wave function,
Eq.~(\ref{eq:8}). One finds that
\begin{align}
  \notag
  D &=  \frac{2\pi}{\omega}\,\dot{y}_i\, \dot{x}_i \times
  (2\pi i \hbar)^2 \left.\left[\det \frac{\partial
    \dot{\boldsymbol{x}}_f} {\partial\boldsymbol{x}_i}
     \right]^{-1} 
      \times \frac{1}{(2\pi i\hbar)^2}\det {\frac{\partial^2 (S +
          B_i)}{\partial \boldsymbol{x}_i^2}}  
        \right|_{\boldsymbol{x}_f = \mathrm{const}} \\
        \label{eq:12}
 &= \frac{2\pi}{\omega} \,\dot{y}_i\,\dot{x}_i  \cdot  \det \left
 . {\frac{\partial^2 (S + B_i)}{\partial \boldsymbol{x}_i \partial
     \dot{\boldsymbol{x}}_f}} \right|_{\boldsymbol{x}_f = \mathrm{const}},
\end{align}
where the three multipliers in the first line correspond respectively
to the prefactors of the in-state (\ref{eq:47}),
propagator~(\ref{eq:66}) and the determinant due to the saddle--point
integration over initial coordinates $\boldsymbol{x}_i$. 
In the second equality we combined
the multipliers.  

One substitutes the explicit form of $B_i$ into Eq.~(\ref{eq:12}) and
represents the factor $\dot{y}_i\,\dot{x}_i$ as the determinant of 
diagonal matrix. The result is
\begin{equation}
\label{eq:B1}
D = \frac{2\pi}{\omega}\left.\det\left[ \begin{pmatrix} \dot{x}_i & 0 \\ 0 &
    \dot{y}_i \end{pmatrix} 
  \frac{\partial \dot{\boldsymbol{x}}_i}
       {\partial \dot{\boldsymbol{x}}_f} +
\begin{pmatrix} 0 & 0 \\ 0 & \omega^2 y_i \end{pmatrix}
\frac{\partial \boldsymbol{x}_i}
     {\partial \dot{\boldsymbol{x}}_f}\right]
     \right|_{\boldsymbol{x}_f = \mathrm{const}}\;.
\end{equation}
This last determinant can be evaluated by considering the set of linear
perturbations $\delta \boldsymbol{x}(t)$, which satisfy the linearized
equation of motion~(\ref{eq:4}) in the background of complex
trajectory $\boldsymbol{x}(t)$. The two basic perturbations will be
particularly important, $\boldsymbol{\psi}^{(1)}(t) = \partial
  \boldsymbol{x}(t)/\partial \dot{x}_f$ and
  $\boldsymbol{\psi}^{(2)}(t) = \partial \boldsymbol{x}(t)/\partial 
  \dot{y}_f$, where the derivatives  are taken at $\boldsymbol{x}_f =
  \mathrm{const}$. One can explicitly check that
  $\boldsymbol{\psi}^{(n)}(t)$ satisfy Eq.~(\ref{eq:4}). 

Consider linear energy increment due to perturbation
$\delta \boldsymbol{x}(t)$, 
\begin{equation}
  \label{eq:B3}
  \delta E [\delta \boldsymbol{x}] = \dot{x}_i
  \delta\dot{x}_i + \delta E_y[\delta \boldsymbol{x}]\;,
\end{equation}
where the increment $\delta E_y[\delta
\boldsymbol{x}]$ of the initial oscillator energy is given by 
Eq.~(\ref{eq:67}).
Clearly, $\delta E$ is conserved; in particular, 
it can be computed at $t= t_f$. Using the energy increments, one
rewrites the determinant~(\ref{eq:B1}) as
\begin{align}
  \notag
  D &= \frac{2\pi}{\omega}\det \left[ 
    \begin{array}{cc} 
      \delta E[\boldsymbol{\psi}^{(1)}] -  \delta
      E_y[\boldsymbol{\psi}^{(1)}] & 
      \delta E[\boldsymbol{\psi}^{(2)}] -  \delta
      E_y[\boldsymbol{\psi}^{(2)}] \\
      \delta  E_y[\boldsymbol{\psi}^{(1)}] & 
      \delta  E_y[\boldsymbol{\psi}^{(2)}]  
    \end{array}
  \right] \\
  \label{eq:B4}
  &= \frac{2\pi}{\omega}\det \left[ 
    \begin{array}{cc} 
      \dot{x}_f & \dot{y}_f\\
      \delta  E_y[\boldsymbol{\psi}^{(1)}] & 
      \delta  E_y[\boldsymbol{\psi}^{(2)}]  
    \end{array}
  \right] \\
  \notag 
  &= \frac{2\pi}{\omega}\, \delta E_y[\dot{x}_f \boldsymbol{\psi}^{(2)}
  - \dot{y}_f 
  \boldsymbol{\psi}^{(1)}]\;,
\end{align}
where in the last two equalities we added the second row to the first,
computed explicitly $\delta E[\boldsymbol{\psi}^{(1)}] = \dot{x}_f$,
$\delta E[\boldsymbol{\psi}^{(2)}] = \dot{y}_f$ and used linearity of
$\delta E_y$. Let us introduce the linear combination 
$$
\delta \boldsymbol{x}^{(1)}(t) =  -
\dot{y}_f/\dot{x}_f \,\boldsymbol{\psi}^{(1)}(t)
+\boldsymbol{\psi}^{(2)}(t)\;.
$$ 
This vector satisfies
the linearized equations of motion with the Cauchy data at $t = t_f$,
\begin{equation}
  \label{eq:22}
  \delta\dot{\boldsymbol{x}}^{(1)}_f = (-\dot{y}_f/\dot{x}_f,\; 1)\;, \qquad
  \delta\boldsymbol{x}_f^{(1)} = 0\;. 
\end{equation}
Note that $\delta E[\delta\boldsymbol{x}^{(1)}] = 0$. Using 
$\delta \boldsymbol{x}^{(1)}$, one finds,
\begin{equation}
  \label{eq:B5}
  D = \frac{2\pi \dot{x}_f}{\omega}\, \delta E_y[\delta\boldsymbol{x}^{(1)}]\;.
\end{equation}
So, we have obtained the final state prefactor $D^{-1/2}$ in
Eq.~(\ref{eq:8}). 

Let us proceed with the probability prefactor $A_{pot}$. Taking the
saddle--point integral in Eq.~(\ref{eq:19}), one arrives at the
following formula,
\begin{equation}
  \label{eq:17}
  A_{pot} = \frac{ \dot{x}_f \sqrt{\pi\hbar}}{|D| \sqrt{D'} }\;,
\end{equation}
where the factor coming from the integration is
\begin{equation}
  \label{eq:18}
  D' = \left.\mathrm{Im} \frac{\partial \dot{y}_f}{\partial y_f}
  \right|_{E,\,E_y,\,x_f = \mathrm{const}}\;. 
\end{equation}
One evaluates $D'$ considering the perturbation 
$$
\delta \boldsymbol{x}^{(2)}(t) = -\omega^2 \frac{y_f}{\dot{x}_f}
\,\boldsymbol{\psi}^{(1)}(t) +  
 \left.\frac{\partial \boldsymbol{x}(t)}
{\partial y_f}\right|_{\dot{\boldsymbol{x}}_f,\,  x_f = \mbox{\footnotesize
    const}}\;.
$$
It
satisfies the linearized  
equations of motion (\ref{eq:4}) with the final Cauchy data
\begin{equation}
  \label{eq:23}
  \delta \dot{\boldsymbol{x}}^{(2)}_f = (-\omega^2 y_f/\dot{x}_f,\;
  0)\;, \qquad \delta\boldsymbol{x}^{(2)}_f = (0,\; 1)\;,
\end{equation}
so that  $\delta E[\delta \boldsymbol{x}^{(2)}] = 0$. 

One notices that $\delta\boldsymbol{x}^{(1)}(t)$ and
$\delta\boldsymbol{x}^{(2)}(t)$ do not change the values of $x_f$ and 
$E$. Moreover, these two are the only linearly independent
perturbations which have this property. Thus, the perturbation
\begin{equation}
  \label{eq:B6}
  \boldsymbol{\rho} (t) = \left. \frac{\partial
      \boldsymbol{x}(t)}{\partial y_f} \right|_{E, \,E_y, \,x_f =
    \mbox{\footnotesize const}}
\end{equation}
is their linear combination, $\boldsymbol{\rho}(t) =
\alpha\, \delta\boldsymbol{x}^{(1)}(t) + \beta\,
\delta\boldsymbol{x}^{(2)}(t)$. One has,
\begin{eqnarray*}
  \boldsymbol{\rho}_y(t_f) = 1 &\Rightarrow & \beta =
  1\;,\\ 
  \delta E_y[\boldsymbol{\rho}] = \alpha \,\delta E_y[\delta
  \boldsymbol{x}^{(1)}]  +  \beta\, \delta
  E_y[\delta\boldsymbol{x}^{(2)}] = 0 &\Rightarrow & 
  \frac{\alpha}{\beta} 
  = -\frac{\delta E_y[\delta\boldsymbol{x}^{(2)}]}{\delta
    E_y[\delta\boldsymbol{x}^{(1)}]}\;. 
\end{eqnarray*}
From Eqs. (\ref{eq:18}), (\ref{eq:B6}) we deduce the formula
\begin{equation}
  \label{eq:B7}
  D' =  \mathrm{Im}\, \dot{\rho}_y(t_f) =
  \mathrm{Im}\, {\alpha} = \frac{\mathrm{Im}\, \left(
      \delta E_y[\delta\boldsymbol{x}^{(1)}] \cdot\delta
      E_y^*[\delta\boldsymbol{x}^{(2)}]\right)}{|\delta 
    E_y[\delta\boldsymbol{x}^{(1)}]|^2}\;.
\end{equation}
Substituting expressions (\ref{eq:B5}) and (\ref{eq:B7}) into
Eq. (\ref{eq:17}), one obtains Eq.~(\ref{eq:21}).

We finally rewrite the Cauchy data (\ref{eq:22}), (\ref{eq:23}) for
perturbations in canonically covariant form. The new
conditions are:

(i) perturbations $\delta \boldsymbol{x}^{(n)}$ and their momenta
$\delta \dot{\boldsymbol{x}}^{(n)}$ are real at $t = t_f$; 

(ii) they do not perturb the total energy, $\delta E[\delta
  \boldsymbol{x}^{(n)}]=0$;

(iii) their norm is fixed by $\Omega(\delta
    \boldsymbol{x}^{(1)},\delta \boldsymbol{x}^{(2)}) = 1$,  where
    $\Omega = dp_x \wedge dx + dp_y \wedge dy$ is the symplectic form.

Note that the conditions (i)---(iii) do not completely fix the perturbations
$\delta \boldsymbol{x}^{(n)}$. Namely, a linear admixture
of the perturbation $\boldsymbol{\chi}(t) = \dot{\boldsymbol{x}}(t)$, if
added to $\delta \boldsymbol{x}^{(n)}$, does not disturb (i)---(iii)
and the value of the prefactor $A_{pot}$. One shows this exploiting the 
properties of $\boldsymbol{\chi}$, 
$$
\delta E[\boldsymbol{\chi}] = \delta E_y[\boldsymbol{\chi}] = 0\;,
\qquad\qquad \Omega(\delta
\boldsymbol{x}^{(1)},\boldsymbol{\chi}) = \Omega(\delta
\boldsymbol{x}^{(2)},\boldsymbol{\chi}) = 0\;.
$$
In practice one fixes the $\boldsymbol{\chi}$--degeneracy by supplying
some additional Cauchy datum.

It is straightforward to check that the conditions (i)---(iii)
follow from Eqs. (\ref{eq:22}), (\ref{eq:23}). In fact, they are also
{\it sufficient}, i.e. any pair of perturbations $\delta
\boldsymbol{x}^{(1)}$ and $\delta \boldsymbol{x}^{(2)}$  satisfying
(i)---(iii) can be used in the prefactor formula~(\ref{eq:21}).  
One shows this explicitly by decomposing the new perturbations
$\delta \boldsymbol{x}^{(n)}$ in the basis of the old perturbations,
substituting them into Eq.~(\ref{eq:21}) and using the properties
(i)---(iii). 

Let us note the following property of the formula (\ref{eq:21}) for
$A_{pot}$. The perturbations
$\delta \boldsymbol{x}^{(n)}(t)$ leave unchanged all boundary
conditions of the complex trajectory except for one, fixation of
$E_y$. It is precisely the change in this datum which enters the
formula for the prefactor. In the limit of separable system, when $E_y$
becomes a conserved quantity, $\delta E_y[\delta\boldsymbol{x}^{(n)}]$
becomes real due to the condition (i) and 
$A_{pot}$ tends to infinity. This indicates the change of the
$\hbar$-dependence of the prefactor in this limit,
cf. Refs.~\cite{Wilkinson:1986,Creagh:2001,Creagh:2006}.

\section{Numerical method for semiclassical calculations}
\label{sec:numerical-method}
Here we describe the numerical method of finding the trajectories
$\boldsymbol{x}(t)$. The method has 
two useful properties. First, it is applicable to systems with
arbitrary number of degrees of freedom, up to the field theory case 
(${\cal N} =
\infty$). Second, it naturally incorporates the modified semiclassical
technique of Sec.~\ref{sec:unst-traj}. Originally, the method was
proposed in the field theoretical
context~\cite{Kuznetsov:1997az,Bezrukov:2003er}. It was adapted for
quantum mechanical problems in Ref.~\cite{Bonini:1999kj}.

We compute $\boldsymbol{x}(t)$ by solving numerically the classical
equations of motion with the boundary conditions (\ref{eq:6}) and
(\ref{eq:20}) imposed at $t = t_i$ and $t = t_f$ respectively. 
To this end we introduce non--uniform lattice $\{t_k,\; k=1\dots
N_k\}$, where $t_1 = t_i$, $t_{N_k} = t_f$, and discretize the classical
equations of motion and boundary conditions in a straightforward
manner. We find that the second--order discretization works well enough.
As mentioned in appendix~\ref{sec:eval-pre-expon}, the time variable
$t$ runs along the contour in complex time plane, see 
Fig.~\ref{fig:5}. Accordingly, the sites $t_k$ of the lattice belong
to this contour. 

After discretization one obtains the system of $2\times N_k$
complex non--linear algebraic equations for the same number of unknowns
$\boldsymbol{x}_k = \boldsymbol{x}(t_k)$. Let us denote the unknowns
collectively by $z_a$, where $a=1\dots 2N_k$, and equations by ${\cal
  F}_b(z) = 0$. We solve the equations by the Newton-Raphson iterative
method, see 
e.g. Ref.~\cite{NumericalRecipes}. In this method one starts with some
approximation $z = z^{(0)}$ for the solution. Then the approximation
is repeatedly refined by finding corrections $\Delta z$ from the
system of linear equations 
$$
  {\cal F}_a(z^{(0)} + \Delta z) \approx \sum_{b}\frac{\partial {\cal
      F}_a}{\partial 
    z_b}(z^{(0)})\, \Delta z_b  + {\cal F}_a(z^{(0)}) = 0\;. 
$$
Note that the coefficient matrix of this system is
block three--diagonal and can be inverted efficiently. 
At the end of each iteration one redefines the approximation,
$z^{(0)} \to z^{(0)} + \Delta z$. After $3-6$ iterations the method
converges provided the original approximation was good enough.

The drawback of the Newton--Raphson method is the
small radius of convergence. Namely, the method does not produce
correct solution unless the approximation $z^{(0)}$ is
sufficiently close to it. We solve this difficulty in the
following  way. Suppose the solution is known for some values
$(E,\, E_y)$ of the in-state quantum numbers. Then, one finds solution at
$(E + \Delta E, \, E_y + \Delta E_y)$ by the Newton--Raphson
iterations, with the original solution at $(E,\, E_y)$ serving as the
initial approximation. Using this approach one walks in the 
$(E,\, E_y)$ plane by changing the values of quantum
numbers in small steps and finding the respective solutions. Moreover,
in this method one can gradually change any 
parameter of the problem including the regularization  
parameter $\epsilon$ of Sec.~\ref{sec:unst-traj}. 

The final ingredient of our numerical procedure is the method of
finding the semiclassical solution at some special values of $E$ and
$E_y$. We start the procedure by computing the instanton
trajectory which describes tunneling at $E = E_y = 0$. This 
trajectory is real in  Euclidean time and can be obtained by
minimization of Euclidean action\footnote{Say, with the algorithm of
  conjugate gradients~\cite{NumericalRecipes}.}. After finding the
Euclidean  instanton, one bends the time contour in a way shown in
Fig.~\ref{fig:5} and finds the in- and out- parts of the trajectory by
solving the Cauchy problem from points B and C of the contour.

Note that there are several methods of starting the numerical
procedure. In particular, a greater class of Euclidean solutions
(periodic instantons) can be used for that purpose, see
Ref.~\cite{Bonini:1999kj}. In the cases when neither instanton nor
periodic instantons exist one can exploit classical over--barrier
solutions \cite{Levkov:2007e}.

\section{Saddle--point integrals in the modified method}
\label{sec:saddle-integr-modif}
The integral over $\epsilon$ in Eq. (\ref{eq:2}) is
evaluated as follows. One notes that by construction the trajectory
$\boldsymbol{x}_{\epsilon} (t)$ extremizes 
the functional $S_\epsilon + B_i$. Thus,
\begin{equation}
\label{eq:33}
\frac{d}{d\epsilon} (S_{\epsilon} [\boldsymbol{x}_\epsilon] +
B_i[\boldsymbol{x}_\epsilon])  = \frac{\delta (S_\epsilon+B_i)}{\delta
  \boldsymbol{x}_{\epsilon}} \cdot
\frac{d\boldsymbol{x}_{\epsilon}}{d\epsilon} +  
\frac{\partial(S_\epsilon+B_i)}{\partial\epsilon}
 = i \,T_{int} [\boldsymbol{x}_{\epsilon}]\;,
\end{equation}
where in the second equality we used 
$\delta (S_\epsilon+B_i)/\delta
  \boldsymbol{x}_{\epsilon}=0$.
Using this relation, one finds that the saddle point of the leading exponent in
Eq. (\ref{eq:2}) with respect to $\epsilon$ is
achieved when 
\begin{equation}
\label{eq:46}
T_{int}[\boldsymbol{x}_\epsilon] = \tau\;.
\end{equation}
The result for the final wave function is
\begin{equation}
\label{eq:35}
\Psi_f(\boldsymbol{x}_f) =  \int_0^{+\infty}
\frac{d\tau}{\sqrt{2\pi \hbar D_\epsilon}} \sqrt{-\frac{d\epsilon}{d\tau}}
\cdot  \mathrm{e}^{ i(S_\epsilon
  [\boldsymbol{x}_\epsilon] + B_i[\boldsymbol{x}_\epsilon] - i\epsilon\tau)/\hbar +
  i\pi/4} \;. 
\end{equation}
Note that the saddle--point value of $\epsilon$ does not need to be purely
imaginary. 
 
The probability formula (\ref{eq:19}) involves, besides $\Psi_f$, the
complex conjugate out-state $\Psi_f^*$. One derives the analog of
Eq.~(\ref{eq:35}) for $\Psi^*_f$ considering the
path integral, which is  complex conjugate to 
Eq.~(\ref{eq:7}). We substitute expressions for
$\Psi_f$ and $\Psi_f^*$ into Eq.~(\ref{eq:19}) and obtain,
\begin{equation}
  \label{eq:36}
  {\cal P} = \int dy_f \int_0^{+\infty} \frac{ d\tau d\tau' \
    \,\dot{x}_f} 
  {2\pi\hbar\sqrt{D_\epsilon  D_{-\epsilon'}}} 
  \sqrt{\frac{d\epsilon}{d\tau}\frac{d\epsilon'}{d\tau'}}
  \,\cdot \mathrm{e}^{ i(S_{\epsilon}[\boldsymbol{x}_\epsilon] + B_i
    [\boldsymbol{x}_\epsilon] -i\epsilon\tau -
    S_{-\epsilon'}[\boldsymbol{x}_{-\epsilon'}] - 
    B_i[\boldsymbol{x}_{-\epsilon'}]  - i{\epsilon'}
    \tau')/\hbar} \;, 
\end{equation}
where the integral over $\tau'$ comes from the conjugate
out-state. Note the opposite signs of $\epsilon$ and $\epsilon'$ in
Eq.~(\ref{eq:36}); the difference is related to the fact that the
modified action (\ref{eq:5}) depends on the combination $i\epsilon$
which changes the sign under complex conjugation. As a consequence,
the saddle--point 
condition for $\epsilon'$ reads
$T_{int}[\boldsymbol{x}_{-\epsilon'}] = \tau'$, cf. Eq.~(\ref{eq:46}).

The integral over $y_f$ in Eq.~(\ref{eq:36}) is evaluated in the same
way as in appendix~\ref{sec:eval-pre-expon}.
Below we consider the integrals with respect to the interaction times
$\tau$ and 
$\tau'$. One changes the integration variables to 
$\tau_- = \tau - \tau'$ and $\tau_+ = (\tau+\tau')/2$.  
We noted in the main body of the paper that 
fixing $\tau_+$ one stabilizes both trajectories
$\boldsymbol{x}_{\epsilon}$ and
$\boldsymbol{x}_{\epsilon'}$. The integral over $\tau_-$ is
taken by the saddle--point method. One finds the extremum of
the leading exponent with respect to $\tau_-$,
\begin{equation}
\label{eq:42}
\epsilon' = \epsilon\;,
\end{equation}
where the relation (\ref{eq:33}) was used. 
Note that after the integration over $\tau_-$ the value of
$\epsilon$ is defined by the implicit relation
\begin{equation}
\label{eq:27}
T_{int}[\boldsymbol{x}_\epsilon] +
T_{int}[\boldsymbol{x}_{-\epsilon}] = 2\tau_+\;.
\end{equation}
One finds that for real $\tau_+$ the
solution $\epsilon = \epsilon(\tau_+)$ of this equation
is real. To show this we assume that the complex
trajectory $x_\epsilon(t)$ is unique. Then, it is straightforward to
check that the semiclassical equations following
from $S_{\epsilon}$ imply\footnote{If the trajectory is not unique, 
 the relation $\boldsymbol{x}_{\epsilon}^* =
 \boldsymbol{x}_{-\epsilon^*}$  is no longer valid for the terms in
 the tunneling probability which account for the 
interference between different trajectories. 
We do not consider interference effects in  the
present paper.}
that $\boldsymbol{x}_{\epsilon}^* =
\boldsymbol{x}_{-\epsilon^*}$. Therefore the l.h.s. of
Eq. (\ref{eq:27}) is real for real $\epsilon$, and so is the function
$\tau_+(\epsilon)$. This entails the
reality of the inverse function $\epsilon(\tau_+)$. The condition
$\epsilon=\epsilon^*$ and Eq.~(\ref{eq:27}) are 
equivalent to Eqs.~(\ref{eq:10}) of Sec.~\ref{sec:unst-traj}. 
The result of integration over $\tau_-$ is given in the main body of
the paper, Eq.~(\ref{eq:44}). 

\section{Evolution near the sphaleron}
\label{sec:linear}
In this appendix we study the evolution of the system in the vicinity
of the sphaleron orbit. The aim of this analysis is to extract the
behavior of the suppression exponent and prefactor in the regime when
the tunneling trajectory spends a long time near the sphaleron. This
is the case for the (regularized) 
inclusive trajectories at $\epsilon\ll 1$ (Sec.~\ref{sec:unst-traj})
and exclusive trajectories with large topological numbers $m$
(Sec.~\ref{sec:nf}).

Let us start with the limit $\epsilon\to +0$ in the modified
expressions (\ref{eq:14}). We work in the approximation of small sphaleron
amplitude. Though this
approximation is justified only at energies slightly exceeding the
minimum height $V_0$ of the potential barrier, we believe that the
qualitative features of $F_{\epsilon}$ and $A_{pot,\, \epsilon}$
remain the same at higher energies. 

In small vicinity of the saddle point the potential
is approximated by
$$
V(x,\,y) = V_0 + (\omega_+^2 x_+^2 - \omega_-^2  x_-^2)/2 + O(x^3)\;,
$$ 
where the Cartesian coordinates $x_+$  and $x_-$ run along the stable and
unstable directions of the potential, while $\omega_{\pm}$ represent
the respective frequencies.\footnote{In the model (\ref{eq:24}) $V_0 =
  1$,   $\omega_{\pm}^2 = \pm (\omega^2/2 -1)  + \sqrt{\omega^4/4 +
    1}$ and the coordinates $(x_+,\, x_-)$ are rotated with respect
  to $(x,\, y)$ by the angle $\alpha =
  \frac12\mathrm{arcctg}(\omega^2/2)$.} The 
sphaleron orbit describes periodic oscillations along $x_+$,
$$
x^{sph}_+(t) = a_+ \cos(\omega_+ t + \varphi_+)\;, \qquad \qquad
x^{sph}_- (t) = 0\;,
$$
where $a_+$ is related to the sphaleron energy. 

Now consider the modified complex trajectory $\boldsymbol{x}_\epsilon
(t)$. At small $\epsilon$ it has two 
distinctive parts corresponding to the two stages of the tunneling process. 
First, the trajectory arrives into the vicinity of the sphaleron
orbit. Second, it leaves the sphaleron and evolves into the out-region. In
the vicinity of the saddle point one writes
\begin{equation}
\label{eq:25}
x_{+,\epsilon}(t) = x^{sph}_{+}(t)\;, \qquad \qquad 
x_{-,\epsilon}(t) = a_- \mathrm{e}^{-\omega_-t} +
\epsilon \tilde{a}_- \mathrm{e}^{+\omega_- t} \;.
\end{equation}
In writing down Eq.~(\ref{eq:25}) we took into account two
facts. First, at $\epsilon=0$ the exponentially growing term in the
equation for $x_{-,\epsilon}$ vanishes and the trajectory 
 stays forever in the vicinity of the sphaleron. Second, this term is
 proportional to $\epsilon$ due to the linear dependence of
the modified equations of motion on $\epsilon$. At small $\epsilon$
Eq.~(\ref{eq:25}) describes long intermediate stage of modified
evolution near the sphaleron.

The exponentially growing term in Eq. (\ref{eq:25})
destroys the sphaleron orbit within the time interval
\begin{equation}
\label{eq:tint}
\tau_+ = -\frac{1}{\omega_{-}}\mathrm{ln} \,\epsilon + O(1)\;.
\end{equation}
Using the Legendre transformation (\ref{eq:11}) we find,
$$
F_\epsilon = F_{sph} + \int_{\tau_+}^{+\infty}d\tau_+ \,
2\epsilon(\tau_+) = F_{sph} + \frac{2\epsilon}{\omega_{-}} +
O(\epsilon^2)\;. 
$$ 
Therefore, $F_{\epsilon}$ tends linearly to its limiting value.

Consider now the limit $\epsilon\to +0$ of the prefactor. One finds
$A_{pot,\,\epsilon}$ by considering linear perturbations $\delta
\boldsymbol{x}^{(n)}(t)$ in the background of the modified
trajectory. Namely, one fixes the Cauchy data for $\delta
\boldsymbol{x}^{(n)}$ at $t = t_f$ and evolves them backwards in
time. As one reaches the stage of near--sphaleron evolution, the
perturbations start growing: they contain the part
$\boldsymbol{\eta}(t) \sim \mathrm{e}^{-\omega_- t}$ which grows
exponentially as $t$ decreases. One writes,
\begin{equation}
\label{eq:45}
\delta \boldsymbol{x}^{(n)}(t) = d^{(n)} \cdot \boldsymbol{\eta}(t) +
\delta \boldsymbol{x}^{(n)}_{reg} (t)\;,
\end{equation}
where the last term stay bounded as $t$ decreases. The coefficients
$d^{(n)}$ are real due to the final Cauchy data. Consider now the
perturbations 
$\delta \boldsymbol{x}^{(n)}(t)$ at $t = t_i$. One observes that the
two terms in Eq.~(\ref{eq:45}) behave differently in the limit
$\epsilon\to 
+0$. Namely, $\boldsymbol{\eta}(t_i) \sim \mathrm{e}^{\omega_- \tau_+}
\sim O(1/\epsilon)$, while the second term is finite, $\delta
\boldsymbol{x}^{(n)}_{reg} (t_i) \sim O(1)$. Using this dependence, on
finds from Eq.~(\ref{eq:21}) that $A_{pot,\,\epsilon} \sim
O(\epsilon^{1/2})$. This fact and Eq.~(\ref{eq:tint})
imply that the limit (\ref{eq:14b}) for $A_{sph}$ exists.

Our next goal is to prove Eq.~(\ref{eq:appr}). In what follows we drop
the assumption of small sphaleron amplitude. Consider the
sphaleron orbit $\boldsymbol{x}_{sph}(t)$. This orbit is
a periodic solution of the equations of motion with period $T_{sph}$, 
$\boldsymbol{x}_{sph}(t+T_{sph})=\boldsymbol{x}_{sph}(t)$.  
It is completely specified by two parameters, the total energy $E$ and
time origin $t_0$. A small perturbation $\delta\boldsymbol{x}$ around
$\boldsymbol{x}_{sph}$ satisfies the equation
\begin{equation}
\label{eq:pert}
\delta\ddot{\boldsymbol{x}}+V''(\boldsymbol{x}_{sph})
\delta\boldsymbol{x}=0\;.
\end{equation} 
This equation has two obvious solutions, 
$\delta\boldsymbol{x}_1=\d \boldsymbol{x}_{sph}/\d E$, 
$\delta\boldsymbol{x}_2=\dot{\boldsymbol{x}}_{sph}$; these are the
derivatives of the sphaleron orbit with respect to its two
parameters. 
The two remaining solutions of Eq.~(\ref{eq:pert}) describe the
formation and decay of the sphaleron. According to Floquet theorem, 
\begin{equation}
\label{eq:pert+-}
\delta\boldsymbol{x}_-(t)=\delta\tilde{\boldsymbol{x}}_-(t)
\e^{-\tilde\beta t}~,~~~
\delta\boldsymbol{x}_+(t)=\delta\tilde{\boldsymbol{x}}_+(t)
\e^{\tilde\beta t}\;,
\end{equation}
where $\delta\tilde{\boldsymbol{x}}_-(t)$, 
$\delta\tilde{\boldsymbol{x}}_+(t)$ are periodic functions with period
$T_{sph}$ and $\tilde{\beta}$ is the Lyapunov exponent. In what
follows we restrict our attention to the perturbations 
(\ref{eq:pert+-}): we omit the mode $\delta\boldsymbol{x}_1$ because
we are interested in
perturbations preserving the total
energy; the mode $\delta\boldsymbol{x}_2$ is removed by the 
trivial time shift.

The tunneling trajectory $\boldsymbol{x}(t)$ with energy $E$ 
has the following form in the vicinity of the sphaleron,
\[
\boldsymbol{x}(t)=\boldsymbol{x}_{sph}(t)+
C_-\,\delta\boldsymbol{x}_-(t)+C_+\,\delta\boldsymbol{x}_+(t)\;.
\]
The coefficients $C_-$, $C_+$ completely parameterize the
trajectory. Varying these coefficients one goes over the
possible values of the initial and final oscillator energies, $E_y$
and $E_y^f$. Let us consider the suppression $F$ calculated on the
tunneling trajectory as a function of $E_y$, $C_+$. The choice $C_+=0$
corresponds to the trajectory which stays at the sphaleron forever. 
Clearly,
$F(E_y,C_+=0)=F_{sph}$. At small but non-zero values of $C_+$ one has,
\begin{equation}
\label{eq:linear}
F=F_{sph}+C_+\cdot\left.\frac{\d F}{\d C_+}\right|_{E_y=const}  +
O(C_+^2)\;. 
\end{equation} 
Consider now two exclusive trajectories which have the same
$E_y^f$ and the topological numbers\footnote{We restrict our attention
  to  the trajectories of the first class in the terminology of
  Sec.~\ref{sec:nf}.} $m$ and $m+2$. 
As discussed in Sec.~\ref{sec:nf}, the $(m+2)$-th trajectory performs
one additional oscillation in the vicinity of the sphaleron orbit as
compared to the
$m$-th trajectory. This implies that the coefficients $C^{(m+2)}_+$
and $C^{(m)}_+$ corresponding to these trajectories are related by
(see Eq.~(\ref{eq:pert+-}))
\begin{equation}
\label{eq:53}
C^{(m+2)}_+=\e^{-\tilde\beta T_{sph}}C^{(m)}_+\;.
\end{equation}
Note that the Lyapunov exponent $\tilde{\beta}$ in this expression
depends on the properties of the sphaleron orbit only; in particular,
it is independent of the final oscillator energy. 
Substituting the relation~(\ref{eq:53}) into Eq.~(\ref{eq:linear})
yields Eq.~(\ref{eq:appr}). 

Using the results of this appendix, one estimates the parameter
$\beta$ in Eq.~(\ref{eq:appr}): $\beta = T_{sph} \tilde{\beta} \approx
2\pi \omega_-/\omega_+$. In the model (\ref{eq:24}) $\beta \approx
24$. 

\section{Implementation of the uniform formula}
\label{sec:calc-unifrom}
The quantities entering the uniform correction factors~(\ref{eq:37})
are computed as follows. 

\begin{figure}[htb]
\centerline{\includegraphics[width=0.5\textwidth]{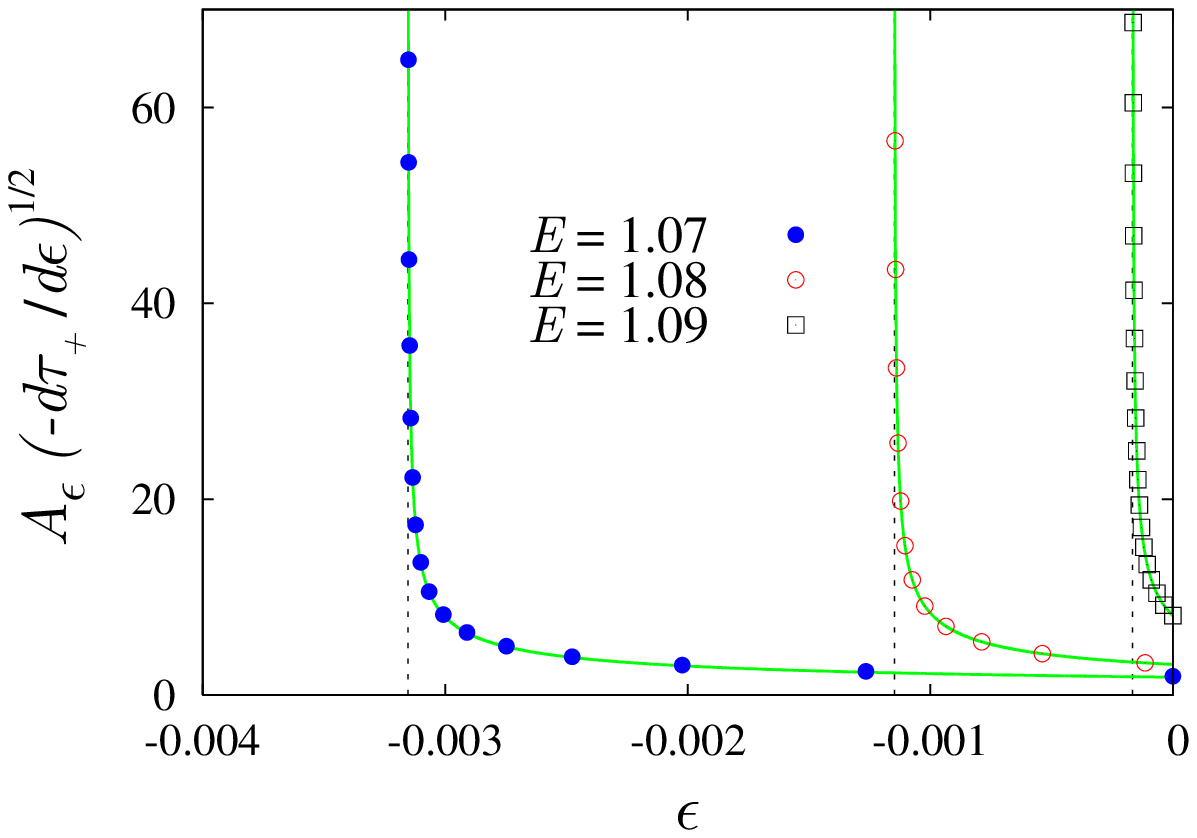}
\includegraphics[width=0.5\textwidth]{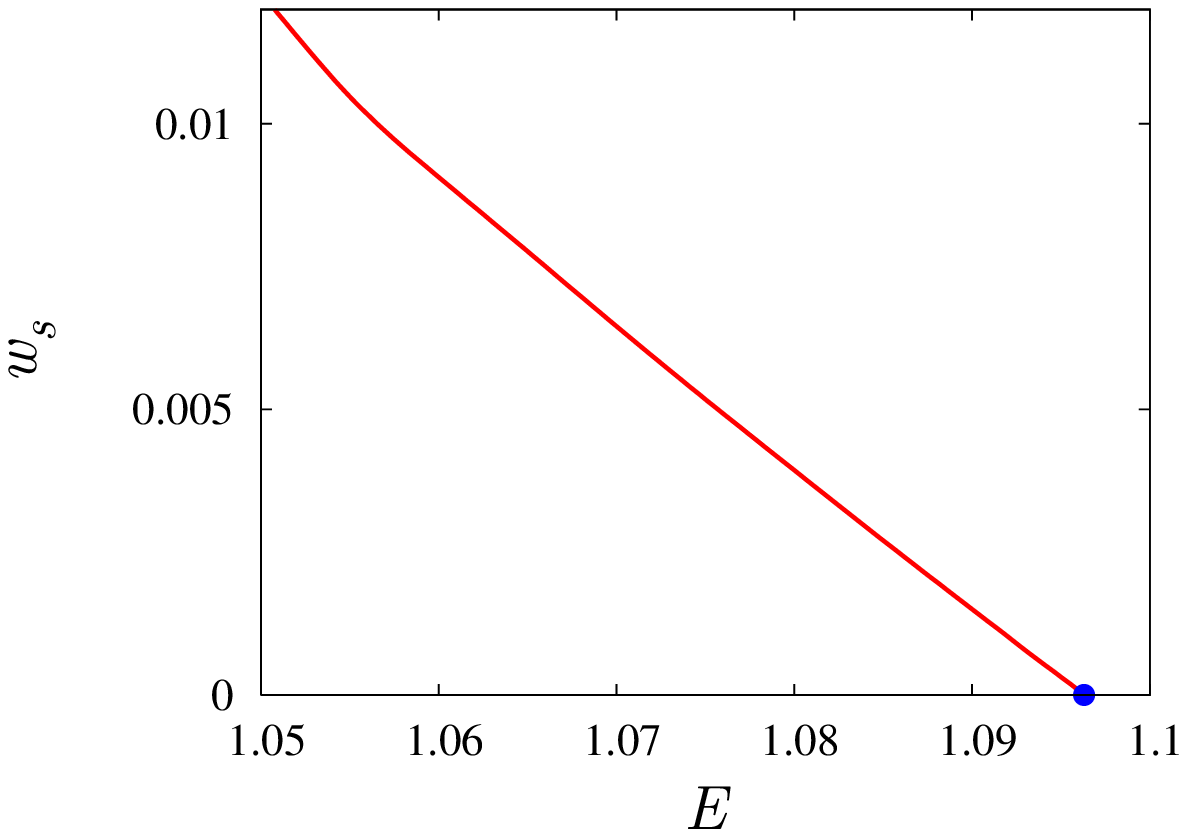}}
\hspace{4.2cm}(a)\hspace{8.2cm}(b)
\caption{\label{fig:11} (a) The integrand in Eq. (\ref{eq:60}) plotted as 
  function of $\epsilon$ for several values of total energy
  $E<E_c(E_y)$ and $E_y = 0.05$. The limiting values $\epsilon =
  \epsilon_i$ are shown by the vertical dotted lines. (b) The saddle
  point $w_s(E)$ at $E_y = 0.05$. } 
\end{figure}

Consider first ${\cal M}_{pot}$. One obtains the saddle--point
value $w_s$ by taking the integral in Eq.~(\ref{eq:16}) where it
is convenient to change the integration variable to $\epsilon$,
\begin{equation}
\label{eq:60}
w_s = \frac{1}{\sqrt{\pi\hbar}} \int_{\epsilon_i}^{0} d\epsilon' \,
 \sqrt{-d\tau_+'/d\epsilon'} \cdot A_{pot,\epsilon'}\;.
\end{equation}
Here $\epsilon = \epsilon_i<0$  corresponds to  $\tau_+ = +\infty$
and we used the fact that the saddle point is achieved at
$\epsilon=0$. Note that the integrand $I(\epsilon) =
\sqrt{-d\tau_+/d\epsilon} \cdot A_{pot,\epsilon}$ of Eq.~(\ref{eq:60}) is
singular at $\epsilon=\epsilon_i$, since it contains the derivative
of $\tau_+$ in the nominator. We plot this integrand (points in
Fig.~\ref{fig:11}a) for several values of energy and $E_y = 0.05$. 
We find that the function $I(\epsilon)$ is well fitted by the formula
$I(\epsilon) \approx A/(\epsilon-\epsilon_i)^{1/2}+B$ (lines in
Fig.~\ref{fig:11}a). We exploited this fact in the numerical
computation of the integral in Eq. (\ref{eq:60}). The relative
numerical error of $w_s$ was always smaller than  $10^{-3}$. 
The result for the saddle point $w=w_s(E)$ is plotted in
Fig.~\ref{fig:11}b. As expected, $w_s$ is positive,
decreases with energy and reaches $w_s = 0$ at $E = E_c(E_y) \approx
1.096$.

The second derivative of the suppression exponent entering
Eq.~(\ref{eq:37a}) is computed using the formula
\begin{equation}
\label{eq:61}
F''(w_s) = \frac{2\pi\hbar}{A_{pot}^2}\;,
\end{equation}
which follows from the definition of $w$, Eq.~(\ref{eq:16}). 
Using $w_s$ and $F''(w_s)$, one finds the argument $\varkappa_{pot}$
of the Fresnel integral
 and thus ${\cal M}_{pot}$, see Eq.~(\ref{eq:37a}). 

Consider now ${\cal M}_{sph}$. From Eqs.~(\ref{eq:16}), (\ref{eq:11}) 
one derives in a straightforward way the
following expressions for the derivatives of the suppression 
exponent at $w=0$,
\begin{equation}
  \label{eq:63}
  F'(0) = \frac{\hbar}{A_{sph}} \;,
  \qquad\qquad
  F''(0) = \frac{1}{2} \frac{d}{dF_\epsilon} 
  \left[\frac{\hbar}{A_{sph,\epsilon}}\right]_{\epsilon = 0}^2\;.
\end{equation}
\begin{figure}[htb]
\centerline{\includegraphics[width=0.5\textwidth]{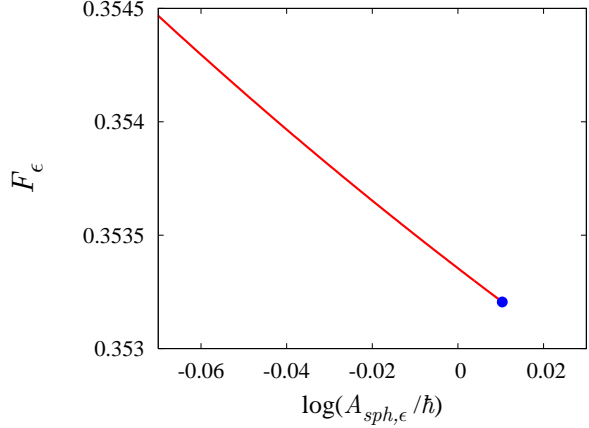}}
\caption{\label{fig:13} Suppression exponent $F_\epsilon$ as
  function of $\log A_{sph,\, \epsilon}$. The graph is plotted at
  $E = 1.12 > E_c(E_y)$, $E_y = 0.05$.}
\end{figure}
Here we denote by $A_{sph,\,\epsilon}$ the value of the r.h.s. of
Eq.~(\ref{eq:14b}) 
at finite $\epsilon$. Combining the multipliers (\ref{eq:63}),
one obtains for the variable entering Eq.~(\ref{eq:37b}),
\begin{equation}
\label{eq:64}
\varkappa^2_{sph} =  \left.
- \frac{1}{2\hbar} \cdot \frac{dF_\epsilon}{d\log A_{sph,\, \epsilon}}\right
|_{\epsilon = 0}\;.
\end{equation}
Note that $\varkappa_{sph}$ is real and positive. One finds it
by plotting $F_{\epsilon}$ as a function of $\log
A_{sph,\, \epsilon}$ at small $\epsilon$ (Fig.~\ref{fig:13}) and
fitting the graph with the linear function. 

\section{Semiclassical probability at small $E_y$} 
\label{sec:state-parameters}
We showed in Sec.~\ref{sec:limit-small-quantum} that the probability
of tunneling from the ground state of $y$-oscillator is given (up to
the overall factor $(\pi/\mathrm{e})^{1/2}$)
by the standard semiclassical expression, 
where one should use $E_y = \hbar
\omega/2$. Let us determine explicitly the $\hbar$--dependence of this
expression, i.e. extract the leading exponent and prefactor of the
probability.

First, we rewrite $F_{pot}$ using the parameters $T$ and
$\theta$. Namely, we evaluate the in-state term $B_i$ by 
taking the integral in Eq.~(\ref{eq:3}),
\begin{align*}
  2\mathrm{Im}\, B_i[\boldsymbol{x}] &=
  \mathrm{Im}\,(y_i \dot{y}_i +
  x_i \dot{x}_i) + \mathrm{Im}\,
  \left(x_i\dot{x}_i  + \left(2E_y/\omega\right)
    \mathrm{arccos}(\omega y_i/\sqrt{2E_y})\right) \\&=
  \mathrm{Im}(y_i \dot{y}_i +
  x_i \dot{x}_i) - 2ET - E_y \theta/\omega\;,
\end{align*}
where in the last equality the asymptotic form (\ref{eq:6.2}),
(\ref{eq:6.3})
 of
the trajectory was used. For the suppression exponent (\ref{eq:26}) one
obtains,
\begin{equation}
\label{eq:49}
F_{pot} = 2\mathrm{Im}\tilde{S}[\boldsymbol{x}]  - 2ET - 
E_y \theta/\omega\;,
\end{equation}
where 
\begin{equation}
\label{eq:48}
\tilde{S}[\boldsymbol{x}] = \int
dt\left[-\boldsymbol{x}\ddot{\boldsymbol{x}}/2 -
  V(\boldsymbol{x})\right]\; 
\end{equation}
is the classical action integrated by parts. 

The second observation is as follows. Consider the differential
of the action (\ref{eq:48}) with respect to the in-state quantum
numbers  $E$, $E_y$. One writes,
\begin{align*}
d2\mathrm{Im}\, \tilde{S} &= d\mathrm{Im}\, (2S + x_i
\dot{x}_i + y_i \dot{y}_i) = \mathrm{Im}\,( x_i d\dot{x}_i -
\dot{x}_i dx_i+ y_i d\dot{y}_i - \dot{y}_i dy_i
) \\&= 2 E dT + E_y d\theta/\omega\;.
\end{align*}
This equation together with Eq.~(\ref{eq:49}) show that the parameters
$(T,\, \theta)$ are related to $(E,\, E_y)$ by the Legendre
transformation. Consequently,
$$
dF_{pot} = -2TdE - \theta dE_y/\omega\;.
$$
Thus, $(-2T)$ and $(-\theta/\omega)$ are equal to the derivatives of
the suppression exponent with respect to $E$ and $E_y$. 

Using the above observation, we expand the suppression exponent around
the point $E_y = 0$,
$$
F_{pot}= F_{pot}\Big|_{E_y = 0} -
\frac{1}{\omega} \int_0^{E_y} \theta(E_y') \, dE_y' \;. 
$$
The function $\theta(E_y)$ at small $E_y$ is determined by noting that
\begin{equation}
\label{eq:6.4}
E_y = 2\omega^2 |\bar{a}|^2 \mathrm{e}^{ - 2\omega T - \theta}\;,
\qquad\Rightarrow \qquad
\theta = -\mathrm{ln} (2E_y/\omega) + \theta_0  + O(E_y)\;,
\end{equation}
where we used the fact that $|\bar{a}|$ has a well--defined limit as $E_y\to
0$. This yields the expression (\ref{eq:50}) for the suppression
exponent. 
Finally, substituting Eq.~(\ref{eq:50}) into
Eq.~(\ref{eq:9}) and recalling the additional factor
$(\pi/\mathrm{e})^{1/2}$ one arrives at the expressions (\ref{eq:6.5}). 




\begin{thebibliography}{99}
\bibitem{Creagh:1998}
  S.~C.\ Creagh, in {\it Tunneling in complex systems}, ed. by
  S.\ Tomsovic (World Scientific, Singapore, 1998).
\bibitem{Tomsovic:2001}
  S.\ Tomsovic, Physica Scripta {\bf T90}, 162 (2001).
\bibitem{Miller:2001}
  W.~H.\ Miller, J.\ Chem.\ Phys. {\bf 48}, 1651 (1968); E.~L.\ Sibert
  III, J.~T.\ Hynes and W.~P.\ Reinhardt, {\it ibid.} {\bf 77}, 3595
  (1982).
\bibitem{Meyer:1991}
  R.~E.\ Meyer, SIAM J.\ Appl.\ Math. {\bf 51}, 1585 (1991); {\it ibid.}
  {\bf 51}, 1602 (1991).
\bibitem{Creagh:1994}
  S.~C.\ Creagh, J.\ Phys.\ A {\bf 27}, 4969 (1994).
\bibitem{Wilkinson:1986}
  M.\ Wilkinson, Physica\ D {\bf 21}, 341 (1986);
  J.\ Phys.\ A {\bf 20}, 635 (1987).
\bibitem{Takada:1994}
  S.\ Takada and H.\ Nakamura, J.\ Chem.\ Phys. {\bf 100}, 98 (1994);
  S.\ Takada, P.~N.\ Walker and M.\ Wilkinson, Phys.\ Rev.\ A {\bf 52},
  3546 (1995); S.\ Takada, J.\ Chem.\ Phys. {\bf 104}, 3742 (1996).
\bibitem{Creagh:2001}
  S.~C.\ Creagh and M.~D.\ Finn, J.\ Phys.\ A {\bf 34}, 3791
  (2001). 
\bibitem{Creagh:2006}
  G.~C.\ Smith and S.~C.\ Creagh, J.\ Phys.\ A {\bf 39},
  8283 (2006).
\bibitem{Bohigas:1993}
  O.\ Bohigas, S.\ Tomsovic and D.\ Ullmo, Phys.\ Rept.\ {\bf 223}, 43
  (1993).
\bibitem{Doron:1995}
  E.\ Doron and S.~D.\ Frischat, Phys.\ Rev.\ Lett. {\bf 75}, 3661 (1995);
  S.~D.\ Frischat and E.\ Doron, Phys.\ Rev.\ E {\bf 57}, 1421 (1998).
\bibitem{Shudo:1995}
  A.\ Shudo and K.~S.\ Ikeda, Phys.\ Rev.\ Lett.\ {\bf 74}, 682 (1995); {\it
    ibid.} {\bf 76}, 4151 (1996); Physica\ D { \bf 115}, 234 (1998).
\bibitem{Creagh:1999}
  S.~C.\ Creagh and N.~D.\ Whelan, Phys.\ Rev.\ Lett. {\bf 77}, 4975
  (1996); {\it ibid.} {\bf 82}, 5237 (1999).
\bibitem{Mouchet:2001}
  A.\ Mouchet, C.\ Miniatura, R.\ Kaiser, B.\ Gr\'{e}maud and D.\ Delande,
  Phys.\ Rev.\ E {\bf 64}, 016221 (2001).
\bibitem{Ribeiro:2004}
  A.~D.\ Ribeiro, M.~A.~M.\ de\ Aguiar and M.\ Baranger, Phys.\ Rev.\ E
  {\bf 69}, 066204 (2004).
\bibitem{Levkov:2007e}
  D.~G.\ Levkov, A.~G.\ Panin and S.~M.\ Sibiryakov, Phys.\ Rev.\ E {\bf 76}
  046209 (2007).
\bibitem{Backer:2008}
  A.\ B\"acker, R.\ Ketzmerick, S.\ L\"ock and L.\ Schilling,
  Phys.\ Rev.\ Lett. {\bf 100}, 104101 (2008).
\bibitem{Dembowski:2000}
  C.\ Dembowski {\it et al}, Phys.\ Rev.\ Lett. {\bf 84}, 867 (2000);
  R.\ Hofferbert {\it et al}, Phys.\ Rev.\ E {\bf 71}, 046201 (2005).
\bibitem{Hensinger:2001}
  W.~K.\ Hensinger {\it et al}, Nature {\bf 412}, 52 (2001);
  W.~K.\ Hensinger {\it et al}, Phys.\ Rev.\ A {\bf 70}, 013408
  (2004).
\bibitem{Steck:2001}
  D.~A.\ Steck, W.~H.\ Oskay and M.~G.\ Raizen, Science {\bf 293}, 274
  (2001); Phys.\ Rev.\ Lett. {\bf 88}, 120406 (2002).
\bibitem{Backen:2008}
  A.\ B\"acker {\it et al}, Phys.\ Rev.\ Lett.\ {\bf 100}, 174103
  (2008).
\bibitem{Miller}
  W.~H.\ Miller, Adv.\ Chem.\ Phys. {\bf 25}, 69 (1974).
\bibitem{Heller:1981}
  M.~J.\ Davis and E.~J.\ Heller, J.\ Chem.\ Phys. {\bf 75}, 246 (1981);
  E.~J.\ Heller and M.~J.\ Davis, J.\ Phys.\ Chem. {\bf 85}, 307 (1981).
\bibitem{Onishi:2003}
  T.\ Onishi, A.\ Shudo, K.~S.\ Ikeda and K.\ Takahashi, Phys\ Rev.\ E
  {\bf 64}, 025201 (2001); {\it ibid} {\bf 68}, 056211 (2003).
\bibitem{Takahashi:Ikeda}
  K.~Takahashi and K.S.~Ikeda, J.\ Phys.\ A {\bf 36},
  7953 (2003); Europhys. \ Lett. \ {\bf 71}, 193 (2005); {\it
    erratum-ibid} {\bf 75}, 355 (2006).
\bibitem{Bezrukov:2003yf}
  F.~Bezrukov and D.~Levkov, arXiv:quant-ph/0301022;
  J.\ Exp.\ Theor.\ Phys.\  {\bf 98}, 820 (2004)
  [Zh.\ Eksp.\ Teor.\ Fiz.\  {\bf 125}, 938 (2004)].
  
\bibitem{Levkov:2007a}
  D.~G.\ Levkov, A.~G.\ Panin and S.~M.\ Sibiryakov, Phys. Rev. A {\bf
    76}, 032114 (2007).
\bibitem{Levkov:2007prl}
  D.~G.\ Levkov, A.~G.\ Panin and S.~M.\ Sibiryakov,
  Phys.\ Rev.\ Lett.\ {\bf 99}, 170407 (2007).
\bibitem{Takahashi:2006}
  K.\ Takahashi and K.~S.\ Ikeda, Phys.\ Rev.\ Lett. {\bf 97}, 240403
  (2006). 
\bibitem{Takahashi:2008}
  K.\ Takahashi and K.~S.\ Ikeda, J.\ Phys.\ A {\bf 41},
  095101 (2008).
\bibitem{Shudo:2008}
  A.\ Shudo, Y.\ Ishii and K.~S.\ Ikeda, Europhys.\ Lett. {\bf 81}, 50003
  (2008).
\bibitem{Bezrukov:2003er}
  F.\ Bezrukov, D.\ Levkov, C.\ Rebbi, V.\ Rubakov and P.\ Tinyakov,
  Phys.\ Rev.\ D {\bf 68}, 036005 (2003); Phys.\ Lett.\  B {\bf 574},
  75 (2003).
\bibitem{Levkov:2004tf}
  D.~G.\ Levkov and S.~M.\ Sibiryakov,
  Phys.\ Rev.\ D {\bf 71}, 025001 (2005); 
  JETP Lett.\  {\bf 81}, 53 (2005) [Pisma Zh.\ Eksp.\ Teor.\ Fiz.\
  {\bf 81}, 60 (2005)].
\bibitem{Klinkhamer:1984di}
  F.~R.~Klinkhamer and N.~S.~Manton, Phys.\ Rev.\ D {\bf 30}, 2212
  (1984).
\bibitem{Wiggins:2001}
  S.\ Wiggins, L.\ Wiesenfeld, C.\ Jaff\'e and T.\ Uzer,
  Phys.\ Rev.\ Lett. {\bf 86}, 5478 (2001).
\bibitem{Rubakov:1992ec}
  V.~A.\ Rubakov, D.~T.\ Son and P.~G.\ Tinyakov,
  Phys.\ Lett.\  B {\bf 287}, 342 (1992).
\bibitem{induced}
  M.~P.\ Mattis, Phys.\ Rept.\  {\bf 214}, 159 (1992);\\
  P.~G.\ Tinyakov, Int.\ J.\ Mod.\ Phys.\ A {\bf 8}, 1823 (1993);\\
  V.~A.\ Rubakov and M.~E.\ Shaposhnikov, Phys.\ Usp.\  {\bf 39}, 461
  (1996) [Usp.\ Fiz.\ Nauk {\bf 166}, 493 (1996)].
\bibitem{Bonini:1999kj}
  G.~F.~Bonini, A.~G.~Cohen, C.~Rebbi and V.~A.~Rubakov,
  Phys.\ Rev.\ D {\bf 60}, 076004 (1999); quant-ph/9901062.
\bibitem{f90code}
  http://solver.inr.ac.ru
\bibitem{Affleck:1980mp}
  I.~Affleck, Nucl.\ Phys.\ B {\bf 191}, 429 (1981).
\bibitem{Levkov:2008}
  D.~G.\ Levkov, A.~G.\ Panin, {\it to be published}.
\bibitem{Berry}
  M.~V.~Berry, K.~E.~Mount, Rept. Prog. Phys. {\bf 35}, 315 (1972).  
\bibitem{VanVleck}
  J.H.~Van Vleck, 
  Proc.\ Natl.\ Acad.\ Sci.\ USA {\bf 14}, 178 (1928).
\bibitem{Kleinert}
  H.~Kleinert, {\it Path Integrals in Quantum Mechanics, Statistics,
  Polymer Physics, and Financial Markets}, World Scientific,
  Singapore, 2006.
\bibitem{Kuznetsov:1997az}
  A.~N.\ Kuznetsov and P.~G.\ Tinyakov, Phys.\ Rev.\ D\ {\bf 56}, 1156
  (1997).
\bibitem{NumericalRecipes}
  W.~H.\ Press {\it et al}, {\it Numerical recipes in C: the art of
  scientific computing}, Cambridge University Press, 1992. 
\end{thebibliography}
\end{document}